\documentclass[twocolumn]{aastex62}
\usepackage{natbib, chngcntr}
\usepackage[version=3]{mhchem}
\usepackage{stmaryrd}
\usepackage{enumitem}
\renewcommand{\tabcolsep}{3.2pt}
\usepackage{mathtools, amsmath}
\usepackage{color}
\usepackage{hhline}
\usepackage{placeins}
\usepackage{graphicx}
\usepackage{physics}

\begin{document}

\title{Self-luminous and irradiated exoplanetary atmospheres explored with \texttt{HELIOS}}

\author{Matej Malik}
\affiliation{Center for Space and Habitability, University of Bern, Gesellschaftsstrasse 6, 3012, Bern, Switzerland}
\affiliation{Department of Astronomy, University of Maryland, College Park, MD 20742, USA}

\author{Daniel Kitzmann}
\affiliation{Center for Space and Habitability, University of Bern, Gesellschaftsstrasse 6, 3012, Bern, Switzerland}

\author{Jo\~{a}o M. Mendon\c{c}a}
\affiliation{Center for Space and Habitability, University of Bern, Gesellschaftsstrasse 6, 3012, Bern, Switzerland}
\affiliation{Astrophysics and Atmospheric Physics, National Space Institute, Technical University of Denmark, Elektrovej, 2800 Kgs. Lyngby, Denmark}

\author{Simon L. Grimm}
\affiliation{Center for Space and Habitability, University of Bern, Gesellschaftsstrasse 6, 3012, Bern, Switzerland}

\author{Gabriel-Dominique Marleau}
\affiliation{Physikalisches Institut, Universit\"{a}t Bern, Gesellschaftsstrasse 6, 3012 Bern, Switzerland}
\affiliation{Institut f\"{u}r Astronomie and Astrophysik, Eberhard Karls Universit\"{a}t T\"{u}bingen, Auf der Morgenstelle 10, 72076 T\"{u}bingen, Germany}

\author{Esther F. Linder}
\affiliation{Physikalisches Institut, Universit\"{a}t Bern, Gesellschaftsstrasse 6, 3012 Bern, Switzerland}

\author{Shang-Min Tsai}
\affiliation{Center for Space and Habitability, University of Bern, Gesellschaftsstrasse 6, 3012, Bern, Switzerland}
\affiliation{Atmospheric, Oceanic, and Planetary Physics Department, Clarendon Laboratory, University of Oxford, Sherrington Road, Oxford OX1 3PU, UK}

\author{Kevin Heng}
\affiliation{Center for Space and Habitability, University of Bern, Gesellschaftsstrasse 6, 3012, Bern, Switzerland}

\correspondingauthor{Matej Malik}
\email{malik@umd.edu}

\shorttitle{Exoplanetary atmospheres with \texttt{HELIOS}}
\shortauthors{M. Malik et al.}

\begin{abstract}
We present new methodological features and physical ingredients included in the 1D radiative transfer code \texttt{HELIOS}, improving the hemispheric two-stream formalism. We conduct a thorough inter-comparison survey with several established forward models, including \texttt{COOLTLUSTY}, \texttt{PHOENIX}, and find satisfactory consistency with their results. Then, we explore the impact of (i) different groups of opacity sources, (ii) a stellar path length adjustment, and (iii) a scattering correction on self-consistently calculated atmospheric temperatures and planetary emission spectra. First, we observe that temperature--pressure (T-P) profiles are very sensitive to the opacities included, with metal oxides, hydrides, the alkali atoms (and ionized hydrogen) playing an important role for the absorption of shortwave radiation (in very hot surroundings). Moreover, if these species are sufficiently abundant, they are likely to induce non-monotonic T-P profiles. Second, without the stellar path length adjustment, the incoming stellar flux is significantly underestimated for zenith angles above 80$^\circ$, which somewhat affects the upper atmospheric temperatures and the planetary emission. Third, the scattering correction improves the accuracy of the computation of the reflected stellar light by $\sim$ 10\%. We use \texttt{HELIOS} to calculate a grid of cloud-free atmospheres in radiative-convective equilibrium for self-luminous planets for a range of effective temperatures, surface gravities, metallicities, and C/O ratios, to be used by planetary evolution studies. Furthermore, we calculate dayside temperatures and secondary eclipse spectra for a sample of exoplanets for varying chemistry and heat redistribution. These results may be used to make predictions on the feasibility of atmospheric characterizations with future observations.

\end{abstract}


\section{Introduction}
\label{sec:intro}

With over 3500 confirmed exoplanet detections, we have a very large pool of objects at our finger tips, waiting to be characterized. This abundance of potential data poses a challenge for numerical climate suites to deliver not only accurately but also within a reasonable time. At the core of the pyramid of atmospheric models sits the one-dimensional radiative transfer forward code, generating physically self-consistent atmospheric temperatures in radiative-convective equilibrium and mock planetary spectra. The one-dimensional format makes it ideally suited to include physics in colorful detail and be more flexible in applications than models of higher dimensions. Hence, it fulfills the role of the work horse of exoplanetary atmospheric characterization. Although in past years pure forward models have given some way to data-driven retrieval suites (e.g., \citealt{line13b, lavie17}), which interpret data with the help of Bayesian statistics, forward models remain indispensable for providing accurate predictions for yet to be assessed atmospheres. Even more, we are perhaps seeing a new dawn of self-consistent radiative transfer modeling as they alone are able to provide the large training set of mock spectral data required by modern machine learning algorithms (e.g., \citealt{marquez18}). Currently, there are several radiative transfer code families used in the exoplanetary field, with one branch of the ancestry originating from cool star studies and the other one from planetary, solar system modeling. The stellar branch is populated by the \texttt{SAM2} code \citep{tsuji67, tsuji78, tsuji02} and the \texttt{PHOENIX} code \citep{allard95, hauschildt99b}. The latter was also adapted for cooler exoplanets and brown dwarfs \citep{barman05, lothringer18}, but is still being used to model stellar atmospheres \citep{husser13}. A branch of this code, named \texttt{DRIFT-PHOENIX}, was extended by a consistently integrated cloud model \citep{helling08}. From a similar heritage comes \texttt{COOLTLUSTY} \citep{hubeny03, sudarsky03}, which is an offshoot of the stellar code \texttt{TLUSTY} \citep{hubeny88, hubeny95}. The most prominent representative on the planetary side of the model tree is the Marley/Fortney code \citep{fortney05, fortney08}, which was originally developed for solar system studies \citep{mckay89, marley99} and modified for extrasolar applications \citep{marley96}. The \texttt{DISORT} codes \citep{stamnes88, meadows96, hamre13} were inherited from the Earth sciences. In recent years, the codes \texttt{ATMO} \citep{amundsen14}, \texttt{petitCODE} \citep{molliere15, molliere17}, \texttt{GENESIS} \citep{gandhi17}, \texttt{Exo-REM} \citep{baudino15} and \texttt{HELIOS} \citep{malik17} emerged, written explicitly to treat exoplanetary conditions.

In this study, we build on \texttt{HELIOS} and present a number of improvements to its methodology. We compare the code in its updated form with other forward models, and explore the importance of several physical and numerical radiative transfer ingredients. 

In addition to the main purpose of this work, which is to present and discuss \texttt{HELIOS}'s updated form, we include two additional packages that may be of use to the community. First, we continue the legacy of past efforts to provide self-consistently calculated atmospheres of self-luminous young planets \citep{burrows97, chabrier00, baraffe03, baraffe08, mordasini12}, which often utilize pre-calculated atmospheric models, such as the BT-Settl grid \citep{allard01, allard11, allard14}. The atmosphere, as the outermost boundary, determines the rate at which a young planet cools over time, gradually releasing its internal heat. Hence, it is imperative to connect the interior evolution model with the correct atmosphere to obtain an accurate description of the cooling process. The interior models and the atmospheres are calculated independently and stitched together at the radiative-convective boundary in the optically thick atmosphere or on top of the outermost layer of the convective interior model. The entropy, providing the connection to the formation process \citep{Marleau14}, in this zone determines which two models fit together. To this end, grids of atmospheres are calculated that commonly span a parameter space in effective temperature, surface gravity and metallicity. In addition to the entropy value at the lower atmospheric boundary, the other important output of atmospheric models is the emission spectrum, which helps assess the planet's observability at various stages during its evolution and gives information on the formation history \citep{spiegel12}.

Even though many pre-calculated model grids exist, continuous progress is being made regarding opacity and chemical data. We hope that our completely new, alternative atmosphere grid may be of use to the community.

Furthermore, we use our modeling machinery to provide the community with self-consistently calculated cloud-free atmospheres for a variety of irradiated planets of current interest, ranging from super Earths, sub-Neptunes, hot Jupiters to ultra-hot Jupiters. Using various C/O ratios, metallicities and heat redistribution efficiencies we generate a suite of predictive models for the dayside temperatures and secondary eclipse spectra for the planets explored.

All our numerical codes used in this work are open-source and publicly available at \url{github.com/exoclime}.

\section{New Features of \texttt{HELIOS}}
\label{sec:features}

In the past year, since its first appearance in \cite{malik17}, \texttt{HELIOS} has undergone a number of methodological improvements. In this section, we elucidate the main new features and sketch out the corresponding algebra and core equations. As an overview, \texttt{HELIOS} includes now the following features:

\begin{itemize}

\item{A {\bf direct irradiation beam} decouples the stellar flux from atmospheric thermal emission. It allows us to model an atmospheric column at a specific latitude and longitude, whereas before one was restricted to average hemispheric or global conditions.}

\item{{\bf Convective adjustment} may be applied to correct atmospheric layers which are convectively unstable. Convection is expected to take over as the main energy transport mechanism in the deep atmospheric layers.}

\item{The {\bf contribution function} for each wavelength bin is now available as output. This makes it possible to identify the wavelength-dependent photosphere.}

\item{A {\bf geometrical correction} to the stellar beam path length is included, which proves important for large zenith angles. Without this correction the attenuation of the stellar irradiation is overestimated in plane-parallel grids.}

\item{We add a {\bf correction factor to the two-stream equations} which improves the accuracy of the scattered flux in the hemispheric two-stream prescription.}

\item{We employ a vastly {\bf expanded list of opacity of sources}, including an ample mixture of infrared absorbers, metal and other hydrides, metal oxides, the alkali metals Na \& K, and  H$^-$.}

\end{itemize}

We now discuss each improvement in turn.

\subsection{Direct Irradiation Beam}

\subsubsection{Definitions}
\label{sec:definitions}

In \texttt{HELIOS}'s original design, the stellar shortwave flux and the planetary thermal flux were treated equally using the two-stream approximation \citep{malik17}. With only the two-stream fluxes it is not possible to set a stellar irradiation angle. Furthermore, the temperatures on the planetary dayside can only be adjusted {\it globally} by the $f$ heat redistribution parameter, which sets the efficiency of heat exchange between the day- and the nightside of the planet \citep{spiegel10}. However, it is not possible to model the {\it local} radiative balance for a given latitude and longitude. We account for these two insufficiencies by extending the two-stream equations to incorporate a direct stellar beam in addition to the two-stream flux expressions. This separation of stellar flux and planetary emission has been well-known in the planetary science literature \citep{toon89}, however, in contrast to earlier studies, we do not differentiate between shortwave and longwave radiation. Our flux expressions hold irrespective of the explored wavelength. 

In the following, we define the relevant physical quantities, which deviate slightly from the definitions given in \cite{heng18}. Nonetheless, the subsequent derivations and the final flux expressions are identical.

Let us start with the plane-parallel radiative transfer equation
\begin{equation}
\mu \frac{d I^{\rm tot}}{d \tau} = I^{\rm tot} - S ,
\end{equation}
which provides the change of the total intensity $I^{\rm tot}$ with optical depth $\tau$. The intensity generally depends on the zenith and azimuth angles $\theta$ and $\phi$, while $\tau$ simply measures the vertical change in optical depth, increasing in downward direction. By convention $\theta$ is measured from the upward pointing plane normal vector, which makes $\mu = \cos \theta$ negative for downwards and positive for upwards pointed radiation rays. The function $S$ incorporates all radiation sources that add to the intensity $I^{\rm tot}$. It writes \citep{chandrasekhar60, mihalas70, mihalas78}
\begin{equation}
S = (1 - \omega_0)B + \frac{\omega_0}{4\pi}\int_0^{4\pi} \mathcal{P}(\theta, \phi; \theta^\prime, \phi^\prime) I^{\rm tot}(\theta^\prime, \phi^\prime) ~d \Omega^\prime ,
\end{equation}
where the first term is the thermal blackbody emission with the Planck function $B$ and the single-scattering albedo $\omega_0$. The latter describes the relative strength of atmospheric extinction due to scattering only to the total extinction, which is scattering and absorption. The second term considers the incoming rays from direction ($\theta^\prime, \phi^\prime$) and calculates, by multiplying them with the scattering phase function $\mathcal{P}$,  whether they are scattered into the line of sight ($\theta, \phi$). The integral over all possible incoming angles provides the total addition of scattered intensity to the source function $S$.

In addition to our previous theoretical excursion in \cite{malik17} we divide here the total intensity
\begin{equation}
I^{\rm tot} = I^{\rm dir} + I^{\rm diff}
\end{equation}
into the direct stellar beam $I^{\rm dir}$ and the diffuse intensity $I^{\rm diff}$.\footnote{We use the label {\it diffuse} for radiation which has been either scattered or emitted by the planetary atmosphere.} The direct beam changes with the optical depth as
\begin{equation}
I^{\rm dir} = F_{\ast, {\rm TOA}} e^{\tau  / \mu} \delta(\mu - \mu_\ast) \delta(\phi - \phi_\ast) ,
\end{equation}
propagating on a straight line through the atmosphere downward along the stellar angles $\theta_\ast = \cos^{-1} \mu_\ast$ and $\phi_\ast$. 

The incoming stellar flux at the top of the atmosphere (TOA) is given by 
\begin{equation}
F_{\ast, {\rm TOA}} = \left(\frac{R_\ast}{a}\right)^2 F_\ast,
\end{equation}
where $R_\ast$ is the stellar radius, $a$ is the orbital distance, and $F_\ast$ is the stellar surface flux, which can be either approximated by the blackbody Planck function $B$ for the stellar temperature $T_\ast$, i.e., $F_\ast = \pi B(T_\ast)$, or a synthetic model spectrum, e.g., \texttt{PHOENIX} \citep{husser13} or Kurucz/ATLAS \citep{kurucz79, murphy04, munari05}. The flux associated with the direct beam is 
\begin{equation}
\begin{split}
\label{eq:Fdir}
F^{\rm dir} &\equiv - \int_0^{2\pi} \int_{-1}^1 \mu I^{\rm dir} ~d\mu ~d\phi , \\
		&= - \mu_\ast F_{\ast, {\rm TOA}} e^{\tau / \mu_\ast} ,
\end{split}
\end{equation}
also simultaneously leading to the upper boundary condition of $F_{\rm TOA, \downarrow} = - \mu_\ast F_{\ast, {\rm TOA}}$, where $\tau = 0$. The leading minus sign in eq.~(\ref{eq:Fdir}) is needed because we define the radiative flux to be always a positive quantity.

The diffuse up- and downward fluxes are defined as 
\begin{equation}
\begin{split}
\label{eq:Fdiffuparrow}
F^{\rm diff}_\uparrow &\equiv \int_0^{2\pi} \int_0^1 \mu I^{\rm diff} ~d\mu ~d\phi , \\
F^{\rm diff}_\downarrow	&\equiv - \int_0^{2\pi} \int_{-1}^0 \mu I^{\rm diff} ~d\mu ~d\phi,
\end{split}
\end{equation}
from which we immediately obtain the total up- and downward fluxes, $ F_\uparrow = F^{\rm diff}_\uparrow$ and $ F_\downarrow = F^{\rm diff}_\downarrow + F^{\rm dir}$. Note, that the planetary thermal emission is included in the diffuse component of the flux.

\subsubsection{Interface Flux Expressions}
\label{sec:fluxes}

For the in-depth derivation of the diffuse fluxes we refer the reader to \cite{heng18}. Here, we specifically present and write the diffuse flux expressions as included in \texttt{HELIOS}. From here on, we omit the superscript ``diff'' and the subscript $i$ for certain quantities, e.g., $\omega_0$, for better readability. The fluxes at the $i$-th interface in the staggered numerical grid (see \citealt{malik17}, Fig. 2) read
\begin{equation}
\begin{split}
\label{eq:Fiuparrow}
F_{i,\uparrow} =& \frac{1}{\chi}(\psi F_{i-1,\uparrow} - \xi F_{i,\downarrow} +  2\pi \epsilon{\mathcal B_\uparrow} + \frac{1}{\mu_\ast}{\mathcal I}_\uparrow) , \\
F_{i,\downarrow} =& \frac{1}{\chi}(\psi F_{i+1,\downarrow} - \xi F_{i,\uparrow} + 2\pi \epsilon{\mathcal B_\downarrow} + \frac{1}{\mu_\ast}{\mathcal I}_\downarrow)) ,
\end{split}
\end{equation}
where
\begin{equation}
\begin{split}
\label{eq:Buparrow}
{\mathcal B_\uparrow} &\equiv (\chi + \xi) B_i - \xi B_{i-1} + \frac{\epsilon}{1-\omega_0 g_0} (\chi - \xi - \psi) B^\prime , \\
{\mathcal I}_\uparrow &\equiv \psi{\mathcal G}_+ F^{\rm dir}_{i-1} - (\xi {\mathcal G}_- + \chi {\mathcal G}_+)F^{\rm dir}_ i , \\
{\mathcal B_\downarrow} &\equiv (\chi + \xi) B_i - \psi B_{i+1} +  \frac{\epsilon}{1-\omega_0 g_0} (\xi - \chi + \psi) B^\prime , \\
{\mathcal I}_\downarrow &\equiv \psi{\mathcal G}_- F^{\rm dir}_{i+1} - (\chi {\mathcal G}_- + \xi {\mathcal G}_+)F^{\rm dir}_ i , 
\end{split}
\end{equation}
and $B_i = B(T_i)$ is the Planck function evaluated for the temperature at interface $i$. Further, we use above
\begin{equation}
\begin{split}
\label{eq:chi}
\chi &\equiv \zeta^2_- \mathcal{T}^2 - \zeta^2_+ , \\
\xi &\equiv \zeta_+ \zeta_- (1-\mathcal{T}^2) , \\
\psi &\equiv (\zeta^2_- - \zeta^2_+) \mathcal{T} , \\
\mathcal{\zeta_\pm} &\equiv \frac{1}{2} \left[ 1 \pm \left(\frac{1 - \omega_0}{1 - \omega_0 g_0}\right)^{1/2} \right] , \\
\mathcal{G}_\pm &\equiv \frac{1}{2}\left[ {\mathcal L} \left( \frac{1}{\epsilon} \pm \frac{1}{\mu_\ast(1-\omega_0 g_0)} \right) \pm \frac{\omega_0 g_0 \mu_\ast}{1-\omega_0 g_0}   \right] , \\
{\mathcal L} &\equiv \frac{(1-\omega_0)(1-\omega_0 g_0) - 1}{1/\mu_\ast^2 - 1/\epsilon^2 (1-\omega_0) (1-\omega_0 g_0) } , \\
B^\prime &\equiv \frac{\Delta B}{\Delta\tau} ,
\end{split}
\end{equation}
where the last equation in (\ref{eq:chi}) shows our choice of expanding the Planck function linearly in optical depth, with $\Delta X$ representing the difference of any quantity $X$ across a layer. Setting $B^\prime = 0$ and $B_i = B_i+1$ (or $B_i = B_i-1$, resp.) in eqs. (\ref{eq:Buparrow}) translates into a model with isothermal layers. Equations (\ref{eq:Fiuparrow}) are valid for non-isotropic coherent scattering with the direction of scattering given by the asymmetry parameter $ g_0 \in [-1,1]$, as defined e.g., in \cite{goody89} or \cite{pierrehumbert10}. The transmission function is given as 
\begin{equation}
{\mathcal T} \equiv e^{-1/\epsilon \sqrt{(1 - \omega_0 g_0)(1 - \omega_0)} \Delta\tau} .
\end{equation}
We make use of the first Eddington coefficient $\epsilon$, which is defined as the assumed (constant) ratio between the first and second moments of the intensity, and is the inverse of the commonly known diffusivity parameter ${\mathcal D}$ \citep{armstrong68, heng14}. The latter effectively determines how diffuse the radiation behaves (${\mathcal D} = 2$: fully isotropic radiation, ${\mathcal D} = 1$: fully vertically oriented radiation). 

In the case of $\omega_0 = 1$ (pure scattering), equations (\ref{eq:Fiuparrow}) need to be replaced by
\begin{equation}
\begin{split}
F_{i, \uparrow} &= F_{i-1, \uparrow} + \frac{(F_{i, \downarrow} - F_{i-1, \uparrow})(1-g_0)\Delta\tau}{(1-g_0)\Delta\tau+2\epsilon} + {\mathcal J}_\uparrow , \\
F_{i, \downarrow} &= F_{i+1, \downarrow} + \frac{(F_{i, \uparrow} - F_{i+1, \downarrow})(1-g_0)\Delta\tau}{(1-g_0)\Delta\tau+2\epsilon} + {\mathcal J}_\downarrow ,
\end{split}
\end{equation}
where
\begin{equation}
\begin{split}
{\mathcal J}_\uparrow &\equiv \frac{1}{(1-g_0)\Delta\tau+2\epsilon}{\mathcal K}_\uparrow , \\
{\mathcal K}_\uparrow &\equiv \left[\mu_\ast + (1-g_0)\Delta\tau+\epsilon\right]F^{\rm dir}_i - (\mu_\ast+\epsilon)F^{\rm dir}_{i-1} , \\
{\mathcal J}_\downarrow &\equiv \frac{1}{(1-g_0)\Delta\tau+2\epsilon}{\mathcal K}_\downarrow , \\
{\mathcal K}_\downarrow &\equiv \left[\mu_\ast - (1-g_0)\Delta\tau-\epsilon\right]F^{\rm dir}_i - (\mu_\ast-\epsilon)F^{\rm dir}_{i+1} .
\end{split}
\end{equation}

Although in real atmospheres the limit $\omega_0 = 1$ is strictly academic, we include those expressions in the code for numerical stability reasons when $\omega_0 > 1 - \vert\epsilon\vert$, with $\vert\epsilon\vert \ll$ 1.

\subsection{Convective adjustment}
\label{sec:conv}

We account for atmospheric convection by adjusting super-adiabatic lapse rates back to the dry adiabatic lapse rate \citep{manabe65, manabe67}. In the model grid, we check for each pair of adjacent layers $i$ and $i+1$ (the latter being above the former) whether
\begin{equation}
 \label{eq:test}
T_{i+1} \geqslant T_i \left(\frac{P_{i+1}}{P_i}\right)^{\kappa}
\end{equation}
is satisfied, where $T$ is the temperature, $P$ is the pressure and $\kappa$ is the adiabatic coefficient, for which we use the definition from the planetary climate literature
\begin{equation}
\label{eq:kappadef}
\kappa \equiv \left(\frac{d \ln T}{d \ln P}\right)_S ,
\end{equation}
with $S$ being the entropy. In stellar astrophysics $\kappa$ corresponds to $\nabla_{\mathrm{ad}} \equiv (\Gamma_2 -1) / \Gamma_2$ with
\begin{equation}
\Gamma_2 \equiv \left[1 - \frac{P}{c_P \rho T}\frac{\chi_T}{\chi_\rho}\right]^{-1}
\end{equation}
where $c_P$ is the specific heat capacity, $\rho$ is the density, $\chi_T \equiv (\partial \ln P / \partial \ln T)_\rho$, and $\chi_P \equiv (\partial \ln P / \partial \ln \rho)_T$ \citep{hansen04}. Invoking condition (\ref{eq:test}) is equivalent to testing for Schwarzschild's criterion. If the condition is not satisfied, we correct the temperatures back to the adiabatic lapse rate.

To find the correct adiabat in the convective region we check for one additional criterion. As convection does not create or eliminate energy, we demand that the enthalpy, which is the total internal energy plus the work done to reach equilibrium with the surrounding medium, must be conserved within a convective region. In our one-dimensional case, the quantity of interest is the product of the specific enthalpy, i.e., enthalpy per unit mass, and $m_{\rm col}$ the column mass.
\begin{equation}
\label{eq:condition}
\int_{\rm conv.region} \widetilde{H} ~d m_{\rm col} = \int_{\rm conv.region} \widetilde{H}^\prime ~d m_{\rm col},
\end{equation}
where we evaluate the integral over the whole convective region. The prime superscript denotes a quantity after the adjustment. From thermodynamics and the ideal gas law, 
\begin{equation}
dH = dU + d(PV) = C_V dT + nRdT ,
\end{equation}
with the volume $V$, the heat capacities at constant volume and pressure, $C_V$ and $C_P$, respectively, and the number of gas particles $N$. With $C_V + nR = C_P$, we obtain $dH = C_P dT$. Integrating, dividing by mass and inserting in eq.~(\ref{eq:condition}) leads to 
\begin{equation}
\int_{\rm conv.region}  c_P T  ~d P = \int_{\rm conv.region}  c_P^\prime T^\prime ~d P ,
\end{equation}
with $c_P$ being the specific heat capacity per unit mass. We have further substituted $m_{\rm col} = P / g$ and used the common assumption of treating $g$ as constant throughout the atmosphere. As the solution $T^\prime$ is following an adiabat it is convenient to express it via the potential temperature $\Theta$, which is constant along the adiabat, so that
 \begin{equation}
 \label{eq:tprime}
T^\prime (P) = \Theta \left(\frac{P_0}{P}\right)^{-\kappa} ,
\end{equation}
where we denote $P_0$ as a reference pressure (e.g., the bottom of the convective region). The potential temperature satisfying eq.~(\ref{eq:condition}) is 
\begin{equation}
\label{eq:theta}
\Theta = \frac{\int_{\rm bot}^{\rm top} c_P T ~d P}{\int_{\rm bot}^{\rm top} c_P \left(P_0/P\right)^{-\kappa} ~d P} ,
\end{equation}
where ``top'' and ``bot'' mark the top and bottom boundaries of the convective region \citep{mendonca18b}.

Numerically it is necessary to iterate between radiative convergence and convective adjustment. In practice the following procedure is executed.

\begin{enumerate}

\item{\texttt{HELIOS} iterates until the atmosphere is in radiative equilibrium. See \cite{malik17} for more details.}

\item{All layers are checked for unstable lapse rates with inequality (\ref{eq:test})}.

\item{If unstable layers are found, they are corrected as described by eqs.~(\ref{eq:tprime}) and (\ref{eq:theta}).}

\item{Pairs of layers whose in-between lapse rate has been corrected are now called convective layers.}

\item{One forward step with the radiative iteration is executed. Since this changes again the lapse rates, Steps  2 to 4 need to be repeated.}

\item{In general Steps 2 to 5 are repeated until a converged solution is found. Such a solution needs to satisfy (i) convective stability, (ii) local radiative equilibrium in all radiative layers, and (iii) global energy equilibrium. Whereas criteria (i) and (ii) are inherent to the convective adjustment mechanism, the satisfaction of the global energy criterion is somewhat more tricky. We discuss our method on how to solve it in Appendix \ref{app:conv}.}

\end{enumerate}

The net convective flux, which is non-zero in the convective layers, is a by-product of the convective adjustment, and reads $\mathcal{F}_{\rm conv, -} = \mathcal{F}_{-} - \mathcal{F}_{\rm rad, -}$, where $\mathcal{F}_{-}$ is the net flux and $\mathcal{F}_{\rm rad, -}$ is the net radiative flux. We denote a wavelength-independent flux by $\mathcal{F}$.

In this work, we employ the equation of state of \citet{saumon95} to obtain the adiabatic coefficient. A simple ideal-gas extension toward lower pressures ($P<10^{-2}$ bar) and temperatures ($T\lesssim100$ K) was included by \citet{alibert05} and \citet{mordasini12}. Since the species other than hydrogen and helium (the metals) represent a small contribution by mass, we include them only approximately in the calculation of the adiabatic gradient: As in \citet{baraffe08}, we compute the effective helium fraction $Y_{\rm eff} = Y + Z$, with $X+Y+Z=X+Y_{\rm eff}\equiv1$, where $X$, $Y$, and $Z$ are the mass fractions of elemental hydrogen, helium, and the mass fraction of the metal elements. We then use $Y_{\rm eff}$ to interpolate within the \citet{saumon95} tables.

\begin{figure}
\begin{center}
\begin{minipage}[h]{0.49\textwidth}
\includegraphics[width=\textwidth]{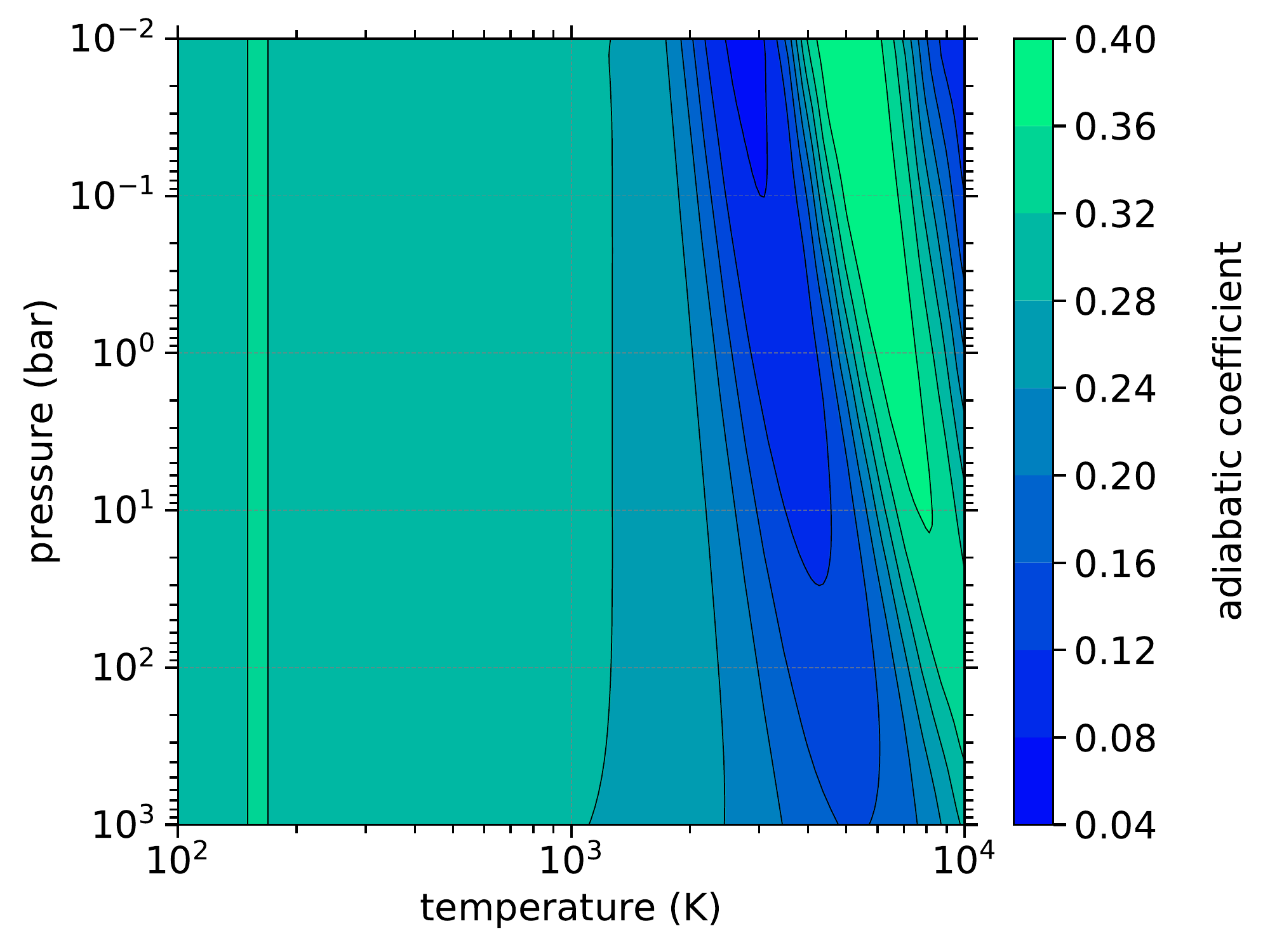}
\end{minipage}
\vspace{-0.3cm}
\caption{The adiabatic coefficient versus temperature and pressure for lower atmospheres. The black contour lines are separated by 0.04. The values depicted are for solar metallicity and carbon-to-oxygen ratio.}
\label{fig:kappa}
\end{center}
\end{figure}

Fig.~\ref{fig:kappa} shows $\kappa$ in the $T$--$P$ regime of interest. The low values $\kappa\sim 0.04$ around 3000 K come from the dissociation of molecular hydrogen, and near $10^4$ K from the ionization of atomic hydrogen. 

To obtain the heat capacity, we use eq.~(2.11) in \citet{pierrehumbert10}, $c_P = k_{\rm B}/(\mu m_{\rm u} \kappa)$, with the Boltzmann constant $k_{\rm B}$, the mean molecular weight $\mu$ and the atomic mass unit $m_{\rm u}$. In the last stages of this work, we realized that this expression is valid only in the limit of a perfect gas (i.e., with a constant number of degrees of freedom) or when the number of degrees of freedom changes but independently of pressure. Since this is not given at dissociation or ionization of hydrogen, for instance, the heat capacity turns out to be underestimated by a factor $\sim3$ in those narrow temperature regions. Explictly, one can show easily (see Appendix \ref{app:kappa}) that the correct equation is
\begin{equation}
\label{eq:cp}
 c_{\rm P} = \frac{k_{\rm B}}{m_{\rm H} \kappa} \left(-\frac{\partial \tilde{S}}{\partial \ln P}\right)_T,
\end{equation}
where $\tilde{S}\equiv S/(k_{\rm B}/m_{\mathrm{u}})$ is dimensionless. As the Sackur--Tetrode equation reveals and we verified numerically, the point is that the factor $\left(-\partial \tilde{S}/\partial \ln P\right)_T$ is equal to $1/\mu$ everywhere in the perfect-gas portion of the $P$--$T$ plane, recovering the limit of \citet{pierrehumbert10}, except where for instance the dissociation and ionization of hydrogen and, to a lesser extent, the ionization of helium or metals occur. For our models, we verified that changing $c_{\rm P}$ does not lead to any noticeable differences in the obtained T-P profile (not shown).

\break

\subsection{The contribution function}

Another addition to \texttt{HELIOS} is the output of the {\it contribution function} and the {\it transmission weighting function}. They both serve the purpose to find the photosphere with and without consideration of the atmospheric emission, respectively. Considering a simple two-stream model with isothermal layers and pure absorption, the upwards flux at $i$-th interface reads
\begin{equation}
F_{i, \uparrow} = \mathcal{T}_{i-1} F_{i-1, \uparrow} + 2 \pi \epsilon B_{i-1} (1 - \mathcal{T}_{i-1}) .
\end{equation}
Being interested in the TOA flux, we set the index $i$ in the previous expression to the TOA and take all layers below into account. This leads to
\begin{equation}
\begin{split}
\label{eq:F_TOA}
F_{{\rm TOA}, \uparrow} = \;&F_{{\rm BOA}, \uparrow} \prod_{i=0}^{n-1} \mathcal{T}_i  \\
&+ \sum_{i=0}^{n-1} \left[ 2 \pi \epsilon B_{i} (1 - \mathcal{T}_i) \prod_{\substack{j=i+1 \\ i \neq n-1}}^{n-1} \mathcal{T}_j \right] ,
\end{split}
\end{equation}
with $n$ being the number of layers. Using the altitude $z$ as the vertical grid coordinate, eq.~(\ref{eq:F_TOA}) is equivalent to
\begin{equation}
\begin{split}
F_{\rm TOA} = \;&F_{\rm BOA} \mathcal{T}(z_{\rm BOA}, z_{\rm TOA}) \\
&+ \int_{z_{\rm BOA}}^{z_{\rm TOA}} B(z) \frac{d \mathcal{T}(z, z_{\rm TOA})}{d z} ~dz ,
\end{split}
\end{equation}
where $\mathcal{T}(z, z_{\rm TOA})$ is the total transmission between the atmosphere at altitude $z$ and TOA. In a discrete layer model the integral term directly translates to the bracket term in eq.~(\ref{eq:F_TOA}). This term represents the contribution of a specific layer $i$ to the TOA spectral emission - hence the name {\it contribution function} \citep{irwin09}. Physically, it corresponds to the atmospheric location where the transmission function $\mathcal{T}$ exhibits the steepest gradient, weighted by the layer's emission. Without the consideration of the emission one obtains the {\it transmission weighting function} $\Psi$, which simply reads
\begin{equation}
\Psi_i = (1 - \mathcal{T}_i) \prod_{j=i+1}^{n-2} \mathcal{T}_j .
\end{equation}
Usually, the contribution function is the quantity analyzed to determine the origin of the planetary emission for a given wavelength. The according atmospheric layer is given by the location, where the contribution function peaks.

\subsection{Stellar Path Length Correction}
\label{sec:stellarpath}

\begin{figure}
\begin{center}
\begin{minipage}[h]{0.48\textwidth}
\includegraphics[width=\textwidth]{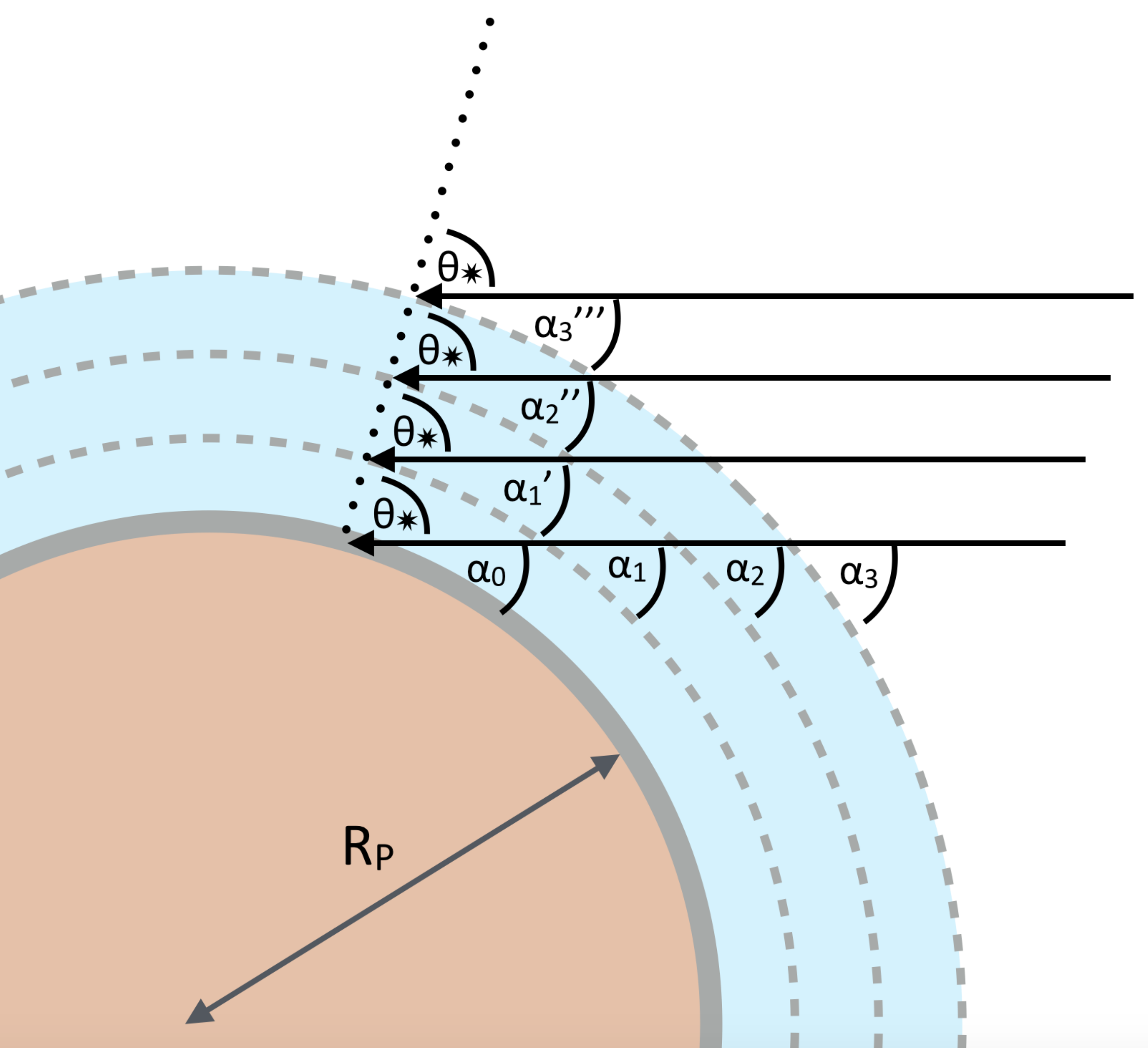}
\end{minipage}
\vspace{-0.1cm}
\caption{Applying a vertical column model (located at dotted line) to spherical geometry requires a correction of the stellar zenith angle $\theta_\ast$, depending on the location in the atmosphere. First, the zenith angle decreases with altitude. In terms of $\alpha_0 = 90^\circ - \theta_\ast$, this means that $\alpha_0 < \alpha_1 < \alpha_2 < \alpha_3$ for layers 0 to 3. Second, each model layer is reached by stellar photons which have travelled along a different path (solid lines), such that $\alpha_1 \neq \alpha_1^\prime$, $\alpha_2 \neq \alpha_2^{\prime\prime}$, $\alpha_3 \neq \alpha_3^{\prime\prime\prime}$, etc.}
\label{fig:stellarpath}
\end{center}
\end{figure}

In the plane-parallel assumption the stellar photons are assumed to travel along a straight path, which manifests itself in the use of a constant stellar zenith angle $\theta_\ast$ throughout the model grid. However, for large zenith angles\footnote{Here and for the rest of this work, we deviate from our definition of $\theta_\ast$ in Sect.~\ref{sec:definitions} and scan with the zenith angle the range from $0^\circ$ to $90^\circ$ when going from sub-stellar point to the limb of the planet. In this sense, a large zenith angle corresponds to planetary limb regions.} the stellar beam path length, proportional to $1 / \cos(\theta_\ast)$, is substantially overestimated, i.e., the path length exceeds the real atmospheric extent. If taking the spherical geometry into account, the stellar path length needs to be corrected downward. This is done by adjusting the zenith angle $\theta_\ast$, and consequently $\mu_\ast =\cos(\theta_\ast)$, depending on the location in the atmosphere. In Fig.~\ref{fig:stellarpath} the situation is drawn for a vertical column model with three layers, located along the dotted line. First, the zenith angle should depend on the altitude, which means that $\alpha_0 < \alpha_1 < \alpha_2 < \alpha_3$ in Fig.~\ref{fig:stellarpath} for $\alpha_0 = 90^\circ - \theta_\ast$. Second, each layer of the model is reached by stellar photons that travelled along a different path through the planetary hemisphere, i.e., $\alpha_1 \neq \alpha_1^\prime$, $\alpha_2 \neq \alpha_2^{\prime\prime}$, $\alpha_3 \neq \alpha_3^{\prime\prime\prime}$, etc. We rewrite and implement \cite{li05}'s eq.~(2) as
\begin{equation}
\mu_{\ast, ij} = - \sqrt{1 - \left(\frac{R_{\rm pl}+z_i}{R_{\rm pl}+z_j}\right)^2 (1 - \mu_\ast)} \text{ } \forall j \geqslant i ,
\end{equation}
where $z$ is the altitude and $R_{\rm pl}$ is the planetary radius. For each layer $i$, one must consider the stellar pathway through all overlying layers $j$ separately with $\mu_{\ast, ij}$ (c.f. Appendix in \citealt{mendonca18}). We set $R_{\rm pl}$ as the planet's measured white light radius and associate it with a pressure of 10 bar (as done e.g., in \citealt{kreidberg15}).\footnote{We acknowledge it is challenging to relate a pressure with the observed radius \citep{heng17}, thus the indicated pressure is merely a ``best guess'' assumption.} The height difference from layer $j$ to $R_{\rm pl}$ is given by $z_j$ (and analogously for layer $i$). The quantity $\mu_\ast$ corresponds to the stellar zenith angle at the layer of interest $i$. As expected, $\mu_{\ast, ij}$ reduces to $\mu_\ast$, if $i = j$.

Since the stellar path length correction was envisioned and applied for Earth's atmosphere models, we explore in Sect.~\ref{sec:dir_correct} the effect of such a correction to radiative transfer calculations for a case of a hot Jupiter and a super Earth.

\subsection{Scattering Correction for the Two-Stream Method}
\label{sec:E_param}

Two-stream radiative transfer models possess an accuracy deficit compared to {\rm multi-stream} radiative transfer methods. In a meticulous study, \cite{kitzmann13} showed that the hemispheric two-stream radiative transfer method (e.g., \citealt{heng14}), as employed in \texttt{HELIOS}, overestimates the scattering and the transmission of stellar radiation through cloud decks by up to 20\%. More specifically, the two-stream method overestimates the net greenhouse effect of CO$_2$ clouds, which are strongly scattering in the thermal wavelengths. Thus, using a too simplistic radiative transfer method may lead to wrong temperature estimates and corresponding atmospheric conditions e.g., as happened for studies of the early Mars climate \citep{kitzmann16}. Yet, two-stream models are still prevalent, as more sophisticated radiative transfer methods often prove computationally too challenging to be efficiently used in large scale climate simulations. Hence, there exist a number of tweaks to enhance the scattering behavior of two-stream models, like the $\delta$-Eddington \citep{joseph76} or the two-stream source function methods \citep{toon89}. Recently, \cite{heng17} introduced a correction factor $E$ into the two-stream coupling coefficients $\zeta_\pm$ (c.f. Sect.~\ref{sec:fluxes}), which they calibrated to reproduce results of a 32-stream DISORT code \citep{hamre13}. Later, \cite{heng18} developed a new improved formalism consistently embedding the correction factor $E$ into the two-stream equations. This correction factor depends only on the single-scattering albedo $\omega_0$ and the scattering asymmetry parameter $g_0$. Conveniently, all scattering problems can be reduced to those two quantities, which means that a model equipped with $E$ can treat any atmospheric condition imaginable. The latest version of \texttt{HELIOS} includes \cite{heng18}'s version of the correction factor $E$ through the fitting function given by their eq.~(31).

We test the two-stream method including the scattering correction versus the standard method in Sect.~\ref{sec:scat_corr}.

\section{Numerical Set-up}

\subsection{Chemical Abundances}
\label{sec:chem}

The atmospheric mixing ratios are obtained with the \texttt{FastChem} code, which reliably calculates the thermochemical equilibrium abundances of around 550 gas-phase species for temperatures between 100 K and 6000 K \citep{stock18}. We use solar elemental abundances as stated in Table 1 of \cite{asplund09} throughout this study, unless otherwise stated. For non-solar metallicities we adjust all elements heavier than He and for non-solar C/O ratios we keep O at solar value and adjust C accordingly. 

We post-process the equilibrium abundances with the effects of condensation, i.e., we remove species in the parameter space where their gas abundance is expected to decrease drastically due their own condensation or due to participation in condensate dust species. See Appendix \ref{app:removal} for more details. 

\break

\subsection{Opacities}
\label{sec:opac}

\begin{figure*}
\begin{center}
\begin{minipage}[h]{0.99\textwidth}
\includegraphics[width=\textwidth]{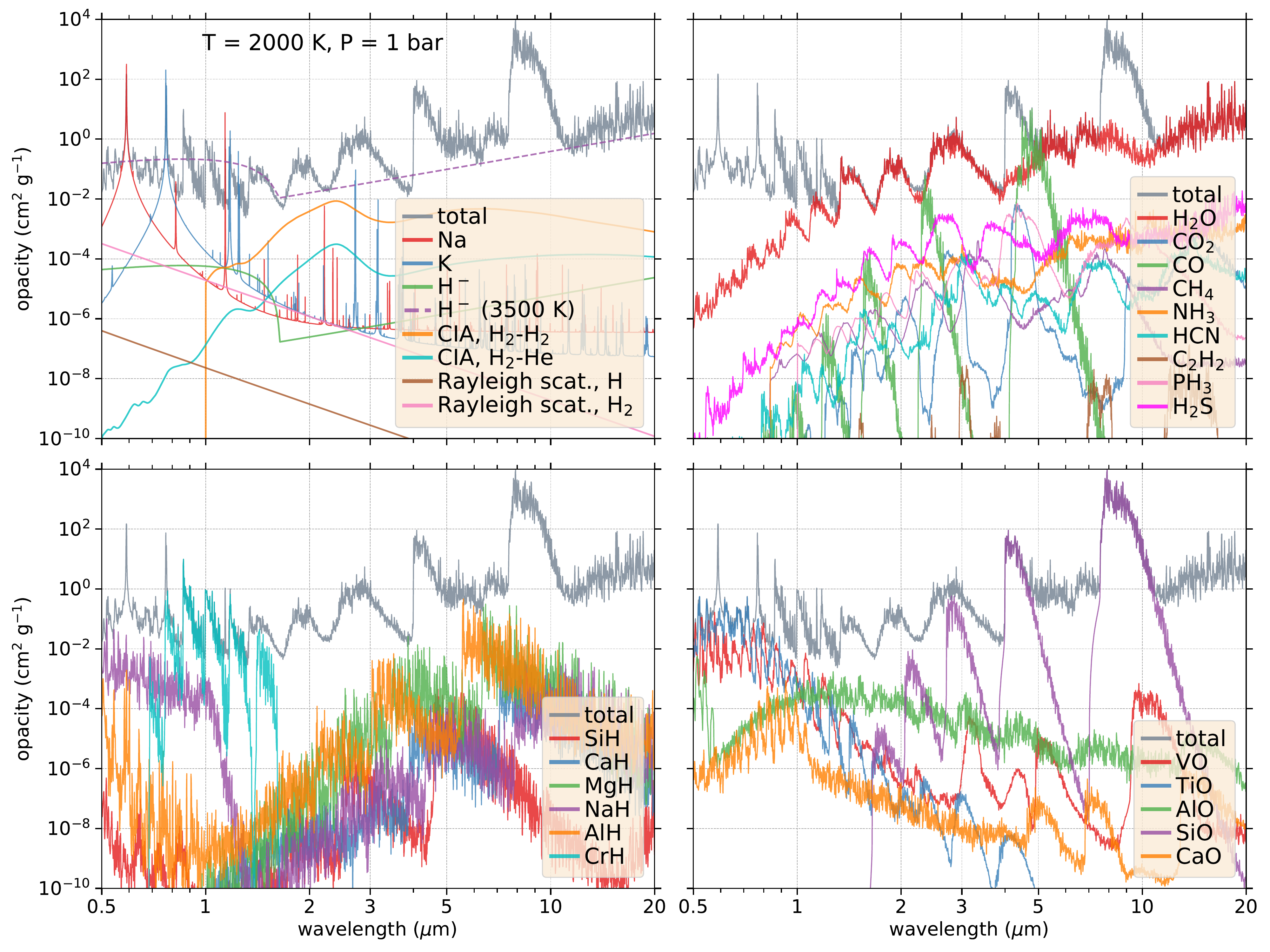}
\end{minipage}
\vspace{-0.3cm}
\caption{Opacity versus wavelength for all the opacity/exctinction sources used in this work. We show an excerpt of the opacity table for one temperature and pressure.  Each opacity is weighted by its chemical equilibrium mass mixing ratio. Depicted are the atomic/ion-, CIA-, and scattering opacities (top left), the main infrared absorbers (top right), metal and other hydrides (bottom left), and metal oxides (bottom right) Each plot also shows the total gas opacity. The displayed opacities are downsampled in resolution for clarity. See Table \ref{tab:opac} for a full list of species.}
\label{fig:opac}
\end{center}
\end{figure*}

\begin{table*}
	\caption{Opacities / scattering cross-sections used in this work.}
	\label{tab:opac}
	\vspace{-0.3cm}
\begin{center}
\bgroup
\def\arraystretch{1.05}
  \begin{tabular}{l l l}
    \hline
name & online source & reference \\ \hhline{===}
{\bf main infrared absorbers} & & \\
H$_2$O & ExoMol\footnote{\url{exomol.com}} & \cite{barber06}  \\
CO$_2$ & HITEMP\footnote{\url{hitran.org/hitemp/}} & \cite{rothman10} \\
CO & ExoMol & \cite{li15} \\
CH$_4$ & ExoMol & \cite{yurchenko14} \\
O$_2$ & HITRAN\footnote{\url{hitran.org/}} & \cite{gordon17} \\
NO & ExoMol & \cite{wong17} \\
SO$_2$ & ExoMol & \cite{underwood16} \\
NH$_3$ & ExoMol & \cite{yurchenko11} \\
OH & HITEMP & \cite{rothman10} \\
HCN & ExoMol & \cite{harris06} \\
C$_2$H$_2$ & HITRAN & \cite{gordon17} \\
PH$_3$ & ExoMol & \cite{sousasilva15} \\
H$_2$S & ExoMol & \cite{azzam16} \\
SO$_3$ & HITRAN & \cite{gordon17} \\
PO & ExoMol & \cite{prajapat17} \\
\hline
{\bf metal / other hydrides} & & \\
SiH & ExoMol & \cite{yurchenko18b}\\
CaH & ExoMol & \cite{yadin12} \\
MgH & ExoMol & \cite{yadin12} \\
NaH & ExoMol & \cite{rivlin15} \\
AlH & ExoMol & \cite{yurchenko18b} \\
CrH & ExoMol & \cite{burrows02b} \\
\hline
{\bf metal oxides} & & \\
VO & ExoMol & \cite{mckemmish16} \\
TiO & VALD\footnote{\url{vald.astro.uu.se/}} & \cite{ryabchikova15} \\
AlO & ExoMol & \cite{patrascu15} \\
SiO & ExoMol & \cite{barton13} \\
CaO & ExoMol & \cite{yurchenko16} \\
\hline
{\bf collision-induced absorption (CIA)} & & \\
H$_2$-H$_2$ & HITRAN & \cite{richard12} \\
H$_2$-He & HITRAN & \cite{richard12} \\
\hline
{\bf atoms \& ions} & & \\
Na \& K & Kurucz\footnote{\url{kurucz.harvard.edu/}} & \cite{kurucz11}, \cite{burrows00}, \\
& & \cite{burrows03}, \\
& & Cubillos et al. (2019, in prep.) \\
H$^-$ & & \cite{john88} \\
\hline
\bf{scattering cross-sections} & & \\
H$_2$ & & \cite{sneep05} \\
H & & \cite{lee04} \\
    \hline
  \end{tabular}
	\egroup
	\end{center}
	\vspace{-0.2cm}
\end{table*}

We have massively extended our list of opacity sources since \citep{malik17}, see Table \ref{tab:opac}. The new version of our in-house opacity calculator \texttt{HELIOS-K} \citep{grimm15} is able to automatically download and calculate the spectral line lists from the ExoMol, HITRAN/HITEMP and Kurucz online data bases, which facilitates the inclusion of new opacities significantly. In this study we include the main spectroscopically active and atmospheric abundant molecules, as expected in atmospheres with temperatures from a few hundred to several thousand Kelvin. We divide our opacities into the main infrared absorbers, metal or other hydrides, metal oxides, Na \& K\footnote{We elaborate on the calculation of the Na and K opacities in App. \ref{app:alkali} in detail.} and H$^-$. We use this division in Sect.~\ref{sec:opac_test} to investigate the spectroscopic impact of these absorbing species. For the atmospheric model grid we use all of the opacity sources listed in Table \ref{tab:opac}. Fig.~\ref{fig:opac} shows the included opacities for one temperature and pressure weighted with their respective equilibrium mixing ratio. The molecular opacities are computed at a resolution of 10$^{-2}$ cm$^{-1}$, the Na and K opacities at 10$^{-1}$ cm$^{-1}$, and H$^-$ at 10 cm$^{-1}$. The calculation of the alkali metals is described in detail in Appendix \ref{app:alkali}. For all the other molecules we use a Voigt profile, and due to the absence of a first-principles theory for pressure broadening, we adopt the community procedure of an ad hoc truncation of the line wings at 100 cm$^{-1}$ from the line center. For the opacities with data from the ExoMol database, we include the default broadening as provided in their online library. For the HITRAN/HITEMP opacities we use their self-broadening parameters.

Lastly, we are aware that we are missing the opacities of CIA H2-H2 blueward of 1 $\mu$m  \citep{borysow01, borysow02}. They play an insignificant role for hot, irradiated atmospheres because other opacity sources present, such as the alkali metals or the metal hydrides and oxides, assume the role of the dominant shortwave absorbers. Furthermore, CIA absorption becomes generally only important in the deep atmosphere, $P \gtrsim 10$ bar, as it scales with the pressure squared. However, at these depths the convective region usually begins. That means that the atmosphere there is optically thick and thus not visible in the planetary emission, and the according temperatures are given by the adiabatic lapse rate and not by radiative equilibrium.

\break

\subsection{Iteration \& Post-processing}
\label{sec:iter}

For the iterative runs, i.e., to find the temperature profile in radiative-convective equilibrium, we employ the $\kappa$-distribution method with a correlated-$\kappa$ approximation as described in \cite{malik17}. We use 300 wavelength bins over a range of 0.33 - 10$^5$ $\mu$m. Each bin contains 20 Gaussian points, where independent flux calculations are performed. With the addition of strong shortwave absorbers, occasional discontinuities in the converged temperature profile may occur. This is countered by a temperature smoothing algorithm (see App. \ref{app:smoothing}). Furthermore, there is now the option to bypass the $\kappa$-distribution method and sample the opacities at the discrete flux wavelength values. This approach, commonly called {\it opacity sampling}, is used in this study to generate high-resolution spectra (8263 wavelength bins) by post-processing a given temperature profile. To elucidate on our terminology, we regard the difference between ``opacity sampling'' and ``line-by-line'' as a matter of sampling resolution.  When the number of sampling points greatly exceeds the number of spectral lines considered, the computation qualifies as being line-by-line. When the number of sampling points is comparable or smaller than the number of lines, as it is in our case, then one is performing opacity sampling.

\section{Results}
\label{sec:res}

In this section we expose \texttt{HELIOS} to a radiative transfer model intercomparison, conduct a number of working tests and present self-consistent atmospheric models for self-luminous and irradiated planets.

\subsection{Model Intercomparison}

\subsubsection{\texttt{Exo-REM}, \texttt{petitCODE} \& \texttt{ATMO}}

\begin{figure*}
\begin{center}
\begin{minipage}[h]{0.49\textwidth}
\includegraphics[width=\textwidth]{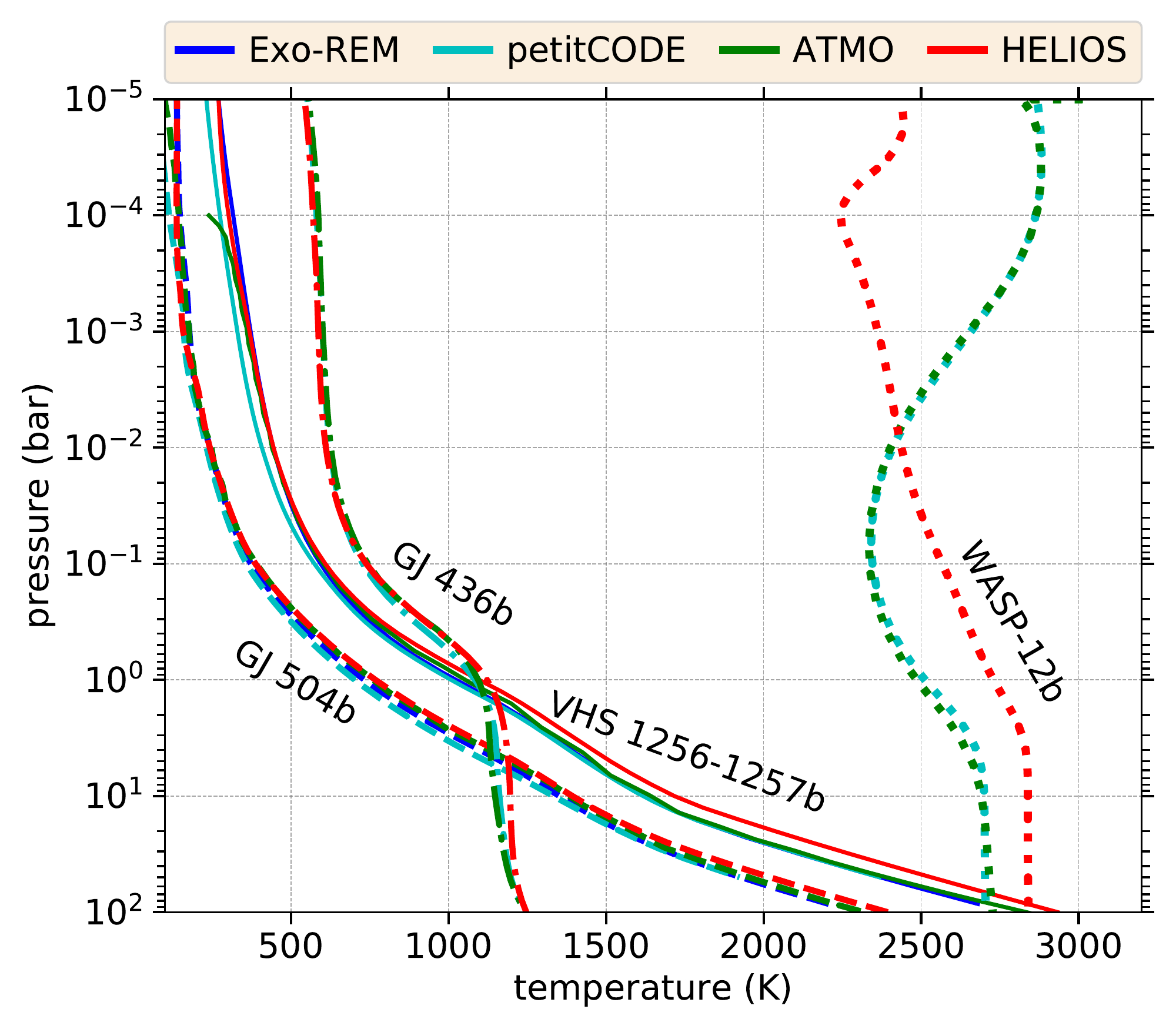}
\end{minipage}
\hfill
\begin{minipage}[h]{0.49\textwidth}
\hspace{-0.57cm}
\includegraphics[width=\textwidth]{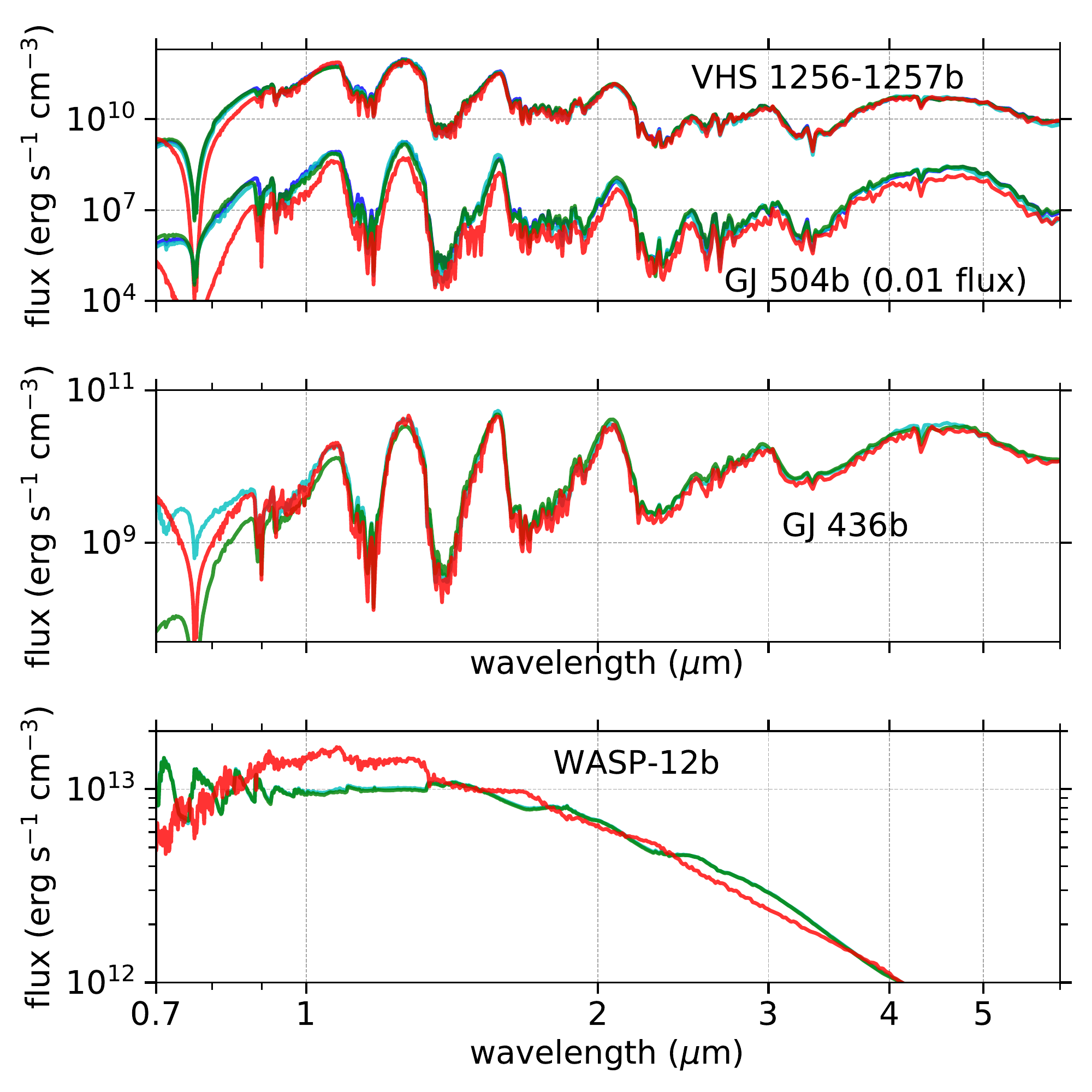}
\end{minipage}
\begin{minipage}[h]{0.99\textwidth}
\vspace{-0.2cm}
\includegraphics[width=\textwidth]{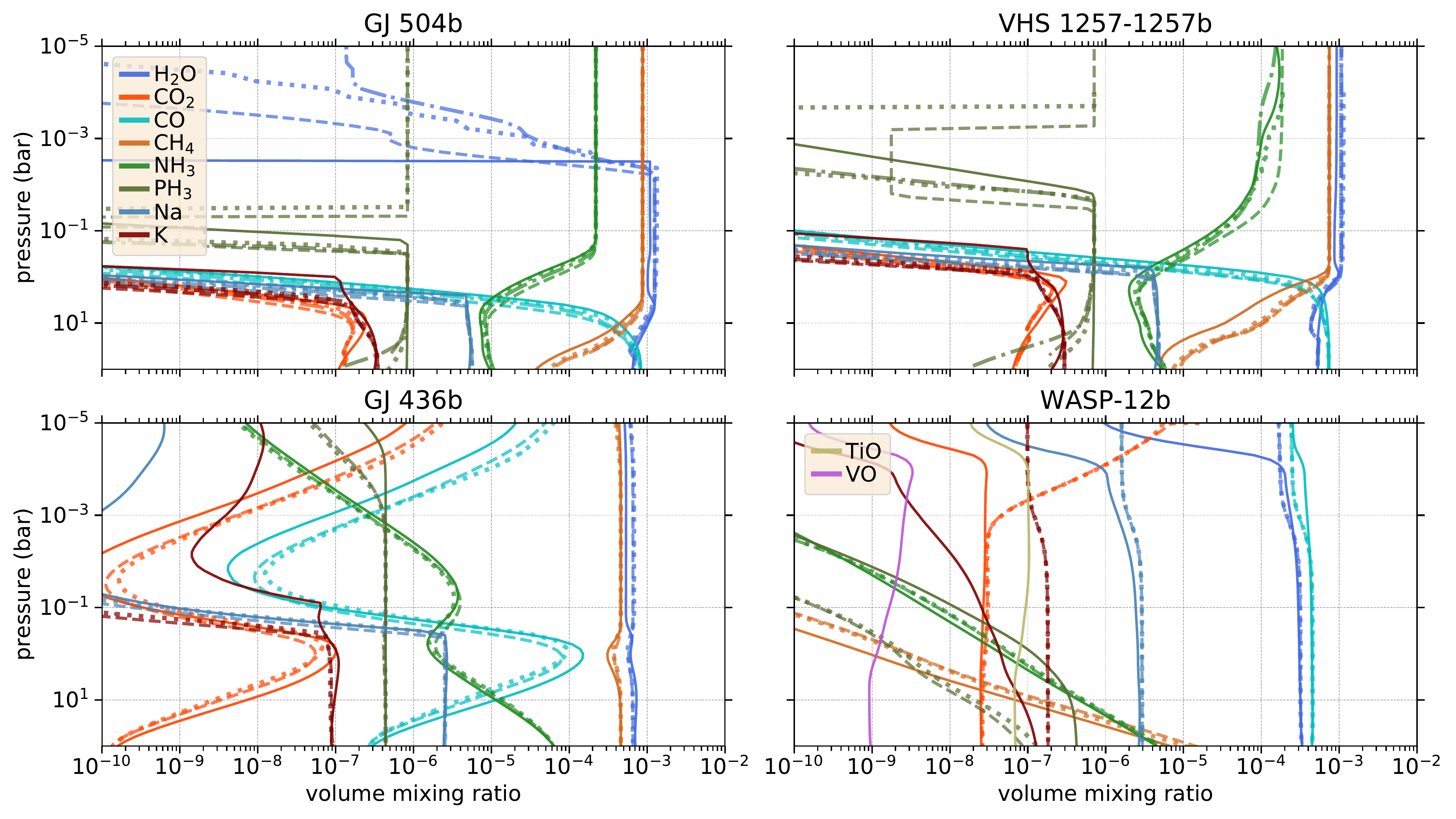}
\end{minipage}
\vspace{-0.1cm}
\caption{Comparison of self-consistently calculated atmospheres by \texttt{Exo-REM}, \texttt{petitCODE}, \texttt{ATMO} and \texttt{HELIOS} for the planets GJ 504b, VSH 1256-1257b, GJ 436b and WASP-12b. {\bf Top left panel}: temperature profiles in radiative-convective equilibrium. {\bf Top right panels}: Emission spectra corresponding to the temperature profiles on the left. The flux of GJ 504b is multiplied by 0.01 to help distinguish between the depicted spectra. {\bf Bottom four panels}: Vertical abundance profiles for the models shown in the top panels. The abundances of \texttt{HELIOS}, \texttt{petitCODE}, \texttt{ATMO}, and \texttt{Exo-REM} are represented by solid, dashed, dotted, and dash-dotted lines, respectively. The color scheme is the same in the four bottom panels. Since \texttt{Exo-REM} only handles non-irradiated planets, its model data are missing for GJ 436b and WASP-12b. Finally, the abundances of TiO and VO, as used for WASP-12b, are not provided in \cite{baudino17} and are shown only for \texttt{HELIOS}.}
\label{fig:baudino}
\end{center}
\end{figure*}

In a series of benchmark tests, \cite{baudino17} compared the three radiative transfer models \texttt{Exo-REM} \citep{baudino15}, \texttt{petitCODE} \citep{molliere15} and \texttt{ATMO} \citep{amundsen14} to identify troublesome model parameters and how these may affect simulated atmospheres. For this purpose they aligned the three codes in terms of chemistry, opacity line lists and the treatment of spectral lines and investigated the effect of each of those aspects. As such an inter-code comparison goes beyond the scope of this study, we limit our comparison to the self-consistent calculation of the four therein explored planets: the self-luminous GJ 504b and VHS 1256-1257b, and the irradiated super Earth GJ 436b and hot Jupiter WASP-12b\footnote{As \texttt{Exo-REM} is not build for external irradiation, its models are missing for the latter two planets.}. In order to mimic their set-up we limit ourselves to the following opacities: H$_2$O, CO$_2$, CO, CH$_4$, NH$_3$, PH$_3$, the alkali metals Na and K and collision-induced absorption from H$_2$-H$_2$ and H$_2$-He. Furthermore, we include Rayleigh scattering for H$_2$ and H. 

Using their planetary parameters and numerical set-up we self-consistently calculate the temperature profiles and emission spectra of aforementioned planets, see Fig.~\ref{fig:baudino}, top left and right panels, respectively. The temperature profiles of the self-luminous planets agree very well. Also, for GJ 436b the temperatures are consistent among all models, although \texttt{HELIOS} leads to slightly warmer deep atmospheric layers (pressure $>$ 1 bar). In general, the temperature differences are minuscule and thus it is not surprising that also the corresponding emission spectra for these three planets match very well. In fact, we are surprised about such a high level of agreement as there are still differences in the treatment of opacities between \texttt{HELIOS} and the other models. Not only do we use different line lists but particularly the calculation of the far wings deviates. They employ a pure Voigt profile for the Na \& K resonance lines whereas our line wings are influenced by \cite{burrows03}'s formalism. This discrepancy is particularly visible in the sub-micron wavelength range, where the far wings of the alkali resonance lines dominate the absorption. Note, that the shown spectra have been downgraded to the same resolution to facilitate a comparison. 

A special case is WASP-12b, because they additionally include TiO, VO as absorbers for this planet. Although we also extend our opacity list with these two molecules, we obtain a substantially different temperature profile and emission spectrum. We cannot reproduce their large temperature inversion and are consequently missing their strong emission features in the optical. We speculate that the difference must be caused by differences in the TiO, VO line lists. Incomplete line lists or erroneous line strengths may both lead to stark differences in shortwave heating of the upper atmosphere.

In Fig.~\ref{fig:baudino}, bottom panels, we show the corresponding vertical abundance profiles for the four planets used in the models. Even though the models are based on the same elemental abundances of \cite{asplund09}, the exact molecular abundances depend on the choice of solving method, the utilized thermochemical data and the size of the chemical network considered. \cite{baudino17} give an overview of the chemistry in \texttt{petitCODE}, \texttt{ATMO} and \texttt{Exo-REM} in their Sect. 2.2. We hypothesize three main reasons for discrepancies in the chemical abundances between the models. First, some differences in the abundances stem directly from differences in the temperature profiles. This is best visible for WASP-12b, for which the found temperatures profiles differ the most. Second, the removal of gas species due to condensation is modeled differently by each of the models. Critical to this issue is at which temperature condensation of the gas species occurs, and whether the species in question might be removed indirectly by the formation of a dust compound. Unlike the other models, \texttt{Exo-REM} includes a ``cold trap'' by removing species located above the one where its condensation curve is crossed.  We describe the method of how we apply condensation in Appendix \ref{app:removal}. Third, we include ions in our thermochemical network, which means that at higher temperatures neutral species are depleted due to ionic transitions. Without this depletion the amount of neutral species may be overestimated. This explains e.g. the discrepant abundances of K for WASP-12b between the models.

\subsubsection{\texttt{COOLTLUSTY} \& \texttt{GENESIS}}

\begin{figure*}
\begin{center}
\begin{minipage}[h]{0.49\textwidth}
\includegraphics[width=\textwidth]{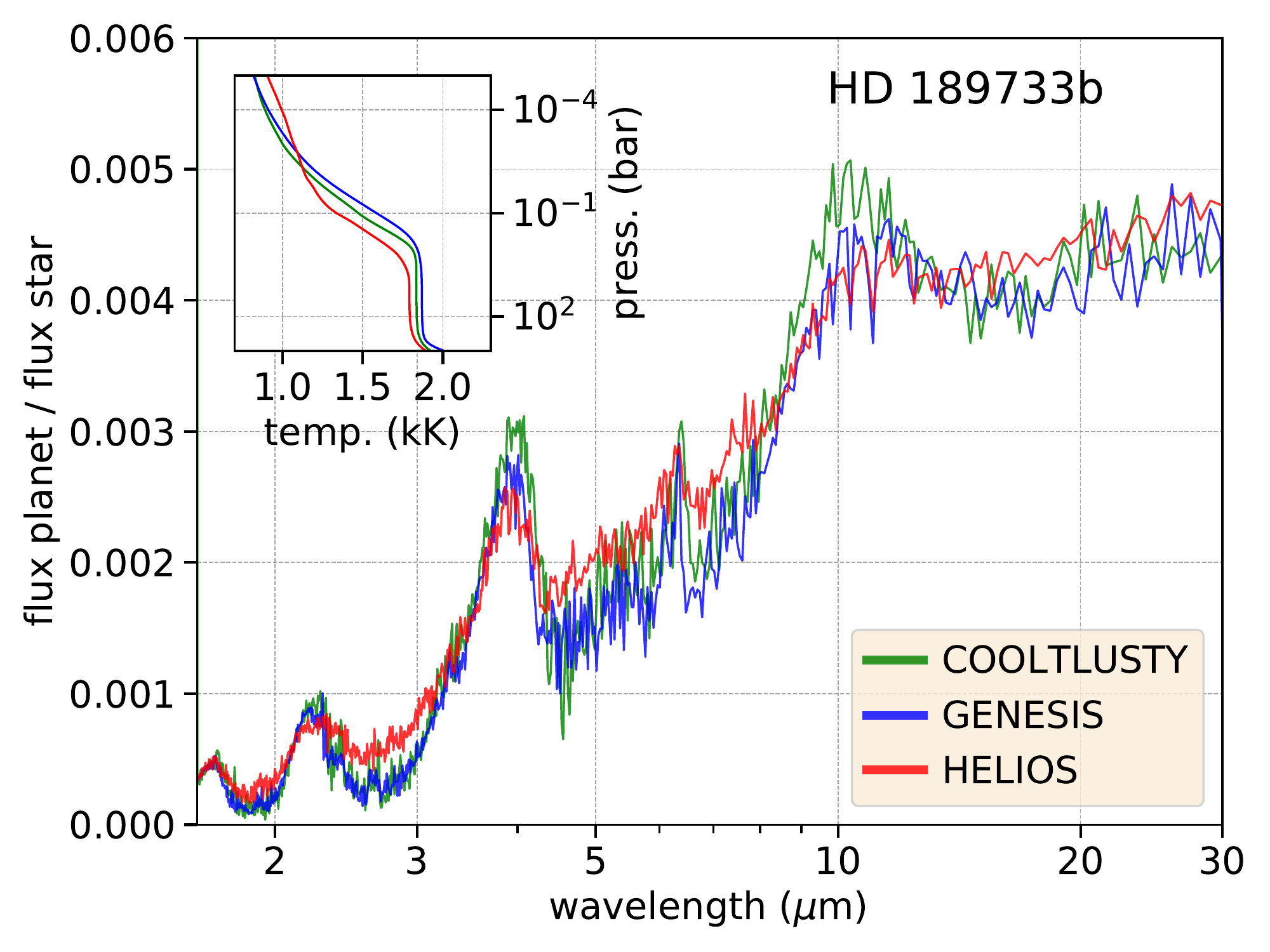}
\end{minipage}
\hfill
\begin{minipage}[h]{0.49\textwidth}
\includegraphics[width=\textwidth]{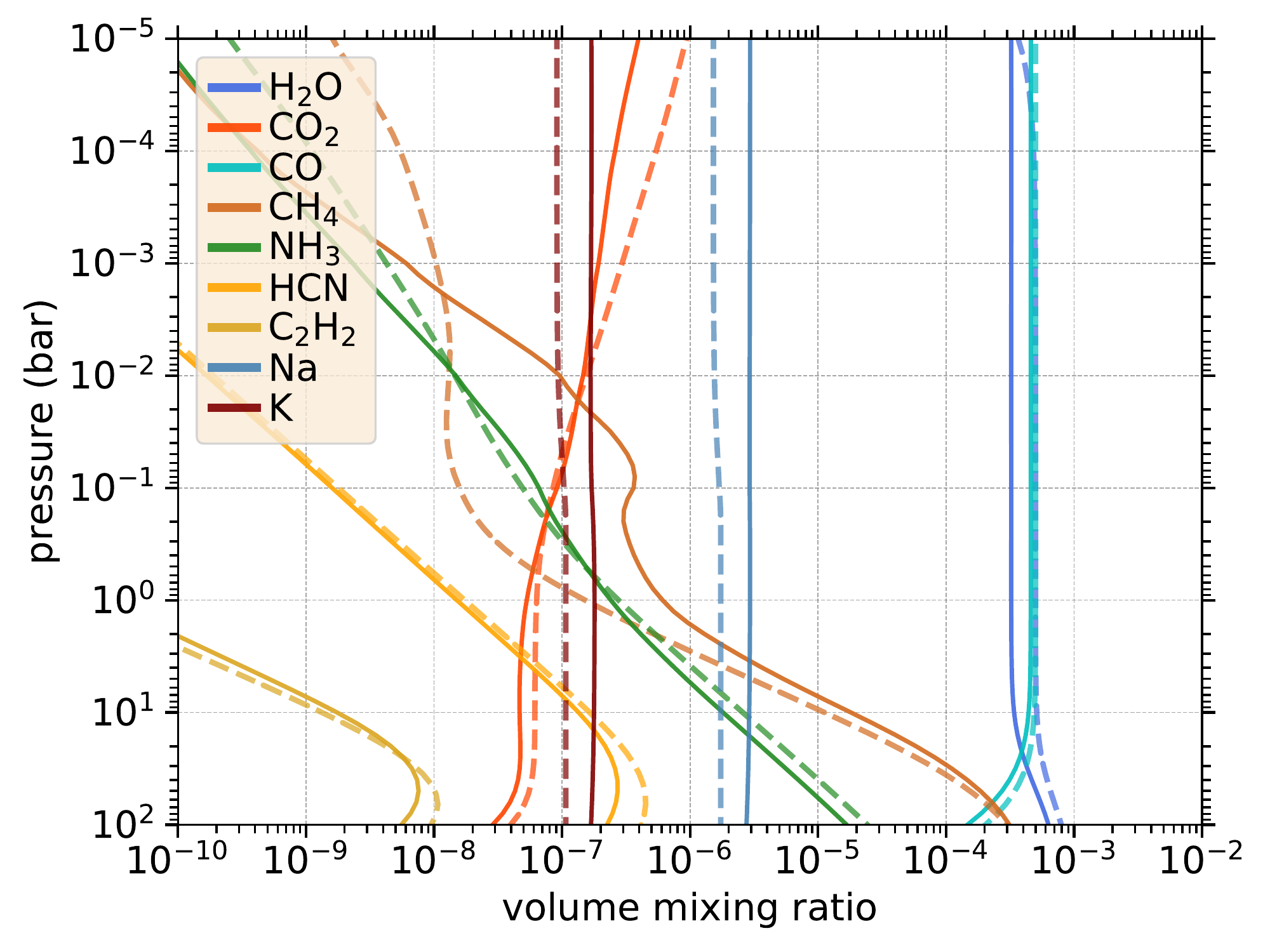}
\end{minipage}
\vspace{-0.2cm}
\caption{{\bf Left panel}: Comparison of the secondary eclipse spectrum and dayside temperatures in radiative-convective equilibrium (inlaid) of HD 189733b modelled with \texttt{COOLTLUSTY}, \texttt{GENESIS} and \texttt{HELIOS}. The spectra are downsampled in resolution for clarity. {\bf Right panel}: The vertical abundances from the \texttt{HELIOS} (solid lines) and \texttt{GENESIS} (dashed lines) models shown on the left. The dominant absorbing species are H$_2$O and CO in the near-infrared, and Na and K at shorter wavelengths.}
\label{fig:burrows}
\end{center}
\end{figure*}

Next we join in \cite{gandhi17}'s model comparison of their new radiative model \texttt{GENESIS} with the results of \cite{burrows08} for HD 189733b. The latter use the \texttt{COOLTLUSTY} model which has a longstanding history of exoplanet applications. The most detailed description of the vast array of opacity sources included in \texttt{COOLTLUSTY} is given in \cite{sharp07}. Much fewer opacities are included in \texttt{GENESIS}, namely H$_2$O, CO$_2$, CO, CH$_4$, NH$_3$, HCN, C$_2$H$_2$, Na, K and CIA H$_2$-H$_2$ and H$_2$-He. We use the latter batch of \texttt{GENESIS}'s opacities for this particular comparison. Furthermore, we use their planetary parameters to obtain the same effective dayside temperature and include a stellar blackbody to avoid potential discrepancies due to a mismatched stellar model.

Fig.~\ref{fig:burrows}, left panel, shows the synthetic secondary eclipse spectra together with the corresponding temperature profiles in radiative-convective equilibrium according to each model. The shown spectra are downsampled to a common resolution to facilitate a qualitative comparison. In general, \texttt{HELIOS}'s spectrum matches well the other two, with the largest difference being somewhat smaller peaks and troughs in the features between 2 and 10 $\mu$m. This is a consequence of the smaller temperature difference between the upper and the lower atmosphere, which directly translates to the magnitude of spectral features. Also, in our case the photosphere is located slightly deeper as compared to the other models. Yet, the global difference between \texttt{HELIOS} and \texttt{COOLTLUSTY} appears only slightly larger than the difference between \texttt{GENESIS} and \texttt{COOLTLUSTY}. As in the last subsection, we argue that the most likely causes for the discrepancies between the models stem from the opacities. An in-depth comparison of the radiative transfer method, which exceeds the scope of this study, would need to eliminate the following problematic factors which are present here: different sets of employed opacities, different line lists, different spectral line-wing treatments (c.f. \citealt{baudino17}) and differences in the stellar spectrum (which directly affect secondary eclipse spectra). Until those factors are brought to a common denominator, an absolute convergence of the models remains elusive or coincidental.

In Fig.~\ref{fig:burrows}, right panel, the vertical abundance profiles as obtained in the \texttt{HELIOS} (solid lines) and \texttt{GENESIS} (dashed lines) models are shown. The main near-infrared absorbers are H$_2$O and CO, and CH$_4$ in the bottom atmosphere. The relatively abundant alkali metals, Na and K, provide the shortwave absorption and thus play an important role for the extinction of star light. We attribute the main discrepancies in the abundance profiles between \texttt{HELIOS} and \texttt{GENESIS} to the size of the employed chemical network, and to differences in the atmospheric temperatures. The latter cause is particularly noticeable in the case of CH$_4$, as its abundance substantially decreases with temperature $T$, if $T \gtrsim 850$ K. The discrepancy in the H$_2$O abundance may stem from using a slightly different value for the solar elemental abundances of carbon and oxygen, C/O$_{\rm \texttt{HELIOS}} = 0.55$ and C/O$_{\rm \texttt{GENESIS}} = 0.5$. Also, in the larger chemical network used in \texttt{HELIOS} the oxygen atoms are divided among more oxygen-bearing species, which leads to a decrease in the H$_2$O abundance. Lastly, since the equilibrium chemistry formulae in \texttt{GENESIS} do not include the alkali metals, they appear to use the solar elemental abundances for Na and K. This approach returns somewhat smaller values than what we have.

\subsubsection{BT-Cond}

\begin{figure*}
\begin{center}
\begin{minipage}[h]{0.49\textwidth}
\includegraphics[width=\textwidth]{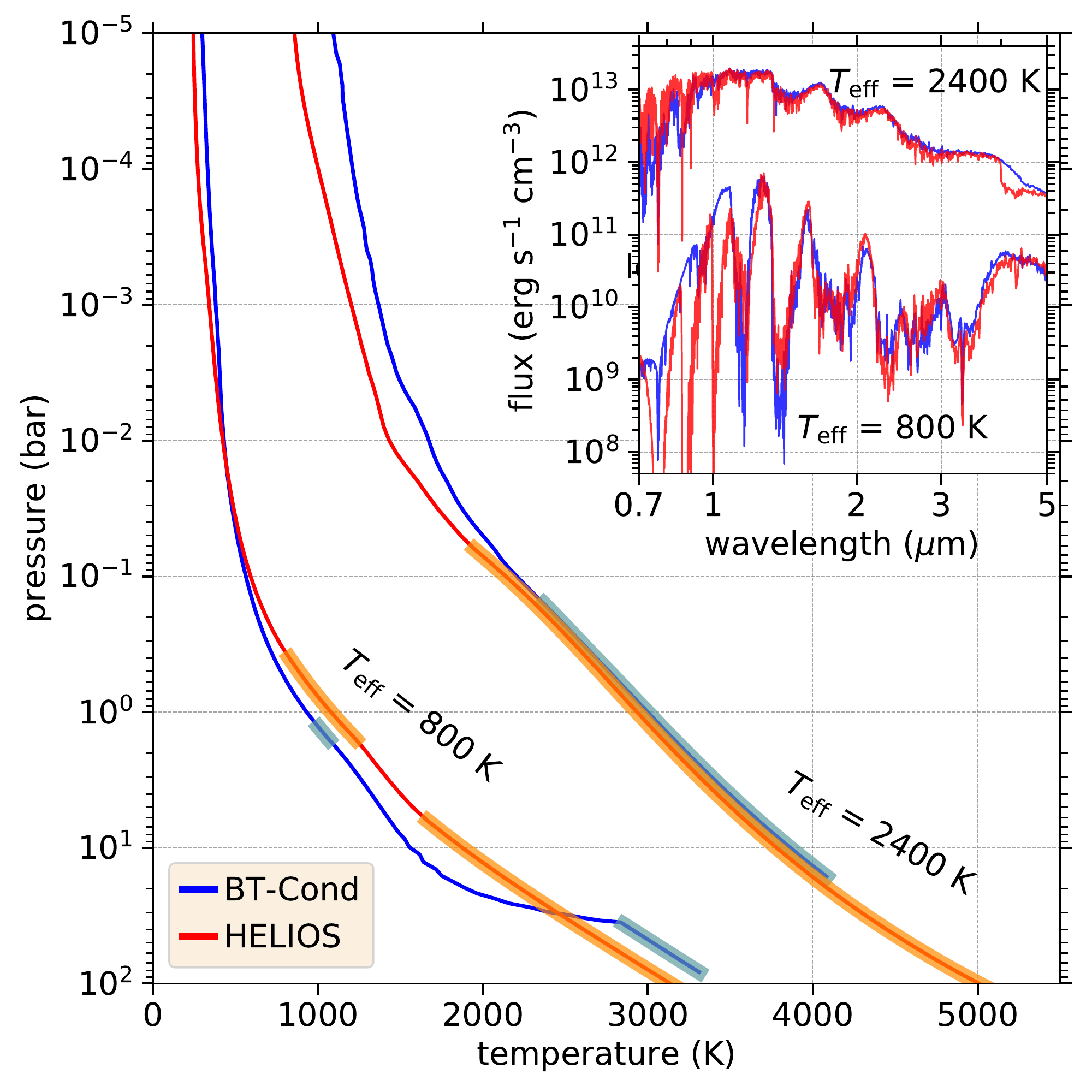}
\end{minipage}
\hfill
\begin{minipage}[h]{0.49\textwidth}
\includegraphics[width=\textwidth]{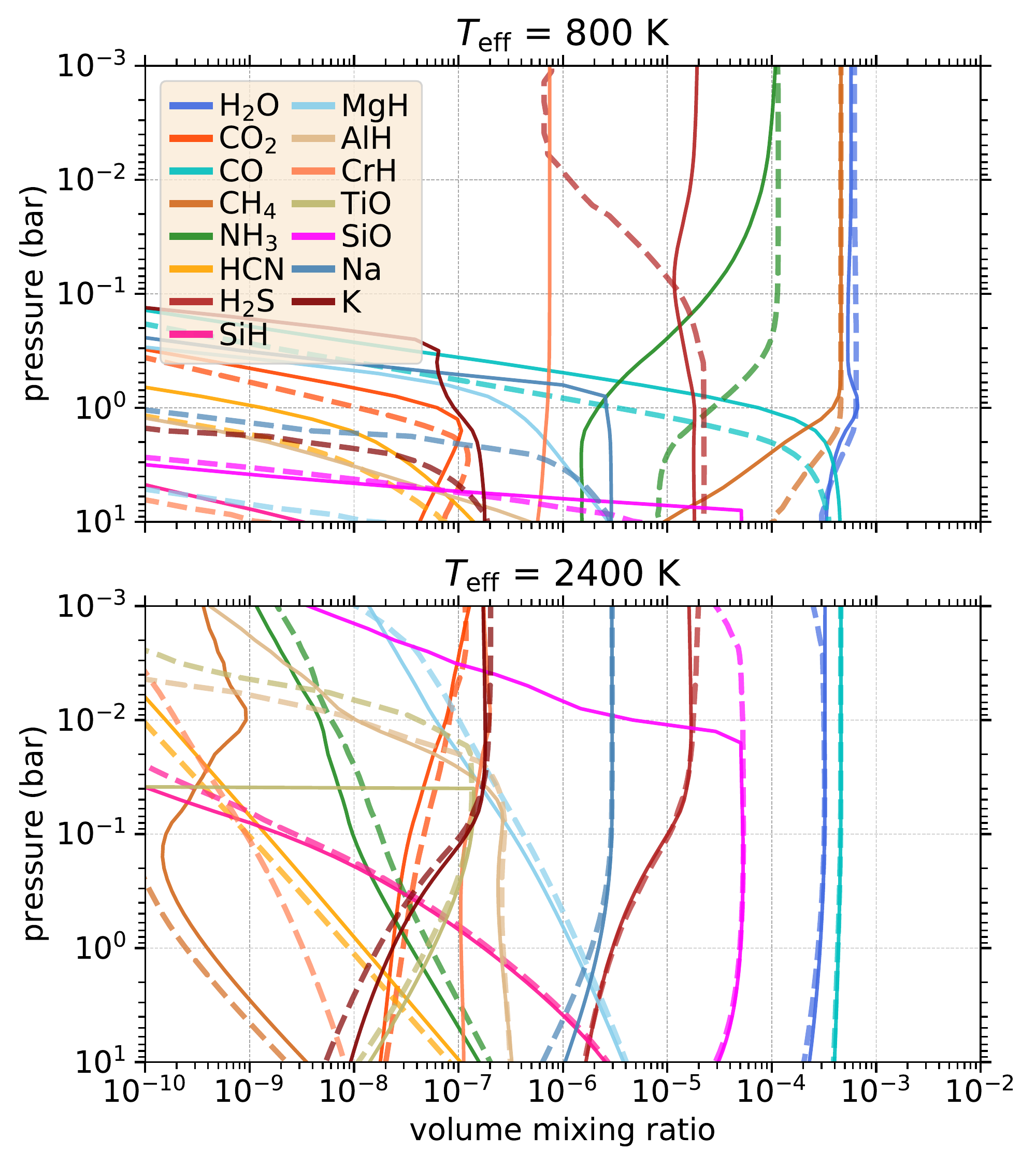}
\end{minipage}
\vspace{-0.2cm}
\caption{{\bf Left panel}: Comparison of temperatures in radiative convective equilibrium and corresponding emission spectra (inlaid) calculated with \texttt{HELIOS} and from the BT-Cond models. Two self-luminous planets with effective temperatures of 800 K and 2400 K and a surface gravity of $10^4$ cm s$^{-2}$ are explored. Convective zones are displayed as broader lines. {\bf Right panels}: The vertical abundances from the \texttt{HELIOS} (solid lines) and BT-Cond (dashed lines) models shown on the left. The color scheme is the same in both right panels. The abundances of the included species VO, NaH, C$_2$H$_2$, PH$_3$, CaO, CaH are not shown due to clarity, since they are comparatively low for both explored cases ($\lesssim 10^{-7}$).}
\label{fig:btcond}
\end{center}
\end{figure*}

There are two reference models that stand out when it comes to atmospheres of self-luminous planets or brown dwarfs: the \texttt{COOLTLUSTY} models and the \texttt{PHOENIX} models. As we compared to \texttt{COOLTLUSTY} in last subsection, we turn here towards \texttt{PHOENIX}. With this code, numerous atmospheric grids have been published with different flavors of ingredients over the past 20 years (e.g., \citealt{allard01, baraffe03, allard11, lothringer18}). The newest and most up-to-date members of this series are the BT-Cond (only gas opacities) and BT-Settl (includes a cloud model) atmospheric grids \citep{fallard12}. We choose BT-Cond as comparison partner, since we do not model clouds. As BT-Cond includes a vast array of opacity sources, we too employ the full array of opacities at our disposal, as listed in Table \ref{tab:opac}. We take as test objects a colder and a hotter non-irradiated planet with effective temperatures of 800 K and 2400 K, respectively. We set the surface gravity to $10^4$ cm s$^{-2}$ and compare the self-consistently calculated radiative-convective temperature profiles and the corresponding emission spectra, see Fig.~\ref{fig:btcond}, left panel. We further analyze the appearance of convective zones (shown as broader lines). We find that the temperatures in the hotter test case agree rather well with the only deviation in the upper atmosphere, where \texttt{HELIOS} predicts colder temperatures. However, being optically thin, this region plays only a minor role for the spectroscopic appearance of the planet and manifests itself as slightly weaker emission at wavelengths larger than 1.3 $\mu$m. The deep convective zone is almost identical for both models, which is the relevant aspect for the coupling to interior models. The colder planet exhibits a somewhat warmer photosphere and colder deeper layers (pressure $>$ 1 bar) in the \texttt{HELIOS} version. Interestingly, the BT-Cond model exhibits a significantly larger temperature gradient in the deep layers, which would be unstable and corrected with our employed equation of state. We attribute this discrepancy to the calculation of atmospheric entropy. On the other hand, the upper atmospheric temperatures are consistent and both models even agree on the detached convective zone, although it is larger in the \texttt{HELIOS} model. Important for the coupling with interior models is the fact that the deep convective zones are somewhat shifted. According to the lower deep temperature in the \texttt{HELIOS} model the spectral emission is somewhat weaker below 1.3 $\mu$m than in the BT-Cond model. Interestingly, we see strong CrH absorption features at 0.9 $\mu$m and 1 $\mu$m that are missing in the BT-Cond model. Indeed, observing the vertical abundance profiles of the two models, Fig.~\ref{fig:btcond}, top right panel, the abundance of CrH is strongly muted for BT-Cond compared to \texttt{HELIOS}. We attribute this difference to a condensation effect, implemented in BT-Cond but missing in \texttt{HELIOS}. Another interesting case is SiO. Namely, the temperatures of the upper atmosphere in the two models are located below (\texttt{HELIOS}) and above (BT-cond) the condensation temperature of MgSiO$_3$, responsible for depleting SiO. Also other species, like NH$_3$, H$_2$S, show considerable differences in their abundance between the two models, which could either be caused by differences in the temperatures, and/or in the thermochemical data and the chemical network utilized.

\subsection{Impact of the Opacities on Temperatures}
\label{sec:opac_test}

We now present a few numerical tests to shed light on the impact of some of the new features included in \texttt{HELIOS}. In this section, we investigate the effects of different opacity sources on converged atmospheric temperature profiles after running the iteration for radiative-convective equilibrium. 

\begin{figure}
\begin{center}
\begin{minipage}[h]{0.49\textwidth}
\includegraphics[width=\textwidth]{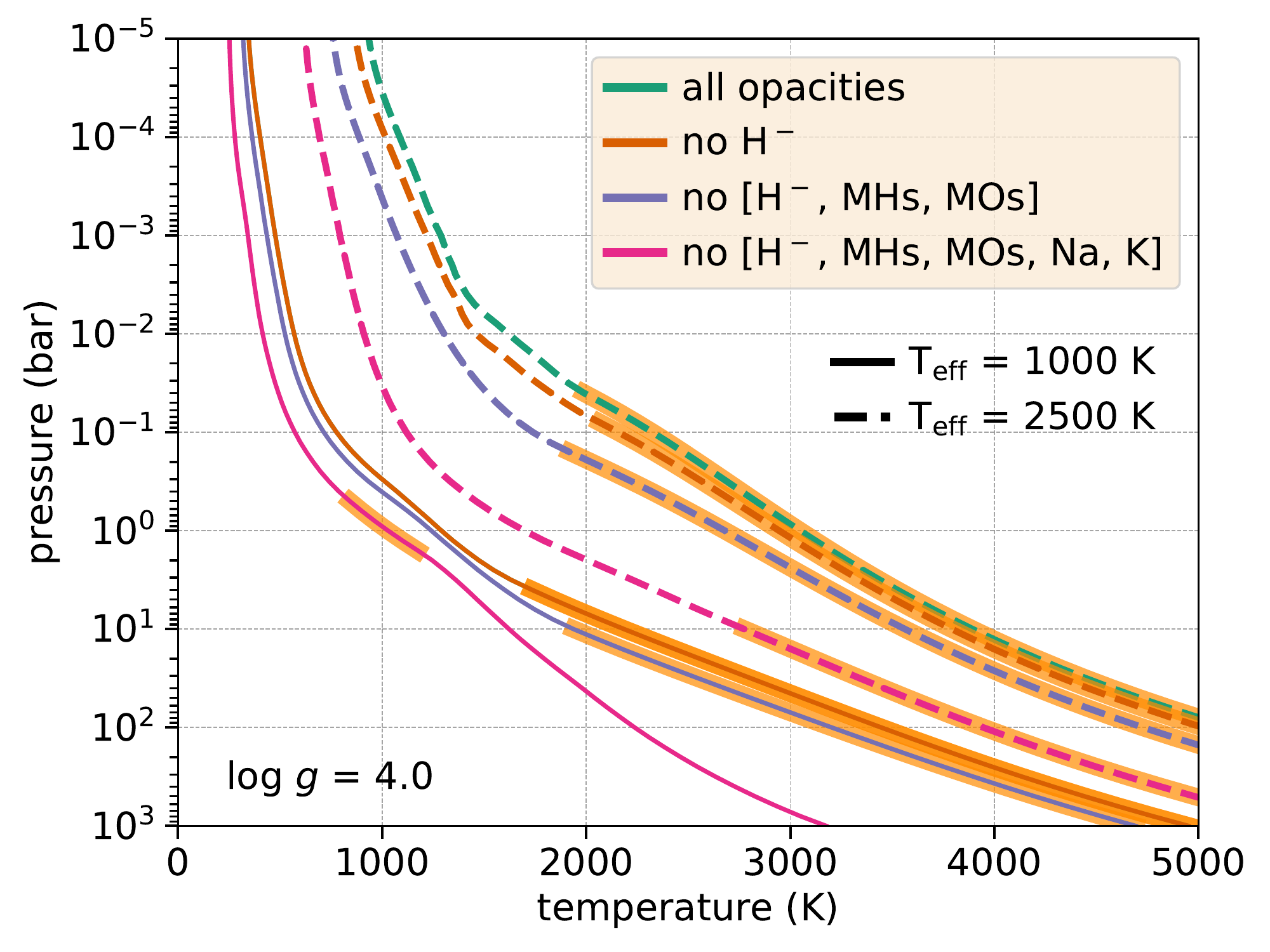}
\end{minipage}
\vspace{-0.3cm}
\caption{Temperature profiles in radiative-convective equilibrium for two self-luminous planets with effective temperatures of 1000 K and 2500 K and a surface gravity of 10$^4$ cm s$^{-2}$. The effect of using different samples of opacity sources on the calculated temperatures is explored. Convective zones are displayed as broader orange lines. The solid green line is perfectly covered by the solid brown line.}
\label{fig:opac_test_nonirrad}
\end{center}
\end{figure}

For these tests we use different opacity tables for our calculations. The tables consist of varying sets of opacities, which are pre-mixed using the same chemical network in order to single out the effect of missing opacities, while keeping the molecular abundances the same.
\begin{itemize}

\item{{\bf Sample A} includes all opacities at our disposal as listed in Table \ref{tab:opac}.}

\item{{\bf Sample B} is equal to sample A minus the H$^-$ continuum opacity.}

\item{{\bf Sample C} is equal to sample B minus the metal and other hydrides and metal oxides as shown in Fig.~\ref{fig:opac}, bottom left and bottom right panels. Effectively, that means we are removing SiH, CaH, MgH, NaH, AlH, CrH, TiO, VO, AlO, SiO, and CaO.}

\item{{\bf Sample D} is equal to sample C minus the Na and K opacities. This removes the last strong shortwave absorbers from the opacity pool.}

\end{itemize}
The first set-up consists of two self-luminous planets with effective temperatures of 1000 K and 2500 K and a surface gravity of 10$^4$ cm s$^{-2}$, see Fig.~\ref{fig:opac_test_nonirrad}. For the colder planet, samples A and B lead to the identical temperature profile. Sample C leads to somewhat cooler temperatures and sample D is markedly different. In the latter case there are no shortwave absorbers left, so the hot bottom atmosphere cools too efficiently to be convectively unstable. Conversely, including all available opacities leads to the hottest temperatures, since having more absorbing species increases the optical depth and moves the photosphere upward. Subsequently, the lower atmosphere emits radiation less efficiently which drives the temperatures up. The hotter planet shows similar trends, even more pronounced. Here, the addition of $H^-$ has a noticeable effect on the temperature. The hotter planet exhibits convective zones extending to lower pressures compared to the colder planet, which is likely due to more spectral lines being active, increasing the overall optical depth. As discussed in Sect.~\ref{sec:intro}, the entropy found in the deep convective zones is of special interest for the coupling with interior models. Hence, we list the obtained values for the entropies $S$ in the conducted runs. For the colder planet the entropy in the model with sample A is $S_{\rm A} = S_{\rm B} = 8.95$ and $S_{\rm C} = 8.74$. There is no deep convective zone when using sample D. The hotter planet models lead to $S_{\rm A} = 11.23$,  $S_{\rm B} = 10.99$, $S_{\rm C} = 10.53$ and $S_{\rm D} = 9.46$, with a modeling accuracy of $\Delta S \approx 0.02$. The values for $S$ are given in units of the Boltzmann constant.

\begin{figure*}
\begin{center}
\begin{minipage}[h]{\textwidth}
\includegraphics[width=\textwidth]{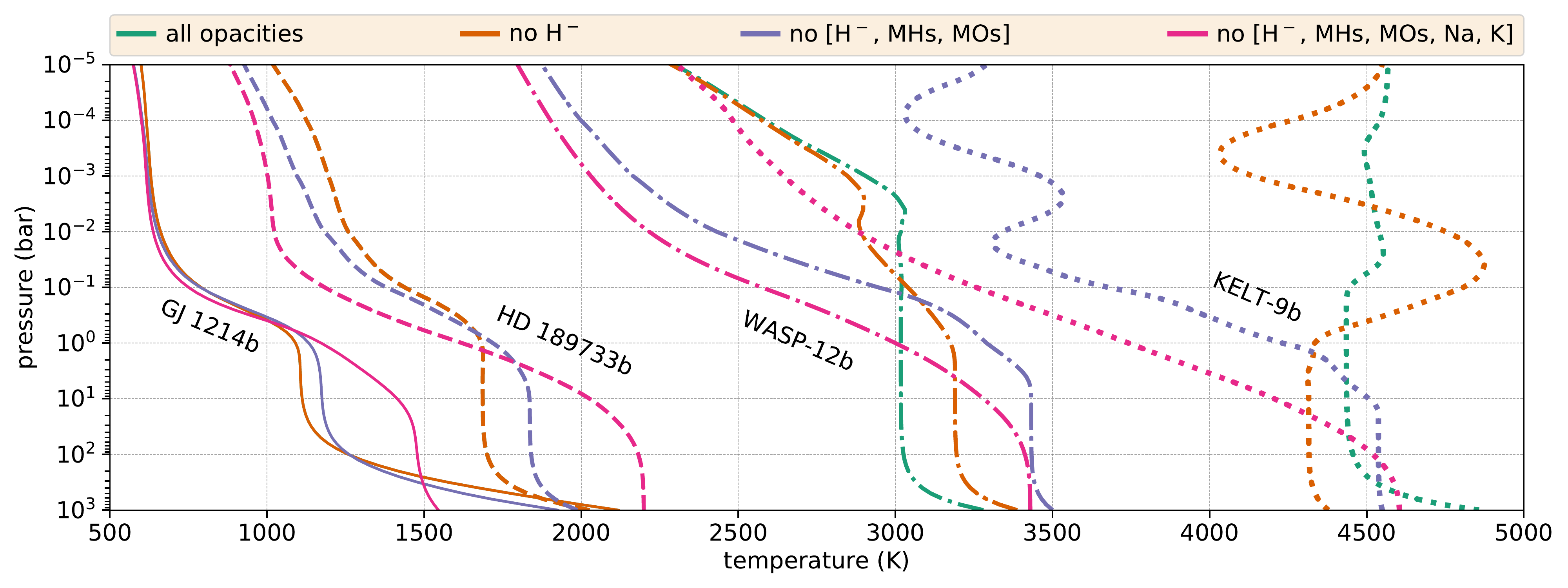}
\end{minipage}
\vspace{-0.3cm}
\caption{Dayside temperature profiles in radiative-convective equilibrium for the planets GJ 1214b ($T_{\rm eff} = 721$ K), HD 189733b ($T_{\rm eff} = 1465$ K), WASP-12b ($T_{\rm eff} = 3013$ K) and KELT-9b ($T_{\rm eff} = 4528$ K). The effect of different samples of opacity sources on the calculated temperatures is shown.  The discrepancy between models increases with planetary effective temperature as the metal oxides, hydrides and H$^-$ starts to dominate the stellar flux absorption. The green profile is perfectly covered by the brown profile in the two colder cases.}
\label{fig:opac_test_irrad}
\end{center}
\end{figure*}

The second set-up consists of the following irradiated planets covering a broad range of effective dayside temperatures: GJ 1214b ($T_{\rm eff} = 721$ K), HD 189733b ($T_{\rm eff} = 1465$ K), WASP-12b ($T_{\rm eff} = 3013$ K) and KELT-9b ($T_{\rm eff} = 4528$ K). We also include an internal temperature of 85 K. For each of these planets we explore the diversity of dayside temperatures obtained, when using the various samples of opacity sources. Our findings are shown in Fig.~\ref{fig:opac_test_irrad}. Two global statements can be made immediately. Firstly, the temperature profiles vary substantially depending on the employed opacities. Secondly, this variance increases with stellar irradiation and effective planetary temperature. For GJ 1214b and HD 189733b samples A and B lead to the same temperatures. Samples C and D lead to cooler upper and warmer lower atmospheres, resulting in a larger atmospheric greenhouse effect. Sample D additionally moves the stellar photosphere downward due to the lack of shortwave absorbers. With WASP-12b and KELT-9b, we move to effective temperatures beyond 3000 K, where H$^-$ opacity starts to have a strong effect. Because H$^-$ introduces a very strong continuum opacity throughout all wavelengths (see Fig.~\ref{fig:opac}), it makes the temperature profile more isothermal. In addition, the metal hydrides and oxides play a strong role in this regime, leading to minor (WASP-12b) or major (KELT-9b) temperature inversions. In the case of KELT-9b, because the stellar irradiation is so strong, even Na and K suffice to lead to two separate temperature inversions. In fact, we expect in very hot atmospheres non-monotonic temperature profiles with multiple inversions to be naturally occurring (see next Sect.~\ref{sec:nonmonotonic}).

In conclusion, we find that missing opacities in models may lead to stark temperature discrepancies. We thus recommend to employ a list of opacity sources as complete as possible to avoid erroneous temperature predictions; an error that directly propagates into the strength of spectroscopic emission features. More specifically, we find the inclusion of metal hydrides, oxides and H$^-$ imperative to obtain accurate temperatures. We remark that in very hot atmospheres atomic or ion species like Fe, Mg, Al or Fe$^+$, Mg$^+$, Ca$^+$ are expected to play a significant role as well, due to their high abundance and vast number of spectral lines in the optical and near-infrared \citep{kitzmann18b}. We further remark that Fe, Fe$^+$ and Ti$^+$ were recently observationally confirmed in the atmosphere of the ultra-hot Jupiter KELT-9b \citep{hoeijmakers18}, reinforcing the notion of the prevalent existence of metals in very hot atmospheres. 

We will investigate the impact of more atoms and ions on self-consistent radiative transfer calculations in a future study.

\subsection{Non-monotonic Temperature Profiles}
\label{sec:nonmonotonic}

As seen in the previous section, we find that for hotter planets with effective temperatures over 2000 K, non-monotonic temperature profiles are not the exception but the norm. In this regime sufficiently high temperatures activate a multitude of rovibronic transitions which generate billions of spectral lines, belonging to many different species, that cause complex extinction patterns for both stellar and planetary radiation. Even more, with planets orbiting cool M-dwarfs or ultra-hot Jupiters exhibiting temperatures like K-stars, the wavelengths of incoming irradiation and local thermal emission may heavily overlap, eroding the classical shortwave and longwave paradigm. In this sense, we feel that the question of whether a planetary atmosphere exhibits a temperature inversion or not (e.g., \citealt{fortney08}) is ill-posed as there may be small or large and one or several inversions, appearing in higher or lower altitudes.

We conduct a numerical test to explore the origins of non-monotonic temperature profiles. As a test set-up we choose to model the dayside of KELT-9b and employ our full opacity list without H$^-$. As seen in Fig.~\ref{fig:opac_test_irrad}, this leads to a temperature profile with two large temperature inversions. In order to better explore the wavelength-dependent opacity effect, we use, for this test only, opacities calculated at a fixed pressure and temperature of $P$ = 1 bar and $T$ = 4500 K. With this simplification, the photospheric pressure at which the optical depth $\tau$ equals unity is given directly by 
\begin{equation}
\label{eq:photo}
P = g / \kappa, 
\end{equation}
where $g$ is the surface gravity and $\kappa$ is the wavelength-dependent opacity function. For the following demonstration it is useful to know that the stellar irradiation of KELT-9 and the planetary thermal emission peak at 0.3 $\mu$m and 0.7 $\mu$m, respectively.

\begin{figure*}
\begin{center}
\begin{minipage}[h]{\textwidth}
\includegraphics[width=\textwidth]{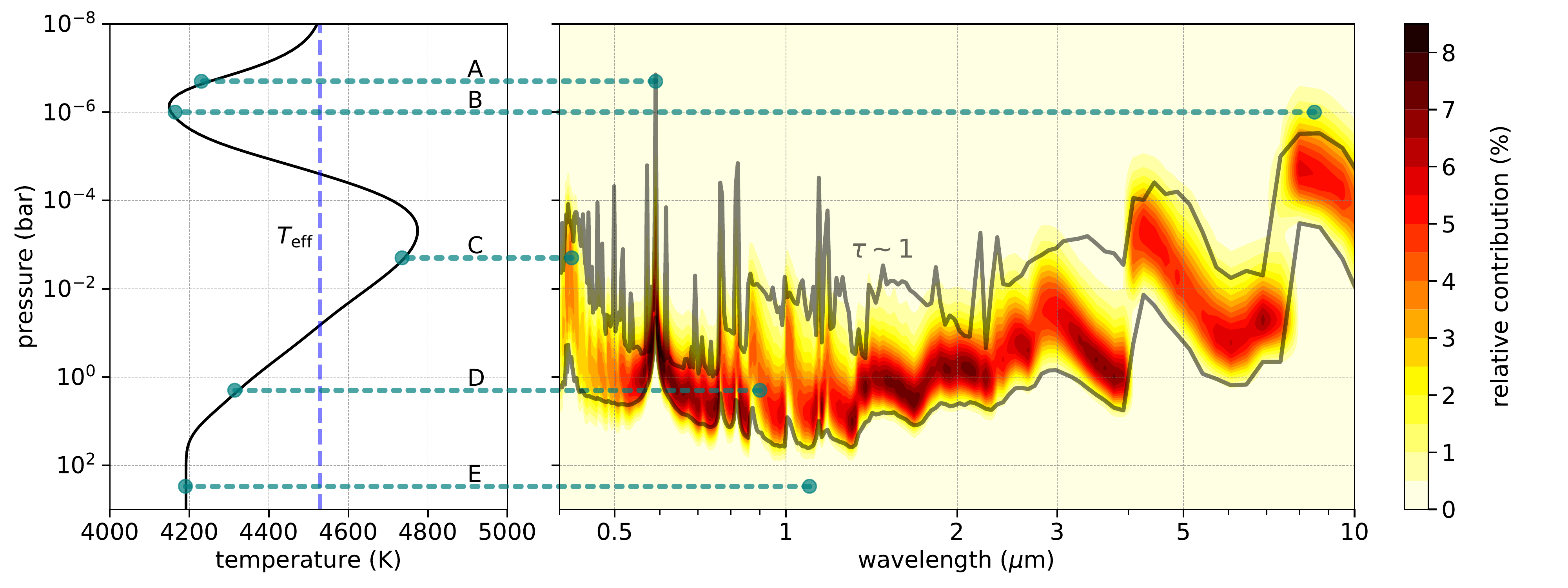}
\end{minipage}
\vspace{-0.3cm}
\caption{{\bf Left:} Converged dayside temperature profile of KELT-9b (left panel) in a simplified test set-up, where the opacities (no H$^{-}$) are held constant with altitude. {\bf Right:} The contribution function and the photosphere (optical depth $\tau \sim 1$) versus wavelength. Due to the use of the k-distribution method, one wavelength bin contains a range of opacities, resulting in a wider photosphere region (enclosed by two gray lines). We discuss the connection between temperatures and the pressure layers and opacity bands they are affected by (lines A to E) in the text.}
\label{fig:app_c}
\end{center}
\end{figure*}

The converged temperature profile is shown in the left panel and the contribution function in the right panel of Fig.~\ref{fig:app_c}. The latter is overlaid by the location of the photosphere, given by eq. (\ref{eq:photo}). Due to the use of the $\kappa$-distribution method, each wavelength bin contains a range of unordered opacities, which returns a range of photospheric pressures. Now, we ask the question which opacity bands and spectral lines are responsible for which features of the temperature profile. To this end, we manually switch off or strongly decrease opacities in certain wavelengths and study the effect on the temperature profiles. We elucidate our findings by focusing on individual temperature regions, which are connected by lines A to E to the pressure layers and opacity bands they are most affected by.

\begin{itemize}

\item{{\bf A:} The temperature inversion above\footnote{``Above'' and ``below'' are meant in spacial sense, i.e., direction lower and higher pressure, respectively.} 10$^{-6}$ bar is caused by the Na resonance doublet. Stellar radiation is absorbed by the line center resulting in an increase of temperature until the thermal radiation emits sufficiently at 0.59 $\mu$m to counter the stellar energy depostion.}

\item{{\bf B:} The dip in temperatures around 10$^{-6}$ bar is mainly caused by efficient cooling through the SiO band at 9 $\mu$m with the band at 4.5 $\mu$m playing a minor part. The layers at this height cool down until the cooling is countered by stellar heating through the other K and Na line peaks at $\lesssim 1$ $\mu$m.}

\item{{\bf C:} The inversion around and below 1 mbar is due to the absorption of stellar radiation by strong shortwave bands of SiH and AlH between 0.4 and 0.5$\mu$m. This region represents the stellar photon deposition depth as the bulk of the stellar flux is absorbed here. The temperatures here increase until the atmosphere can efficiently cool through the SiH and AlH bands as well.}

\item{{\bf D:} At 1 - 10 bars lies the majority of the planetary photosphere. The temperatures are relatively low as the gas is efficient at cooling and only a small fraction of the stellar radiation is left at these depths. The strong emission of thermal radiation is visible by the maxima of the contribution function, which between 0.5 $\mu$m and 2 $\mu$m almost unanimously are located in these layers.}

\item{{\bf E:} Below 100 bar the atmosphere is optically thick to both the thermal as well as the stellar flux. Radiative equilibrium under these isotropic conditions manifests through an isothermal temperature profile.}

\end{itemize}

In conclusion, we find that multiple temperature inversions are caused by combination of the strong spectral bands and lines of metal oxides, metal hydrides and the alkali metals. If these species are present in sufficient atmospheric abundances, as is expected for ultra-hot planets such as KELT-9b, we expect complex temperature profiles with multiple inversions to be the common state. 

However, we note once more that in this test we used opacities calculated at one fixed value for the pressure and temperature, whereas in reality the opacities would change throughout the atmosphere. Finally, we have neglected the H$^-$ continuum opacity, which makes the temperatures substantially more isothermal than presented here (see Fig.~\ref{fig:opac_test_irrad}).

\subsection{Effect of the Stellar Path Length Adjustment}
\label{sec:dir_correct}

As stated in Sect.~\ref{sec:stellarpath}, usual practice of one-dimensional models is to use a plane-parallel grid instead of the accurate spherical geometry of planets. In doing so, the path length of photons in the atmosphere and consequently the optical depth are overestimated. This error can be avoided with a geometrical correction, which shortens the actual stellar path length by adjusting the zenith angle depending on the location in the atmosphere. In the following, we test our implementation of the geometric correction to the stellar path for its influence on modeling exoplanetary atmospheres. We employ our full opacity list and generate temperatures in radiative-convective equilibrium and corresponding mock spectra for the hot Jupiter HD 189733b and the super Earth GJ 1214b. We pick each of those two as poster children for the two prevalent classes of currently characterized exoplanets. We treat the atmosphere of GJ 1214b as hydrogen-dominated, so the effects presented here should be understood as an upper limit on the significance of the path length correction. Heavier atmospheres possess a smaller scale height, diminishing the effect of a path length correction. 

\begin{figure}
\begin{center}
\begin{minipage}[h]{0.49\textwidth}
\includegraphics[width=\textwidth]{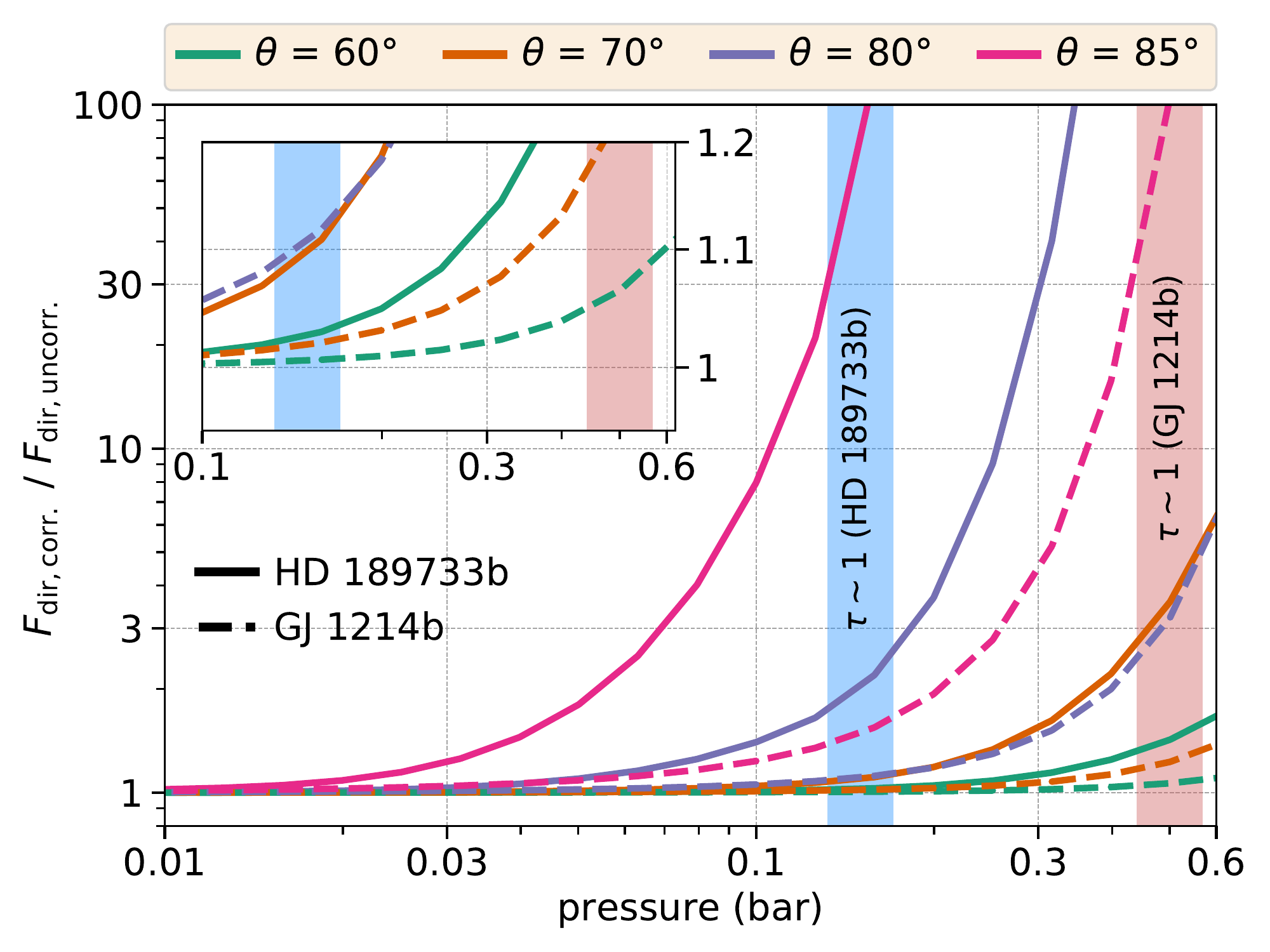}
\end{minipage}
\vspace{-0.3cm}
\caption{The vertical profile of the stellar flux ratio between models  with and without the stellar path length adjustment. The ratio shown for various zenith angles $\theta$ for both HD 189733b and GJ1214b. The approximate photospheres (optical depth $\tau \sim 1$) of these planets are shown at the wavelength of the respective star's peak emission. Pre-calculated dayside temperature profiles are used for this test. Inlaid is a zoom-in of the bottom right area.}
\label{fig:direct_corr}
\end{center}
\end{figure}

As first test, we use a pre-calculated atmospheric temperature profile and let the direct stellar beam propagate downward, once with and once without path length correction. We show in Fig.~\ref{fig:direct_corr} the ratio of the corrected over the uncorrected stellar fluxes with increasing pressure (or atmospheric depth). High up in the atmosphere the ratio is very close to unity, but increases as the stellar photons propagate deeper. As the effect of the geometric correction is relative to the path length, the discrepancy increases with pressure. Also, the error in the uncorrected path length increases with size of the zenith angle. The region of interest is given by the shortwave photosphere (optical depth $\tau \sim 1$), where the stellar photons are deposited and contribute to the atmospheric heating. Layers below the photosphere are left with a vanishingly small stellar flux making an error irrelevant. In Fig.~\ref{fig:direct_corr} the photospheres of HD 189733b and GJ 1214b are shown at the wavelength of the respective star's peak emission. At these depths the ratio of fluxes for HD 18973b and GJ1214b is  $\sim 2$ and $\sim 3$  for zenith angles of 80$^\circ$ and $\sim 1.1$ and $\sim 1.3$ for zenith angles of 70$^\circ$, respectively. We find the effect of the path length correction to be larger for GJ 1214b than for HD 198773b. This is because the path length error, when ignoring the correct geometry, scales with the relative width of the atmosphere, which is given by the ratio of the atmospheric scale height over the planetary radius $H / R_{\rm pl}$. This ratio is larger for GJ 1214b, assuming a hydrogen-dominated atmosphere, then it is for HD 189733b. In fact, we estimate that $(H / R_{\rm pl})_{\rm GJ 1214b} \sim 0.02$, whereas $(H / R_{\rm pl})_{\rm HD189733b} \sim 0.003$. For comparison, the Earth exhibits $(H / R_{\rm pl})_{\rm Earth} \sim 0.001$, meaning the Earth's atmosphere is geometrically thinner, compared to higher-mass planets.

\begin{figure*}
\begin{center}
\begin{minipage}[h]{0.49\textwidth}
\includegraphics[width=\textwidth]{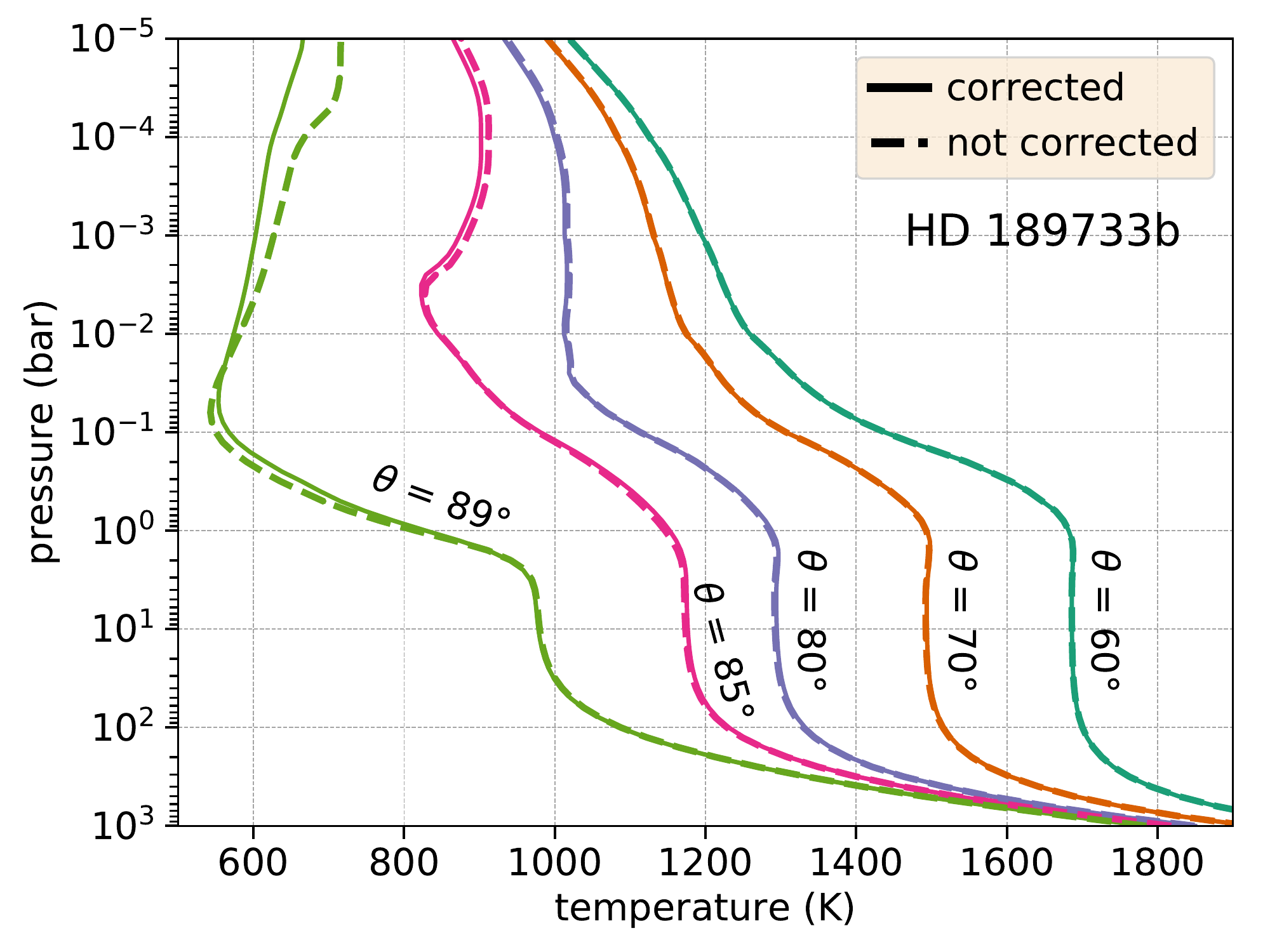}
\end{minipage}
\hfill
\begin{minipage}[h]{0.49\textwidth}
\includegraphics[width=\textwidth]{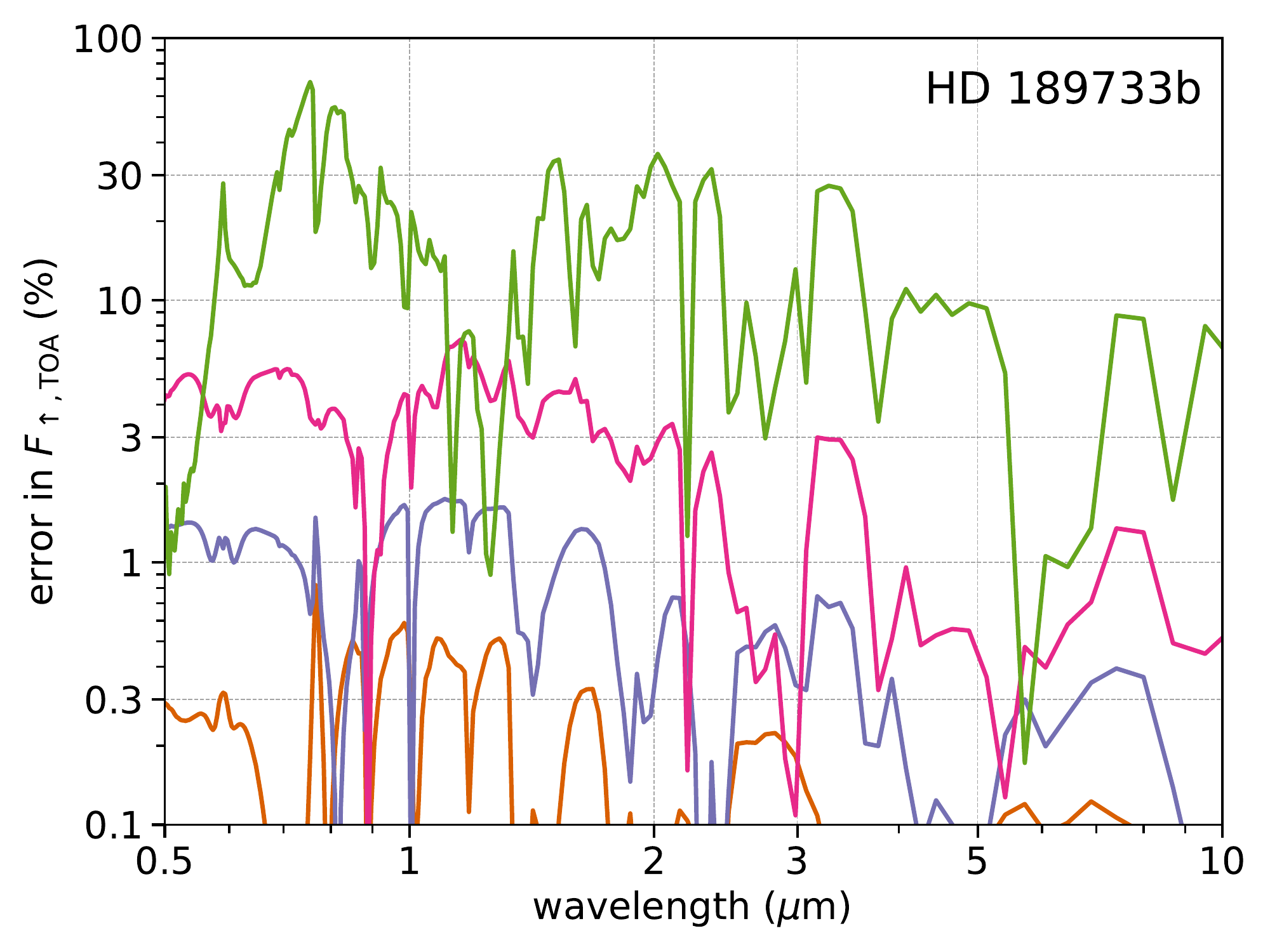}
\end{minipage}
\begin{minipage}[h]{0.49\textwidth}
\includegraphics[width=\textwidth]{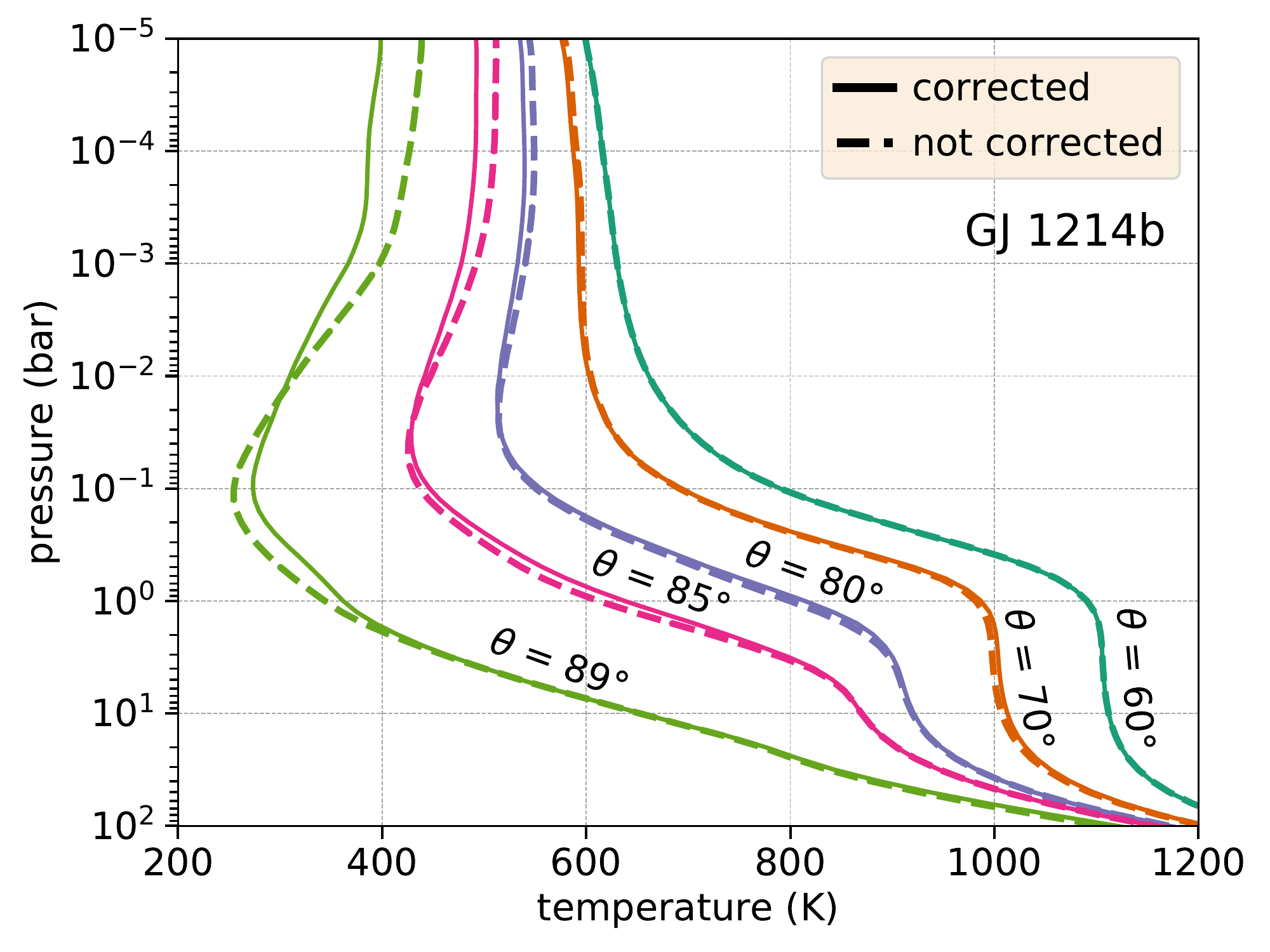}
\end{minipage}
\hfill
\begin{minipage}[h]{0.49\textwidth}
\includegraphics[width=\textwidth]{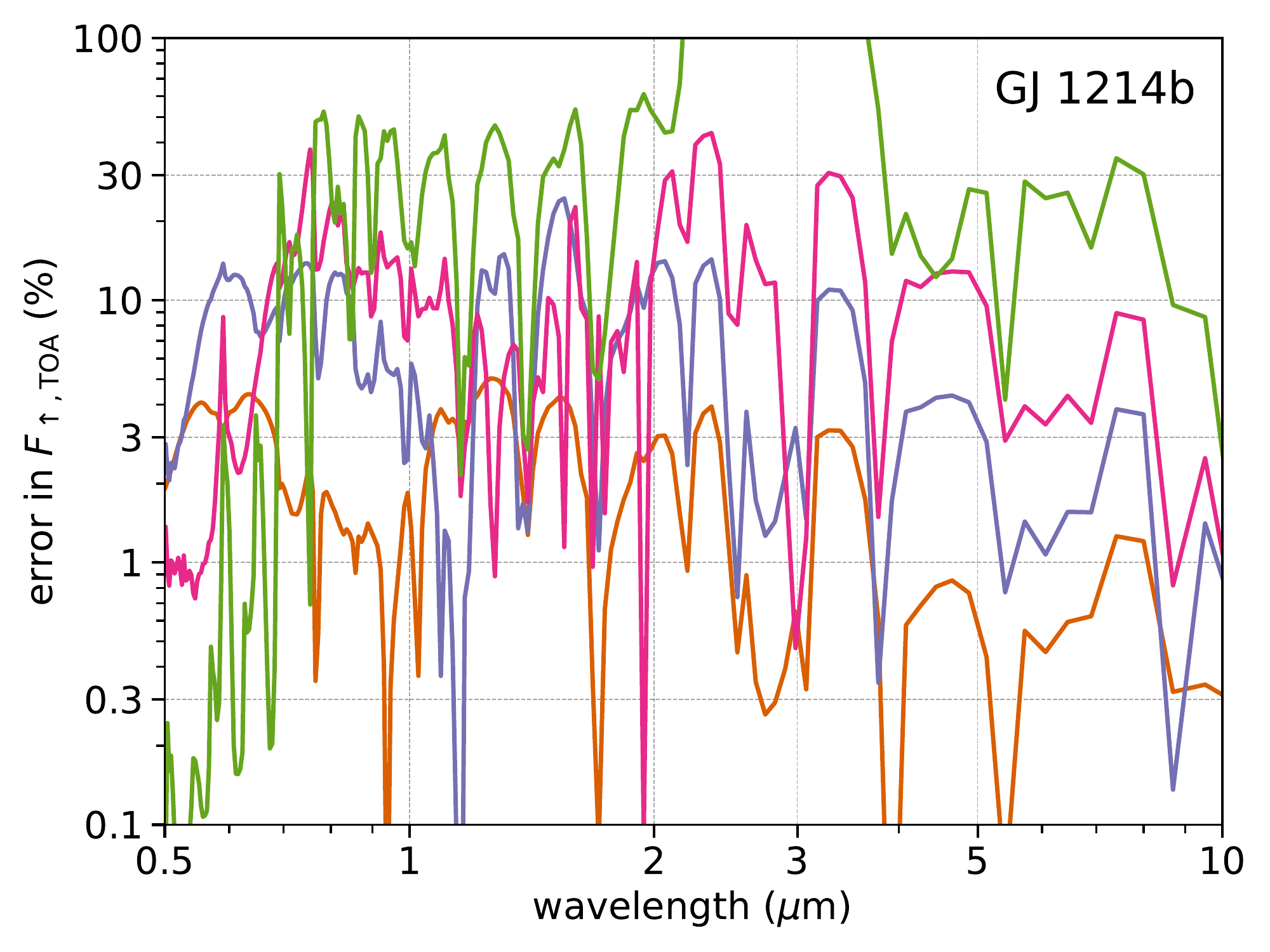}
\end{minipage}
\vspace{-0.3cm}
\caption{{\bf Left panels:} Temperatures in radiative-convective equilibrium for models with and without the stellar path length correction. The effect of the correction is shown for various zenith angles $\theta$ between $60^\circ$ and $89^\circ$. {\bf Right panels:} Error in the outgoing top-of-atmosphere flux versus wavelength when using the uncorrected model. The color-coding matches in the left and right panels. The $\theta = 60^\circ$ case is not shown, as the resulting error is smaller than the depicted values. The top row shows the results for HD 189733b and the bottom row shows GJ 1214b (hydrogen-dominated atmosphere). Due to the strong flux dependence on the temperature $T$, a small relative error $e$ in $T$ translates roughly to the error $4e$ in the emitted flux.}
\label{fig:direct_corr2}
\end{center}
\end{figure*}

As second test we run the fully self-consistent simulation with and without path length correction. We find that the impact of the correction on the temperature profile is negligible for zenith angles up to $70^\circ$, see Fig.~\ref{fig:direct_corr2}, left panels. For zenith angles $\theta > 80^\circ$ the effect becomes noticeable with the uncorrected models overestimating the upper atmospheric heating (pressure $\lesssim$ 0.1 bar). Consequently, due to the larger attenuation of the stellar flux above, the lower atmospheric layers (pressure $\gtrsim$ 0.1 bar) are cooler in the uncorrected models. As in the previous test, the effect is significantly larger for GJ 1214b than for HD 189733b. In Fig.~\ref{fig:direct_corr2}, right panels, we show the effect of the path length correction on the resulting emission spectra. Depicted is the wavelength-dependent percentage error in the outgoing top of the atmosphere (TOA) flux of the uncorrected versus the corrected model for different zenith angles, color-coded analogously to the left panels. For HD 189733b, the errors in the spectrum are exceeding the percent level for zenith angles higher than $80^\circ$. For GJ 1214b even $\theta = 70^\circ$ leads to errors of several percent for some wavelengths.  For both planets, very close at the planetary limb ($\theta = 89^\circ$) the error in the flux is close or exceeds 100 \%. It may be surprising that even though temperatures in the different models are very similar, the error in the flux is significant for high zenith angles. Since the emitted flux depends on the temperature as $T^4$, a small relative error in $T$ translates roughly to four times the error in the flux.

Based on these tests, we conclude that in most cases the path length correction should play a secondary role for zenith angles up to around $70^\circ$. However, if investigating planetary limb regions with zenith angles around $80^\circ$ or higher, we recommend using the path length correction both for the accuracy of the modeled temperatures as well as the spectra. If the atmosphere is expected to be very extended, it may be reasonable to include the correction for zenith angles from $70^\circ$ upward.

\subsection{Impact of the Scattering Correction}
\label{sec:scat_corr}

We explore the effects of the scattering correction, which we have implemented into the hemispheric two-stream formalism of \texttt{HELIOS} (see Sect.~\ref{sec:E_param}). We calculate self-consistently the dayside temperatures and the corresponding spectral emission of HD 198733b, once with and once without the scattering correction. As we do not include clouds in this work, the only present scattering originates from hydrogen Rayleigh scattering. In this case, we would expect isotropic scattering, where $g_0 = 0$. However, to increase the tested parameter range, we artificially vary $g_0$ between 0 and 1 to account for different kinds of scattering particles. The value of $\omega_0$ is given by
\begin{equation}
\label{eq:w0}
\omega_0 = \frac{\sigma_{{\rm H, H_2}}}{\sigma_{{\rm H, H_2}} + \sigma_{\rm mol}},
\end{equation}
where $\sigma_{{\rm H, H_2}}$ is the Rayleigh scattering cross-section due to H and H$_2$, and $\sigma_{\rm mol}$ is the cross-section due to molecular absorption. 

\begin{figure}
\begin{center}
\begin{minipage}[h]{0.49\textwidth}
\includegraphics[width=\textwidth]{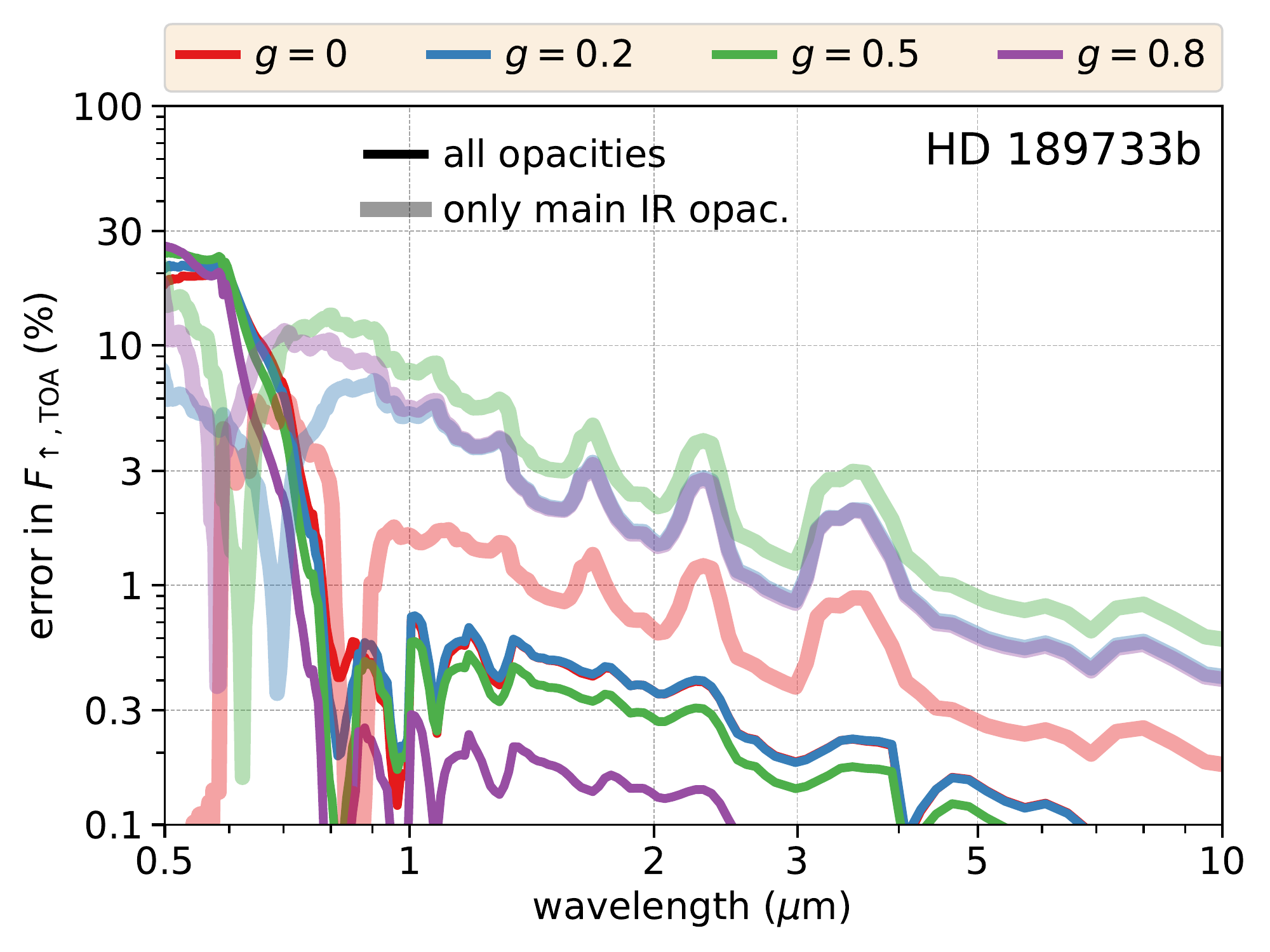}
\end{minipage}
\vspace{-0.3cm}
\caption{Error in the outgoing top-of-atmosphere flux in the hemispheric two-stream model compared to a model which includes the scattering correction. The fluxes are modeled self-consistently for the dayside of HD 189733b. Two tests are conducted, one with all opacity sources, see Table \ref{tab:opac}, and one with only the main infrared absorbers (Fig \ref{fig:opac}, top right panel).}
\label{fig:scat_corr_error}
\end{center}
\end{figure}

In the first test we use our full opacity list and find that the atmospheric temperatures do not change noticeably. Including the scattering correction, the deeper layers (pressure $>$ 1 bar) merely warm about 1 - 2 K compared to the nominal case, regardless of the used $g_0$ (not shown). However, turning our attention to the TOA emitted flux, we find that the correction decreases the backscattered light in the optical wavelengths by about 10 - 30 \%, see Fig.~\ref{fig:scat_corr_error}. For longer wavelengths the discrepancy drops below 1 \%, as here the outgoing flux is dominated by thermal emission. In general we also observe that the discrepancy, i.e., the error in the uncorrected model, becomes smaller as $g_0$ approaches unity. This may appear counter-intuitive as one could think that two-stream methods fare more poorly for larger aerosols, who exhibit forward scattering ($g_0 \approx$ 1). However, our results are consistent with \cite{heng18}, well visualized in their Fig. 1, who show that the scattering correction vanishes, i.e., the factor $E \rightarrow 1$, as $g_0$ goes to unity. Apart from this, we find our reduction of scattered flux consistent with \cite{kitzmann13}'s result that two-stream methods over-predict scattering by up to 20 \%. 

In a second test we include only the main infrared absorbers, shown in Fig.~\ref{fig:opac}, top right panel. In this way, it is the Rayleigh scattering that becomes responsible for the bulk of the stellar flux extinction. This significantly increases the value of $w_0$ in the optical wavelengths, see eq.~(\ref{eq:w0}), making the impact of the scattering correction larger. In this scenario, the bottom atmospheric temperatures diverge between 50 K and 200 K depending on the $g_0$ parameter, and the corrected model is again warmer in the bottom (not shown). Hence, tuning the strength of the scattering directly affects the extent of the greenhouse warming. The thermal emission is subsequently more discrepant, caused by the larger temperature differences between the models. As can be seen in Fig.~\ref{fig:scat_corr_error}, the error exceeds several percent in the near-infrared. The scattered flux discrepancy in the optical is with values around 10 \% on the same order of magnitude as in the previous test.

We conclude that the scattering correction provides a significant improvement for the accuracy of the stellar flux scattering. If scattering dominates over the absorption at shorter wavelengths, any potential errors on the stellar flux propagate to the temperatures and the thermal emission.

\subsection{Atmospheric Model Library}

In addition to the model comparisons and numerical tests presented, we also include self-consistently calculated cloud-free atmospheric models using our improved radiative transfer code \texttt{HELIOS}. We include the full opacity list, as given in Table~\ref{tab:opac}, and the scattering correction, as described in Sect.~\ref{sec:E_param}. Each model output provides amongst others the atmospheric temperature profile in radiative-convective equilibrium and a post-processed high-resolution emission spectrum, as described in Sect.~\ref{sec:opac}. Additionally, it contains information about any other calculated quantity, like the entropy, optical depth, contribution function, etc.

Below, we first describe the parameters of the grid for the self-luminous planets and then we elaborate on the calculated sample of irradiated planets. Finally, we provide information how to download the models.

\subsubsection{Grid of Self-luminous Planets}

The calculated atmosphere grids for self-luminous planets span the following dimensions. The effective temperature range is $T_{\rm eff}$ = [200 K, 3000 K]  with a step size $\Delta T_{\rm eff}$ = 100 K. The surface gravity, $\log_{10} g$ = [1.4, 5.0], with $\Delta \log_{10} g$ = 0.2, [g] = cm s$^{-2}$. The metallicity range is [$M$/H] = [-1, 1] with $\Delta $[$M$/H] = 0.5, where we write
\begin{equation}
\left[\frac{M}{{\rm H}}\right] = \log_{10}\left(\frac{n_M}{n_{\rm H}}\right) - \log_{10}\left(\frac{n_{M, \Sun}}{n_{\rm H, \Sun}}\right),
\end{equation}
with $n_X$ being the number density of element $X$ and $n_{X, \Sun}$ being the solar photospheric value \citep{asplund09}. We write $M$ for every element heavier than Helium. For each metallicity we further calculate models with three values for the C/O ratio: 0.1, solar (0.55), and 1. The temperatures are calculated until a pressure of 1000 bar, so that every model atmosphere couples to the deep convective zone. The limits of the grid are chosen to accommodate low-mass young planets and brown dwarfs and will be extended in future to include older, high-mass brown dwarfs with $\log_{10} g > 5$.

\begin{figure*}
\begin{center}
\begin{minipage}[h]{0.49\textwidth}
\includegraphics[width=\textwidth]{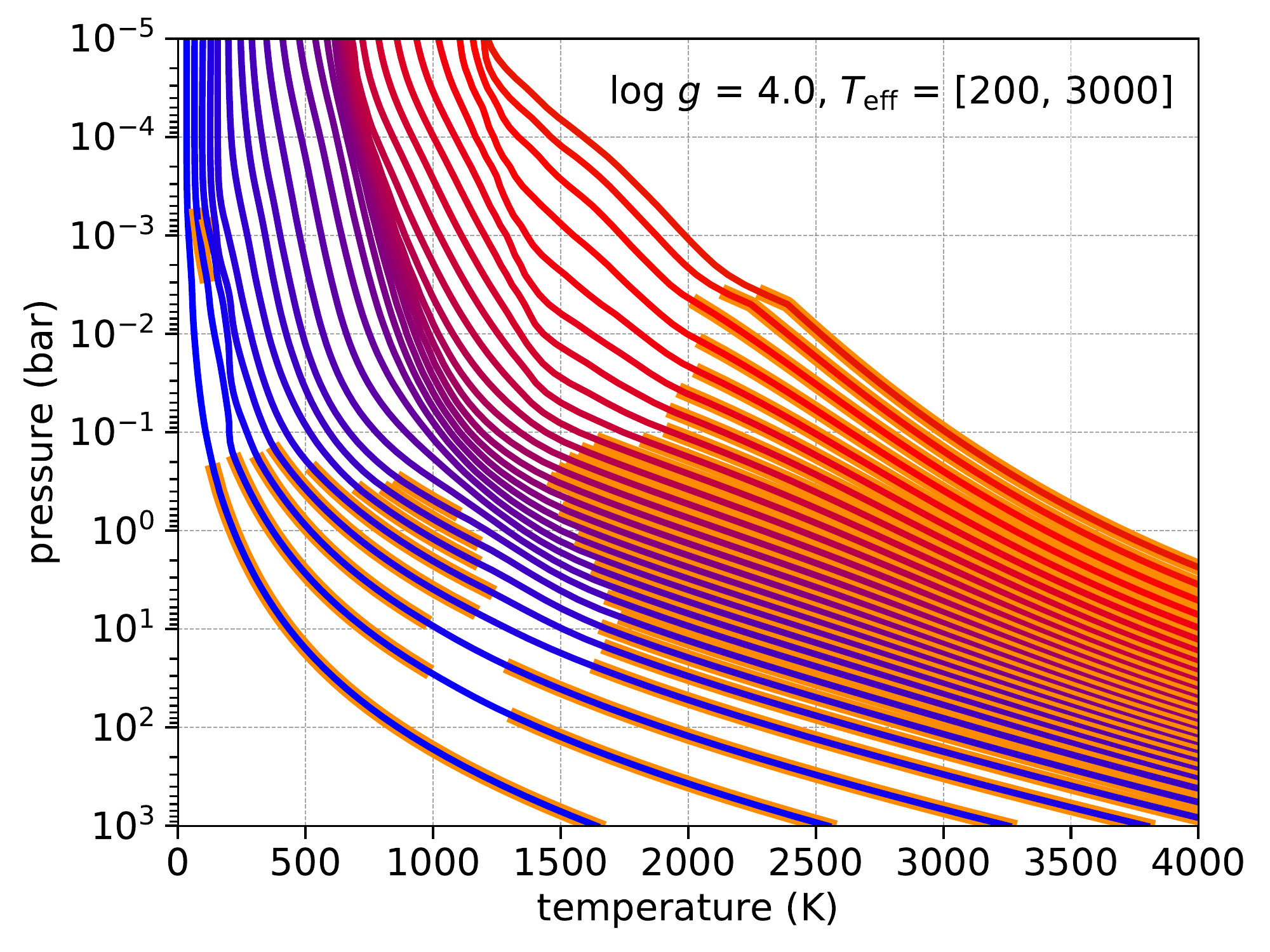}
\end{minipage}
\hfill
\begin{minipage}[h]{0.49\textwidth}
\includegraphics[width=\textwidth]{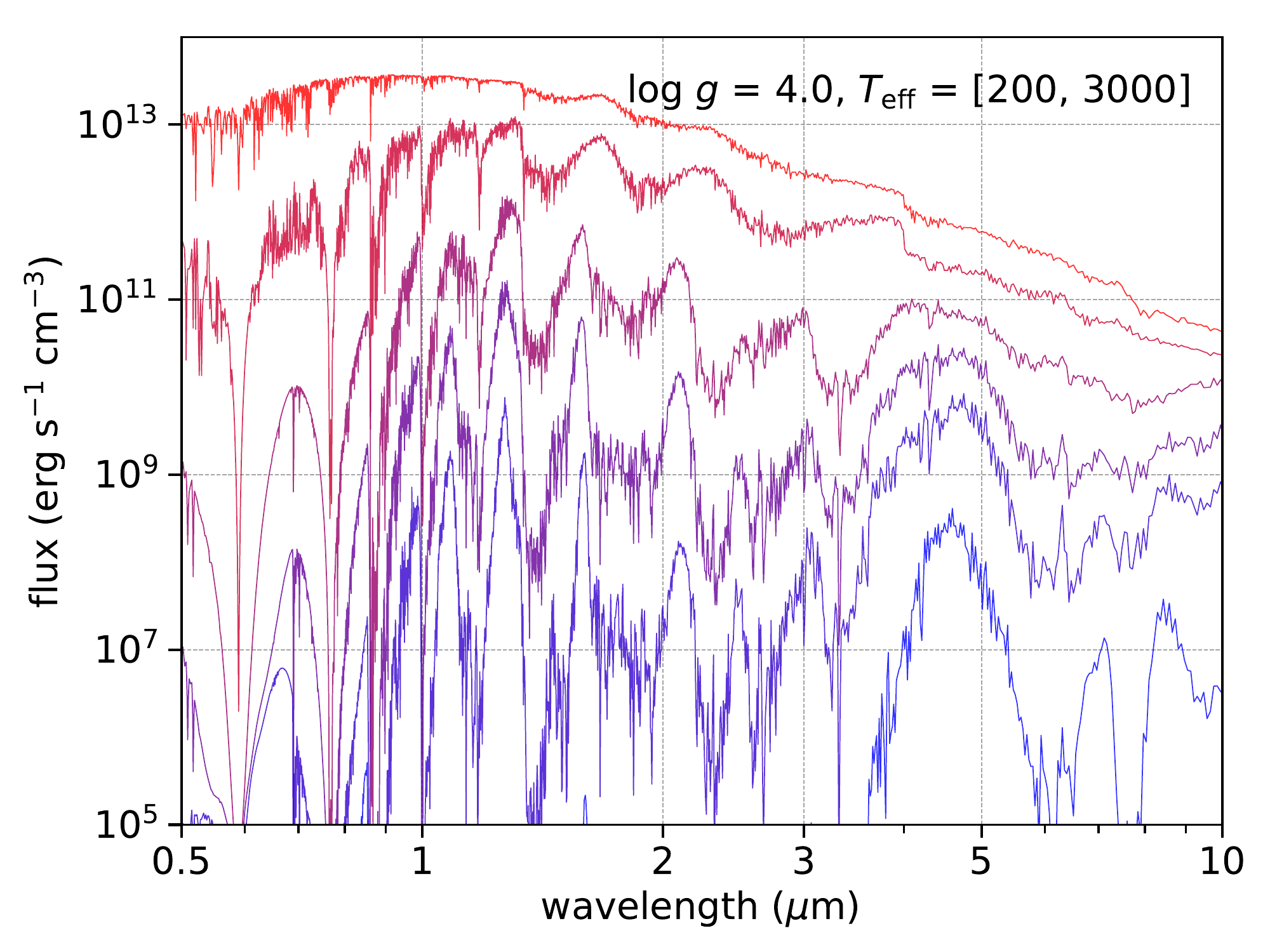}
\end{minipage}
\vspace{-0.3cm}
\caption{{\bf Left:} Grid of temperatures in radiative-convective equilbrium for various effective temperatures ($\Delta T_{\rm eff} = 100$ K) using $\log_{10} g$ = 4.0. Convective zones are marked with wider orange lines. {\bf Right:} Top-of-atmosphere emission spectra for a subset of the calculated models. Depicted are six models with $T_{\rm eff}$ = 200 K, 400 K, 600 K, 1000 K, 2000 K and 3000 K, using solar metallicity and C/O ratio. The spectra are downsampled in resolution for clarity.}
\label{fig:selflum}
\end{center}
\end{figure*}

Fig.~\ref{fig:selflum} shows an excerpt of the grid for self-luminous planets for various effective temperatures using $\log_{10} g$ = 4.0. Displayed are the temperatures in radiative-convective equilibrium and the corresponding TOA emission spectra for some of the models. Note, that the spectra shown are downsampled from the native resolution for clarity.

\begin{figure*}
\begin{center}
\begin{minipage}[h]{0.49\textwidth}
\includegraphics[width=\textwidth]{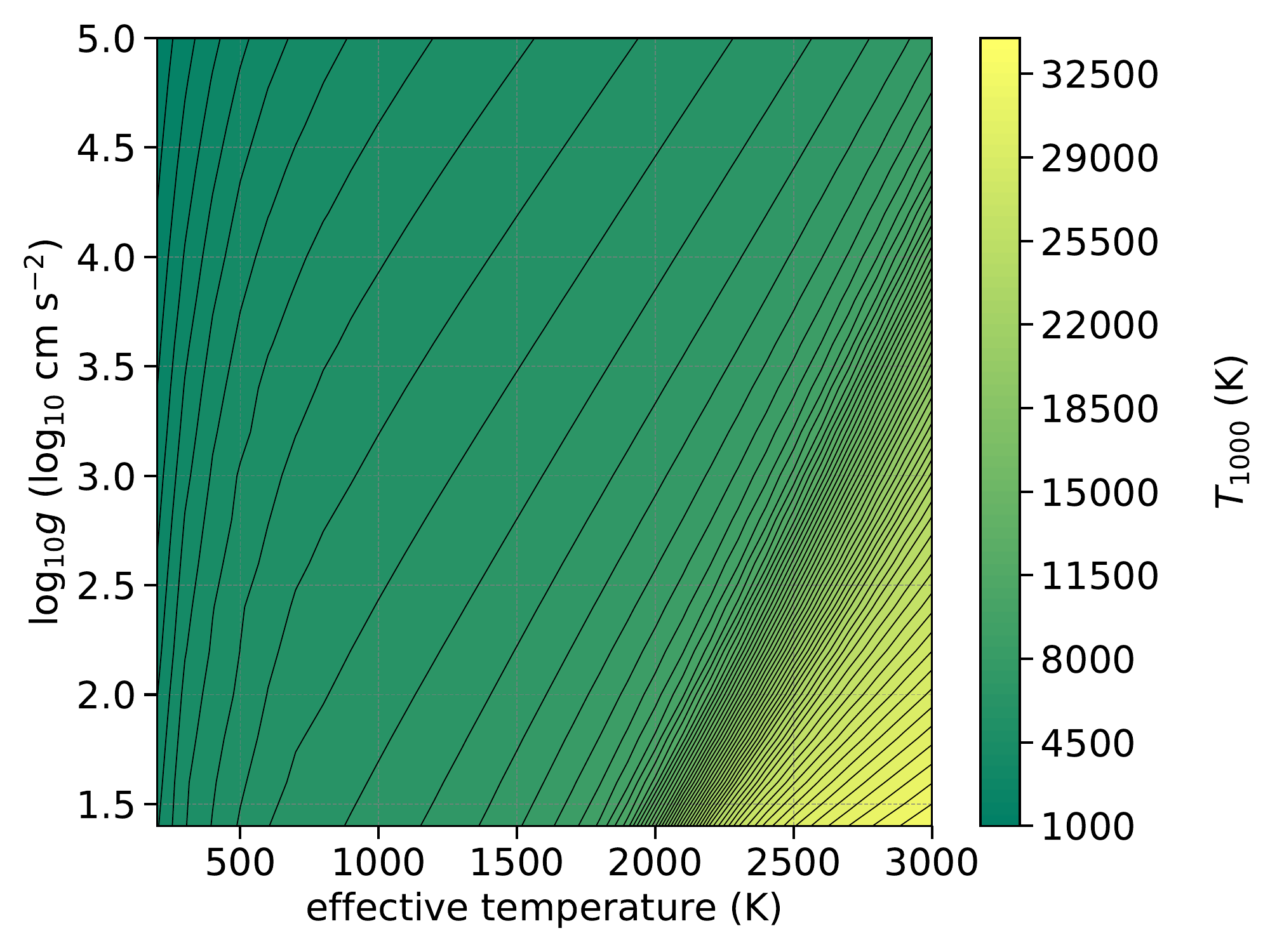}
\end{minipage}
\hfill
\begin{minipage}[h]{0.49\textwidth}
\includegraphics[width=\textwidth]{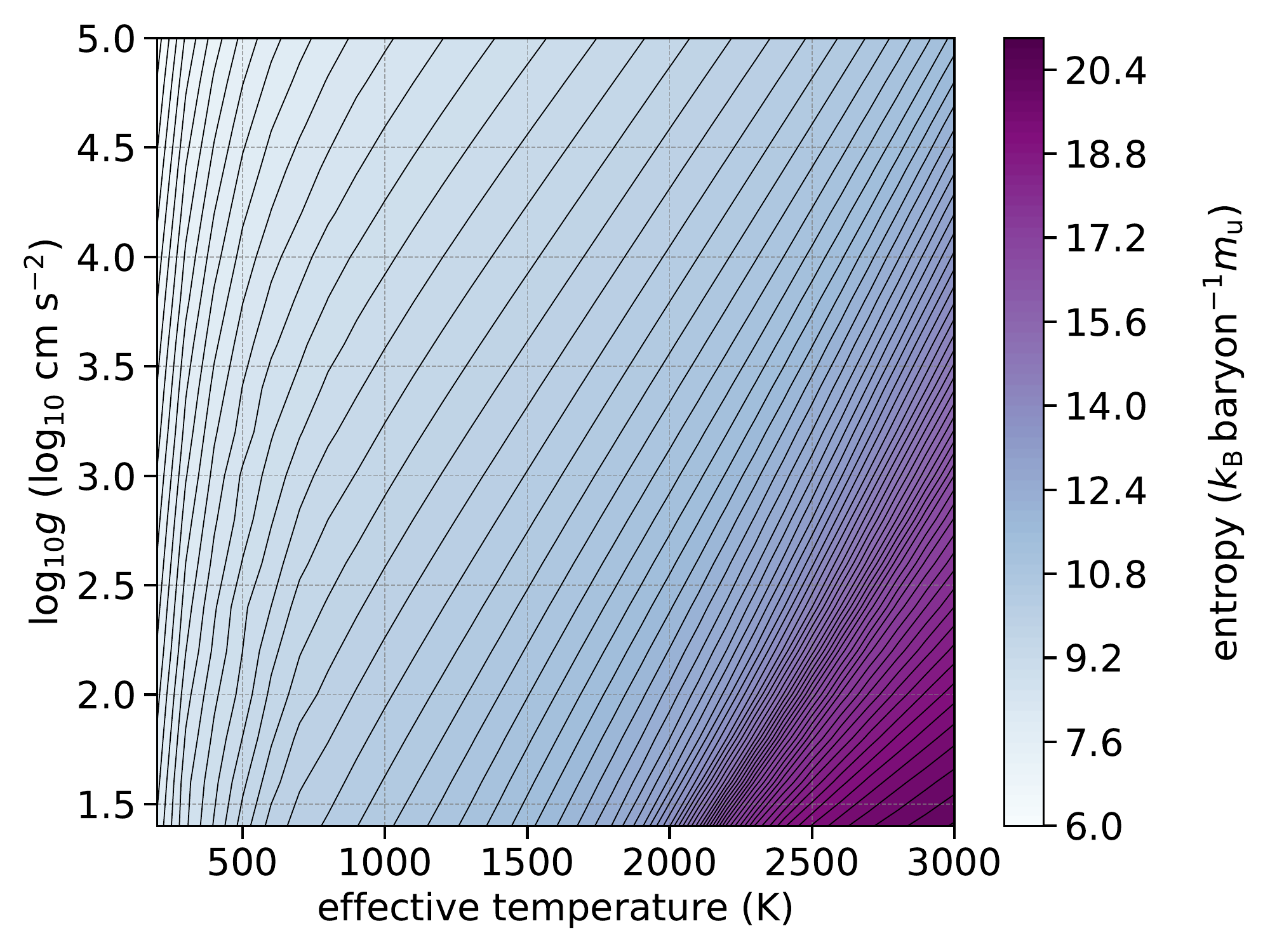}
\end{minipage}
\vspace{-0.3cm}
\caption{The temperature in the deep convective zone at a preFig.~ssure of 1000 bar, $T_{1000}$, (left) and the entropy (right) for the model grid calculated, spanning a range in effective temperature and surface gravity $g$. The contour lines are separated by 500 K on the left, and by 0.2 $k_{\rm B}\,{\rm baryon}^{-1} m_{\rm u}$ on the right, with $k_{\rm B}$ being the Boltzmann constant and $m_{\rm u}$ the atomic mass unit. Depicted are the models for solar metallicity and C/O ratio.}
\label{fig:t1000}
\end{center}
\end{figure*}

Fig.~\ref{fig:t1000}, left panel shows the the temperature at a pressure of 1000 bar, $T_{1000}$, for the model grid calculated. As $T_{1000}$ lies within the deep convective zone for all calculated models, it can be used directly to extrapolate to the interior adiabat (c.f. \citealt{burrows97}). Otherwise, we also provide the entropy for direct coupling, see Fig.~\ref{fig:t1000}, right panel. 

\begin{figure*}
\begin{center}
\begin{minipage}[h]{0.49\textwidth}
\includegraphics[width=\textwidth]{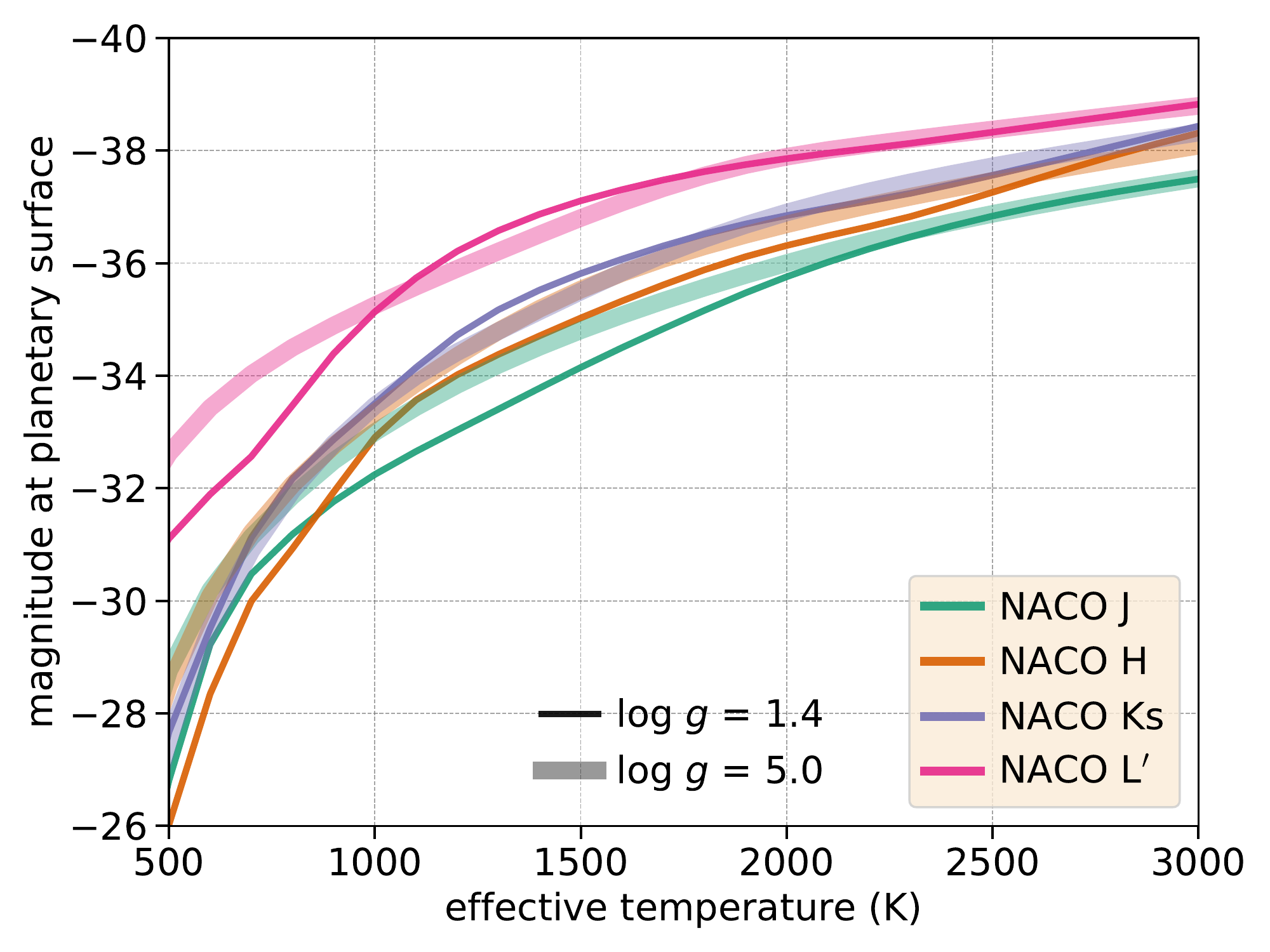}
\end{minipage}
\hfill
\begin{minipage}[h]{0.49\textwidth}
\includegraphics[width=\textwidth]{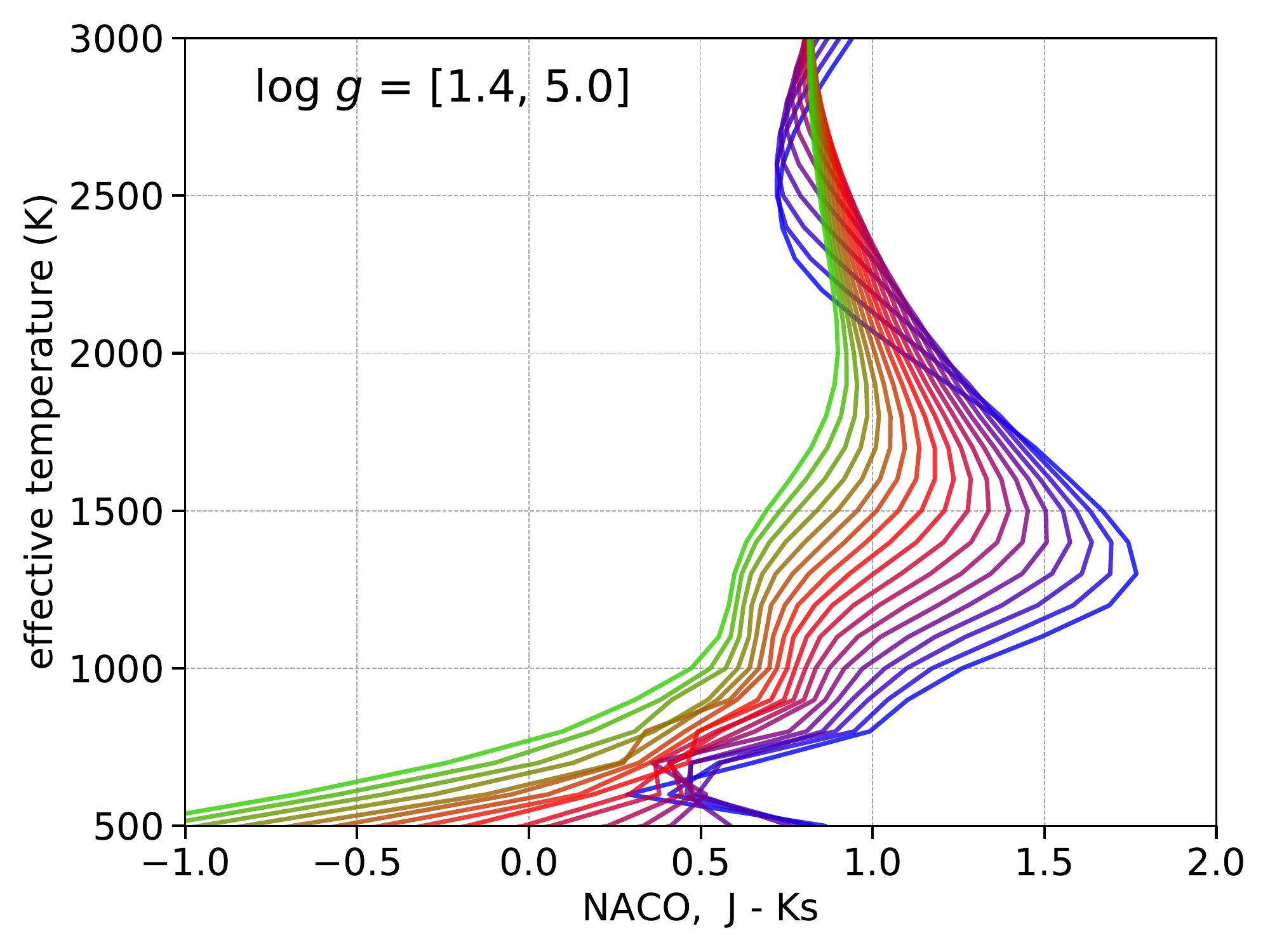}
\end{minipage}
\vspace{-0.3cm}
\caption{{\bf Left:} Magnitude at planetary surface for the \texttt{NACO} J, H, Ks \& L' filters for self-luminous planets of varying effective temperature, shown for two surface gravities. {\bf Right:} Self-luminous planets of varying effective temperature versus \texttt{NACO} J - Ks magnitude. The surface gravities shown range from $\log_{10} g$ = 1.4 (blue) to $\log_{10} g$ = 5.0 (green). Depicted are the models for solar metallicity and C/O ratio.}
\label{fig:magnitudes}
\end{center}
\end{figure*}

In addition to the spectral fluxes, we provide a table with the expected magnitudes at planetary surface for the NACO J, H, Ks, \& L' filter bandpasses, see Fig.~\ref{fig:magnitudes} for an excerpt. They are calculated by convolving the filter profiles with the spectral flux. With a Vega spectrum\footnote{From the Hubble Space Telescope Science Institute web page as alpha\_lyr\_stis\_003.txt.}  the zero points were obtained. For more details see \cite{linder18}.

\subsubsection{Sample of Irradiated Planets}
\label{sec:irradplanets}

In addition to the self-luminous planets, we provide atmospheric models for a sample of irradiated planets of interest. Table \ref{tab:planets} lists the planets currently included in our model library together with the system parameters used. For each planet we calculate 3 $\times$ 3 $\times$ 2 = 18 models, where we vary the metallicity, [$M$/H] = -1, 0, 1, C/O = 0.1, solar (0.55), 1, and the heat redistribution factor $f$ = 1/2, 2/3. The latter parameter accounts for the heat redistribution efficiency from the day- to the nightside \citep{burrows08, spiegel10}. Choosing $f$ = 1/2 is computationally equivalent to setting the zenith angle to 60$^\circ$. For these models we conservatively use a blackbody for the stellar emission and an internal flux consistent with 75 K thermal emission.

\begin{figure*}
\begin{center}
\begin{minipage}[h]{\textwidth}
\includegraphics[width=\textwidth]{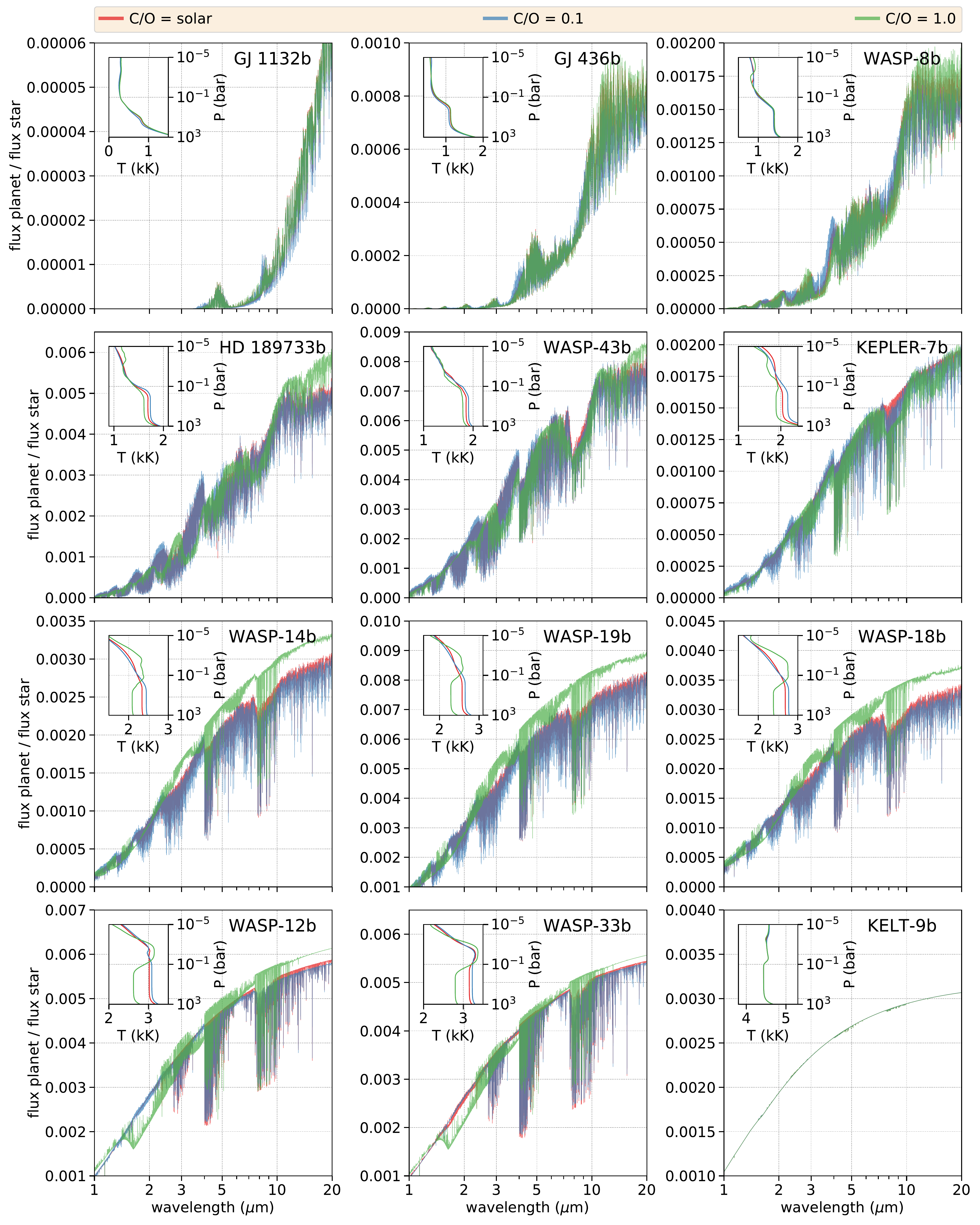}
\end{minipage}
\vspace{-0.3cm}
\caption{Secondary eclipse spectra and dayside temperature profiles for a varying C/O ratio and solar metallicity, shown for the sample of irradiated planets provided. The planets are listed with increasing effective temperature.}
\label{fig:irrad_grid}
\end{center}
\end{figure*}

Fig.~\ref{fig:irrad_grid} displays the dayside temperatures and the secondary eclipse spectra of the calculated planets for f = 1/2, solar metallicity, and three different C/O ratios. The planets are listed with increasing effective temperature.

\begin{table}
	\caption{System parameters used in this study. $R_{\rm pl}$ ($R_{\rm Jup}$): planetary radius, $g$ (cm s$^{-2}$): surface gravity, $a$ (AU): orbital distance, $R_{\ast}$ ($R_{\Sun}$): stellar radius, $T_{\ast}$ (K): stellar temperature.}
	\label{tab:planets}
	\vspace{-0.3cm}
\begin{center}
\bgroup
\def\arraystretch{1.1}
  \begin{tabular}{l r r r r r}
    \hline
name & 												\multicolumn{1}{l}{$R_{\rm pl}$} & \multicolumn{1}{l}{$\log_{10}$ g} & \multicolumn{1}{l}{$a$}  & \multicolumn{1}{l}{$R_{\ast}$}  & \multicolumn{1}{l}{$T_{\ast}$}    \\
\hhline{======}
GJ 1132b\footnote{\cite{berta15, southworth17, bonfils18}} &											0.1058							&	3.068								&	0.0153				 &	0.207						&	3270 \\
GJ 436b\footnote{\cite{butler04}} & 											0.38								& 	3.106								& 	0.04085				 & 	0.464 						&	3684 \\
HD 189733b\footnote{\cite{southworth10, dekok13, boyajian15}} &		1.216							&	3.29									&	0.03142 				 &	0.805 						& 	5050 \\
Kepler-7b\footnote{\cite{latham10, demory11}} & 1.614	&	2.62	&	0.06246	&	1.966	&	5933 \\
KELT-9b\footnote{\cite{gaudi17}} & 1.891	&	3.30		& 0.03462	 & 	2.362	&	9560 \\
WASP-8b\footnote{\cite{queloz10}} & 1.038 	&	3.74	 &	0.0801	&	0.945	&	5600 \\
WASP-12b\footnote{\cite{hebb09, chan11}} & 1.776		 & 3.066	 &	0.02293		& 1.595	&	6300 \\
WASP-14b\footnote{\cite{joshi09}} & 1.281	&	4.01	&	0.036	&	1.306	&	6475  \\
WASP-18b\footnote{\cite{southworth09, hellier09}} & 1.165	&	4.281	&	0.0247	&	1.23	 &	6400 \\
WASP-19b\footnote{\cite{hebb10, hellier11}} & 1.386	&	3.143	 &	0.01655	&	0.99	&	5500 \\
WASP-33b\footnote{\cite{collier10, kovacs13, lehmann15}}  & 1.679	&	3.46	 &	0.0259	&	1.509	&	7430  \\
WASP-43b\footnote{\cite{gillon12}} & 1.036	&	3.672		& 0.0152	&	0.667	&	4520 \\
\hline
\end{tabular}
	\egroup
	\end{center}
	\vspace{-0.1cm}
\end{table}

\subsubsection{Download}

The atmospheric models are available for download on our server. For each model, we provide both the raw \texttt{HELIOS} output files, as well as the temperature profile and the emission spectrum alone. This gives both full access to all the employed model parameters and quick access to the impatient.

The download link is given in the \texttt{readme} file of \texttt{HELIOS} in the {\it GitHub} repository 

\vspace{-0.2cm}
\begin{center}{\url{https://github.com/exoclime/HELIOS}}.\end{center}
\vspace{-0.2cm}
Additionally, the same \texttt{readme} file describes the format of the output files and the directory structure.

The provided atmospheric models are a snapshot of current efforts. We plan to update them in regular intervals, e.g., when we incorporate new opacity data into our code. We also plan to extend the self-luminous planet grids to larger parameter ranges or, if requested, fill out the calculated grids by choosing smaller parameter steps. For the irradiated planets we are going to include more objects as interest arises. Requests may be submitted to this work's lead author.

Finally, as all our modeling tools used in this work, namely \texttt{HELIOS} (\citealt{malik17}; this work), \texttt{HELIOS-K} \citep{grimm15} and \texttt{FastChem} \citep{stock18}, are open-source and publicly available at \url{github.com/exoclime}, the scientific community is encouraged to reproduce and test the model atmospheres provided.

\section{Summary \& Outlook}
\label{sec:sum}

We have presented a number of methodological improvements and new ingredients to the radiative transfer code \texttt{HELIOS}. We have included must-have utilities that any radiative transfer package should contain and were dearly missing in the original version of \texttt{HELIOS} when introduced in \cite{malik17}: the calculation of a {\it direct irradiation beam}, the possibility of {\it convective adjustment} and the output of the {\it contribution function}. Additionally, we have vastly extended our list of opacity sources by including important metal oxides and hydrides, such as TiO, VO, AlO, SiH, the alkali metals, Na \& K and H$^-$, which play an imperative role for very hot atmospheres \citep{sharp07}. Furthermore, we have implemented a geometrical {\it adjustment to the stellar path length} and a {\it scattering correction factor}, new features which bring \texttt{HELIOS} on par with other state-of-the-art radiative transfer solvers for exoplanetary atmospheres on the market.

As part of this study, we have compared \texttt{HELIOS} with other radiative transfer codes and found good consistency with their results, given the diversity of radiative transfer techniques and differences in the opacities between the models. Next, we have checked the impact of different sets of opacities and new incorporated additions to the code on atmospheric properties. We have found that with higher atmospheric temperatures it becomes increasingly important to employ metal hydrides, metal oxides, and H$^-$ opacities as their abundance and thus influence on atmospheric extinction grows. We have found that the solar path length correction leads to significant improvements for zenith angles over 80$^{\circ}$, or even 70$^{\circ}$ for widely extended atmospheres. Finally, the scattering correction improves the accuracy of stellar light scattering compared to the hemispheric two-stream method by 10 - 30 \%, which is consistent with what earlier studies found \citep{kitzmann13}.

With \texttt{HELIOS} we have generated a large grid of self-consistently calculated model atmospheres for self-luminous planets and brown dwarfs. The goal is to provide the community with atmospheres, which can be used by planetary evolution models as constraints on the cooling rate of evolving planets. The corresponding high-resolution spectra help assess the planet's observability during the planet's evolution. We hereby follow-up on previous efforts (e.g., \citealt{burrows97, spiegel12}).

We have further calculated dayside atmospheres for a sample of irradiated planets of interest, which may be used for spectral reconnaissance and to study the influence of various chemical compositions on the dayside temperatures and spectral signatures. In particular, very and ultra-hot Jupiters like WASP-103b, WASP-33b, KELT-20b, and particularly KELT-9b, peak current interests as with dayside temperatures exceeding 3000 K they are expected to offer a view into their atmospheric structure without obstruction by thermodynamically formed and chemical disequilibrium clouds. Self-consistent models, such as \texttt{HELIOS}, may provide useful constraints on the feasibility of atmospheric characterizations with current and future telescopes like the {\sl Hubble} and {\sl James Webb Space Telescopes}.

For the future we see many opportunities on how to expand on the current work. We plan to self-consistently include disequilibrium chemistry, which is important for correct estimates on the strength of spectral features \citep{drummond16}. Furthermore, for colder planets clouds are likely to present a significant hindrance since their strong scattering and absorption may substantially mask spectral signatures. Although we currently do not include clouds in our calculations we are able to predict if an observed atmosphere is consistent with a cloud-free scenario. We may tackle the cloud question in a future study. A cloud scheme sufficient to assess the spectroscopic impact of aerosols could consist of the following components: a cloud structure model \citep{ackerman01, ohno18, gao18}, an estimate of the condensate species and grain composition \citep{lee18} and corresponding extinction efficiencies \citep{kitzmann18a}.

\section*{Acknowledgements}

M.M.,  D.K., J.M., S.G., G.-D.M., E.L., S.-M.T. and K.H. thank the Swiss National Science Foundation (SNF), the Center for Space and Habitability (CSH), the PlanetS National Center of Competence in Research (NCCR) and the MERAC Foundation for partial financial support. G.-D.M. acknowledges the support of the DFG priority program SPP 1992 ``Exploring the Diversity of Extrasolar Planets'' (KU 2849/7-1) and from the Swiss National Science Foundation under grant BSSGI0\_155816 ``PlanetsInTime''. Furthermore, the authors thank P. Cubillos for sharing his recipe for the alkali resonance wings and S. Gandhi for sharing his model data. Lastly, we thank the anonymous referee for improving the quality of this work.

This work has made use of the VALD database, operated at Uppsala University, the Institute of Astronomy RAS in Moscow, and the University of Vienna.

\software{\\
\texttt{HELIOS} (\citealt{malik17}; \url{github.com/exoclime/helios}), \\
\texttt{HELIOS-K} (\citealt{grimm15}; \url{github.com/exoclime/helios-k}), \\
\texttt{FastChem} \citep{stock18}; \url{github.com/exoclime/fastchem}), \\
\texttt{CUDA} \citep{cuda}, \\
\texttt{PyCUDA} \citep{pycuda}, \\
\texttt{python} \citep{python}, \\ 
\texttt{scipy} \citep{scipy}, \\ 
\texttt{numpy} \citep{numpy}, \\
\texttt{matplotlib} \citep{matplotlib}.
}

\clearpage

\appendix
\vspace{-0.1cm}
\section{Calculation of the Alkali Opacities}
\label{app:alkali}

We split the spectral lines of Na and K into two groups: the first consists of the four resonance lines, and the second contains all the other, weaker lines. The lines in the two groups are calculated differently.

\subsection{Resonance lines}

The four resonance lines alone, i.e., the D-doublet for Na at 0.589 $\mu$m and the doublet for K I at 0.77 $\mu$m, account for very strong absorption in the optical and are among the most detectable spectral lines in exoplanets and brown dwarfs \citep{seager00, burrows02a, charbonneau02}. For modeling purposes, they pose a substantial challenge as their line wings deviate significantly from a Lorentzian profile. Attempts to characterize the resonance line wings are motivated by theoretical quantum chemistry and enhanced with laboratory measurements \citep{burrows00, burrows03, nallard12, allard16}. We model the far wings as a composite of the Voigt profile in the line core, stitched together with a damping prescription for the far wings as described in \cite{burrows00}.  

While we are aware of an ongoing debate between Burrows et al. and Allard et al. regarding the correct shape of the sodium and potassium profiles in the line wings, we have chosen to adopt the \cite{burrows00, burrows03} formulation as it is available in analytical form and straightforwardly implementable.  Since the current study is focused on the methodology of \texttt{HELIOS}, rather than the interpretation of measured spectra, we deem it reasonable to defer this issue to a future study.

The line profiles in their core follow a Voigt profile with the collisional broadening half-widths at half maximum (HWHM) $\Gamma$ calculated from impact theory. The used values are $\Gamma_{\rm Na} = 0.071 (T/2000 {\rm K})^{-0.7}$ cm$^{-1}$ atm$^{-1}$ for sodium and $\Gamma_{\rm K} = 0.14 (T/2000 {\rm K})^{-0.7}$ cm$^{-1}$ atm$^{-1}$ for potassium. Using statistical theory, the outer line wings are modeled by a power law, truncated with an exponential Boltzmann factor. The transition between those two regimes, called location of detuning, happens at the frequency shift $\delta\nu$ from the line center. For sodium we use $\delta\nu_{\rm Na} = 30(T/500 {\rm K})^{0.6}$ cm$^{-1}$ and for potassium $\delta\nu_{\rm K} = 20(T/500 {\rm K})^{0.6}$ cm$^{-1}$, where $T$ is the temperature \citep{unsold55, nefedov99, burrows00, Iro05}. The total line profile at frequency $\nu$ is given by
\begin{equation}
\Phi(\nu) =
\begin{cases}
\Phi_{\rm V}(\nu -  \nu_0) & \text{if } \abs{\nu - \nu_0} \leq \delta\nu , \\
\Phi_{\rm V}(\delta\nu)\left(\frac{\delta\nu}{\nu-\nu_0}\right)^{3/2} \exp\left[-\frac{h (\nu-\nu_0)}{k_{\rm B} T}\right] & \text{if } \abs{\nu - \nu_0} > \delta\nu ,
\end{cases}
\end{equation}
where $\Phi_{\rm V}$ is the Voigt profile, $\nu_0$ is the line center frequency, $\delta\nu$ is the distance from the detuning points to the line center, and $h$ and $k_{\rm B}$ are the Planck and Boltzmann constants, respectively.

\subsection{Other lines}

The individual impact of the weaker lines is smaller and their far wings play a subdominant role. Hence, they are sufficiently modelled with a classical Voigt profile. 

In the following, we briefly outline the expressions and parameters used in our nominal method to calculate atomic lines. The frequency $\nu$ dependent cross-section is given by $\sigma_\nu = S \cdot \Phi_\nu$ where the line strength $S$ and the profile $\Phi_\nu$ are calculated independently. The line strength can be expressed as
\begin{equation}
S = \frac{\pi e^2 g_i f_{ij}}{m_e c}\frac{e^{-E_i / k_{\rm B} T}}{Q(T)} \left( 1 - e^{-\Delta E / k_{\rm B} T} \right)
\end{equation}
where $i$ and $j$ denote the lower and upper quantum states, $e$ is the electron charge, $g$ is the statistical weight, $f$ is the oscillator strength, $m_e$ is the electron mass, $c$ is the speed of light, $E$ is the energy, $Q$ is the partition function and the energy difference $\Delta E = E_j - E_i$. Also, $\Delta E = h \nu$ with $\nu$ being the frequency of the absorbed or emitted photon. 

The partition function is defined as
\begin{equation}
Q(T) = \sum_k q_k e^{- E_k / k_{\rm B} T} ,
\end{equation}
where the sum leads over all allowed quantum states $k$. It can be either calculated directly from the energy levels, approximated by fitting functions, or read from pre-calculated tables.

As the ExoMol or the HITRAN databases do not feature atomic species, we have taken the necessary spectral data from the Kurucz line list\footnote{The Kurucz data are found at \url{http://kurucz.harvard.edu/}. We also recommend the NIST spectral database which has a user-friendly online tool (\url{https://physics.nist.gov/PhysRefData/ASD/lines_form.html}).}. With $g_i$, $f_{ij}$, $E_i$, $E_j$ provided in the data base, the line strength is fully determined. Note, that it is common to tabulate $\log_{10}(g_i f_{ij})$ instead of the individual $g_i$ and $f_{ij}$ quantities.

The Voigt profile $\Phi_{\rm V}$ at frequency $\nu$ is defined as the convolution between the Doppler $\Phi_{\rm D}$ and the Lorentz profiles $\Phi_{\rm L}$. Through
\begin{equation}
\label{eq:voigt}
\Phi_{\rm V}(\nu; \Gamma_{\rm D}, \Gamma_{\rm L}) =  \int^\infty_{-\infty} \Phi_{\rm L} (\nu - \nu '; \Gamma_{\rm L}) \Phi_{\rm D} (\nu'; \Gamma_{\rm D}) ~d\nu ' ,
\end{equation}
the Voigt profile depends on the Doppler and Lorentz HWHMs $\Gamma_{\rm D}$ and $\Gamma_{\rm L}$. For thermal velocities, the Doppler HWHM is given by
\begin{equation}
\Gamma_{\rm D} = \frac{\nu}{c} \sqrt{\frac{2 \ln 2 k_{\rm B} T}{m}},
\end{equation}
where $m$ is the particle mass. The Lorentz HWHM is a combination of natural, collisional, Stark and Van der Waals broadening, i.e., $\Gamma_{\rm L} = \Gamma_{\rm nat} + \Gamma_{\rm coll} + \Gamma_{\rm Stark} + \Gamma_{\rm VdW}$. Natural broadening arises due to the limited lifetime of an excited quantum state and the subsequent uncertainty on the energy. Using the Kurucz database we write
\begin{equation}
\Gamma_{\rm nat} = \frac{10^{\Gamma_{\rm R}}}{2\pi},
\end{equation}
where $\Gamma_{\rm R}$ is called the radiative damping coefficient and is a tabulated quantity. Collisional broadening is harder to quantify as it depends on collisions with the surrounding gas and hence is a function of pressure and temperature. At the order-of-magnitude level, we estimate the HWHM due to collisions as \citep{heng15}
\begin{equation}
\Gamma_{\rm coll} = P \sigma_{\rm H_2} \sqrt{\frac{2}{m k_{\rm B} T}},
\end{equation}
where $P$ is the pressure and $\sigma_{\rm H_2} \sim 10^{-15}$ cm$^2$ is the cross-section of molecular hydrogen, which we assume as the dominating collision partner. In this study we neglect the remaining two broadening mechanisms due to Stark line splitting and Van der Waals perturbations. We refer to \cite{sharp07} for an excellent discussion on line profiles and the calculation of opacities.

On a final note, there is no closed analytical solution to calculate the Voigt integral (eq.~\ref{eq:voigt}). However, the problem can be recast in the form of the real part of the complex Faddeeva function with sufficient accuracy \citep{zaghloul12}. The Faddeeva function is included e.g., in the \texttt{scipy} package for \texttt{python}. 

\section{Removal of Gas Species due to Condensation}
\label{app:removal}

As described in Sect.~\ref{sec:chem} we employ a two-step determination of the chemical abundances for some major atmospheric species. There are two options. First, the gas species itself may condensate out. Hence, for each point in temperature $T$ and pressure $P$, after the equilibrium abundance has been determined, we check whether it located above or below the condensation curve for this species. If it is located below the curve, we remove it from our atmospheric catalog. Such a procedure is done for H$_2$O, TiO and VO. 

Apart from condensing out itself, the gas species may also participate in the formation of dust grains. If the species is further the limiting constituent in the grain formation process it is expected to be removed from the gaseous phase. We apply an exponential decline, starting at the dust grain condensation curve and fit then the gradient of the decline as shown in \cite{sharp07}, which results in
\begin{equation}
f(T,P) = f_{\rm EQ}(T_{\rm cond},P) \cdot 10^{-2.2(T_{\rm cond} - T)} ,
\end{equation}
where $f$ is the volume mixing ratio, $f_{\rm EQ}$ is the one obtained from equilibrium chemistry and $T_{\rm cond}$ is the condensation temperature at pressure $P$. We use this approach to remove the following gaseous species: SiO due to its participation in MgSiO$_3$ formation, Na for being in Na$_2$S and K due to formation of KCl. 

The calculation of the condensation curves themselves is based on the condensate data summarized in \citet{woitke18}. We use $\texttt{FastChem}$ to calculate the gas phase composition in chemical equilibrium for a grid of temperature pressure points, assuming solar elemental abundances. As a second step, we then evaluate the potential condensation of various solid and liquid species. In cases, where tabulated saturation vapor pressure data is available, we use this data to check if the partial pressure of a species in the gas phase exceeds its saturation vapor pressure at each T-P point. For other condensates, the Gibbs free energy of formation $\Delta G_f^{\minuso}$ is used to determine their potential presence. For a given condensate, e.g., \ce{A_iB_jC_k}, we calculate the pseudo-activities $a^c$, given by 
\begin{equation}
  a^c = P_{\rm A}^i P_{\rm B}^j P_{\rm C}^k e^{-\Delta G_f^{\minuso} /RT} \ ,
\end{equation}
where $P_{\rm A}$ is the partial pressure of the condensate's constituent A in atomic form within the gas phase. A condensed species is stable if $a^c > 1$. At each T-P point we, therefore, check this condition to evaluate the potential presence of the corresponding solid or liquid.
Note that the approach of using condensation curves treats each condensate individually and independently. In reality, different dust species compete for the various gas phase constituents, such that the resulting, stable condensates might differ from the ones expected by the condensation curve approximation. An overview on how to treat a multi-component gas-solid/liquid system in equilibrium can be found in \citet{gail13}.

\begin{figure}
\begin{center}
\begin{minipage}[h]{0.6\textwidth}
\includegraphics[width=\textwidth]{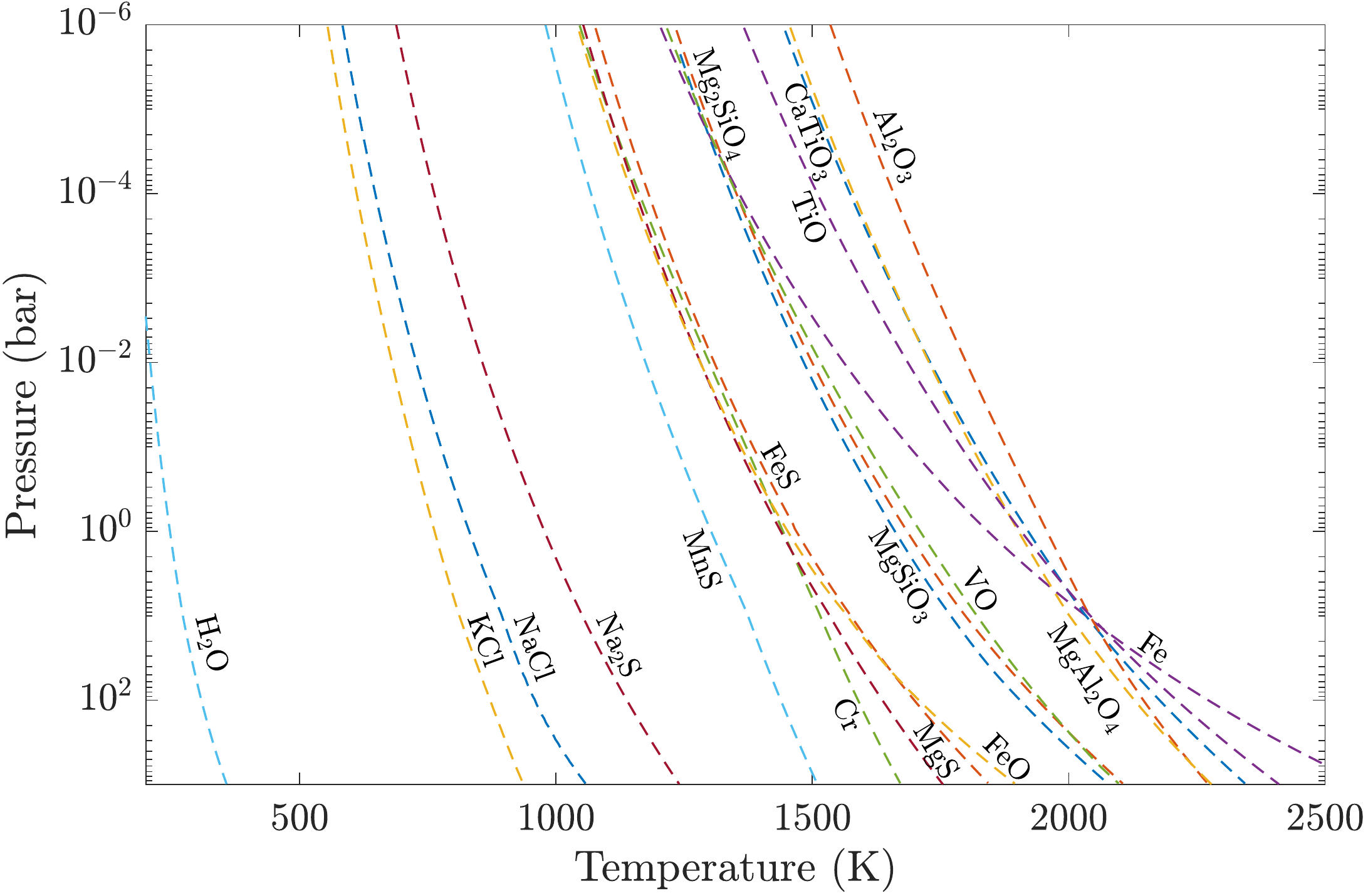}
\end{minipage}
\vspace{-0.0cm}
\caption{Sample of condensation curves for atmospheric gas species and dust grains. The shown curves assume solar elemental abundances and equilibrium chemistry.}
\label{fig:stability}
\end{center}
\end{figure}

A sample of condensation curves are shown in Fig.~\ref{fig:stability}. For completeness, we show not only the species used in this study but also other possible important condensates, which we plan to consider in future works.

\section{Smoothing of Temperatures}
\label{app:smoothing}

With the extension of the opacity list and the inclusion of strong shortwave absorbers, like Na \& K or TiO \& VO, it is possible that discontinuities in the converged radiative equilibrium temperatures emerge. These discontinuities are not a mistake in the numerical code, but an inherent flaw of a discrete one-dimensional grid treatment. Due to the interplay between chemical abundances, opacities and their intricate dependency on temperature and pressure, neighboring layers may be left with large differences in atmospheric abundances even after a converged radiative equilibrium solution has been found. This sometimes leads to jumps in the net flux and in the converged temperatures. However, we expect atmospheres to smooth out chemical abundances across adjacent layers due to small scale dynamics. Hence, we include a smoothing term into our formalism to prevent unphysically large temperature jumps between neighboring grid cells.

For a given layer $i$, we write the arithmetic average of the temperatures of the adjacent layers, $T_{{\rm average}, i}$, as
\begin{equation}
T_{{\rm average}, i} = \frac{T_{i-1} + T_{i+1}}{2} .
\end{equation}
In order to keep the temperature in layer $i$, $T_i$, reasonably close to this average, we set the smoothing term $\mathcal{F}_{{\rm smooth}, i}$ to
\begin{equation}
\mathcal{F}_{{\rm smooth}, i} = \alpha (T_{{\rm average}, i} - T_i)^\beta ,
\end{equation}
with the constant $\alpha$ = 1 erg s$^{-1}$ cm$^{-2}$ K$^{-\beta}$. The exponent $\beta$ sets the strength of the smoothing. The difficulty is to find a value of $\beta$ so that the smoothing is effective, but is not so strong to drive the temperatures too much out of radiative equilibrium. After some trial-and-error, we have found $\beta$ = 7 to offer satisfactory behavior (not shown). Now the smoothing term is added to the regular net flux divergence term $\Delta\mathcal{F}_{ i, -}$,
\begin{equation}
\Delta\mathcal{F}_{i, -}^{\rm new} = \mathcal{F}_{{\rm smooth}, i} + \Delta\mathcal{F}_{ i, -},
\end{equation}
forming the new smoothed flux divergence $\Delta\mathcal{F}_{i, -}^{\rm new}$, which is used in
\begin{equation}
\label{eq:iter}
\Delta T_i = - \frac{1}{\rho_i c_{{\rm p}, i}} \frac{\Delta\mathcal{F}_{i, -}^{\rm new}}{\Delta z_i} \Delta t
\end{equation}
to obtain the temperature step $\Delta T_i$, where $\Delta z_i$ is the layer height, $\rho_i$ is the mass density, $c_{{\rm p}, i}$ is the specific heat capacity and $\Delta t$ is the time step. How eq.~(\ref{eq:iter}) is used in the context of the temperature iteration is explained in detail in \cite{malik17}.

\begin{figure}
\begin{center}
\begin{minipage}[h]{0.5\textwidth}
\includegraphics[width=\textwidth]{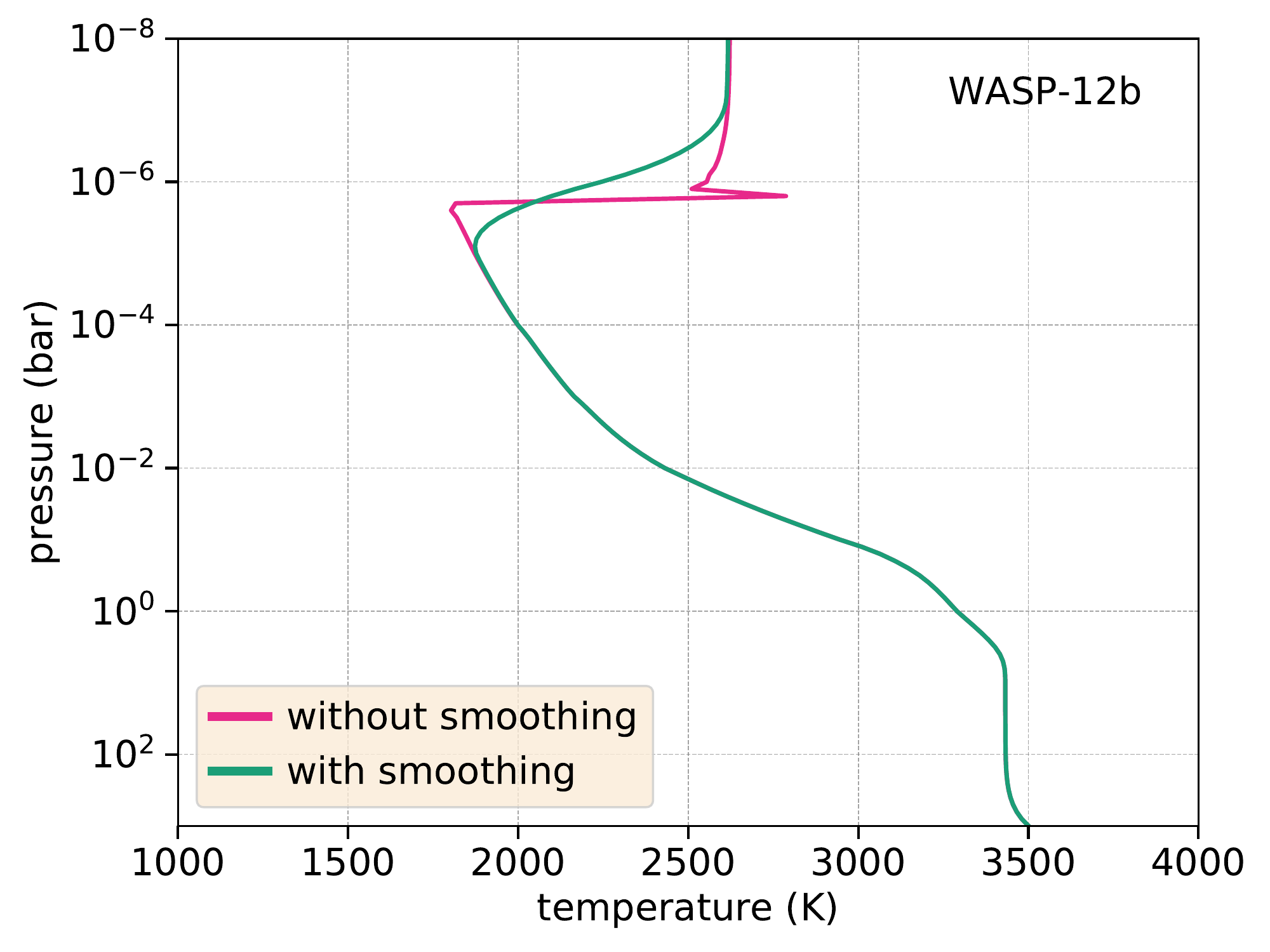}
\end{minipage}
\vspace{-0.1cm}
\caption{Dayside temperature profile of WASP-12b in radiative-convective equilibrium with and without the smoothing term. The Na \& K opacity causes a discontinuity in the temperature profile high up in the atmosphere by absorbing a significant fraction of the incoming stellar radiation.}
\label{fig:smoothing}
\end{center}
\end{figure}

Fig.~\ref{fig:smoothing} shows the dayside temperature profile of WASP-12b in radiative-convective equilibrium with and without the smoothing term. The Na \& K opacity absorbs a significant fraction of the stellar flux high up in the atmosphere and causes a discontinuity in the temperature profile. We find that the inclusion of the smoothing term has no impact on the emitted flux of the planet (not shown), because the discontinuity typically occurs in the optically thin region of the atmosphere. 

\section{Global energy balance for the convection scheme}
\label{app:conv}

As pointed out in Sect.~\ref{sec:conv}, global energy balance is not automatically achieved when using the vanilla convective adjustment scheme. In order to satisfy the global energy balance, the radiative net flux requires to be the same at BOA and TOA. However, due to the convective zones, the radiative layers in the upper atmosphere do not ``feel'' the interior heat flux directly. Although they perfectly adjust to local radiative balance, it proves difficult to establish radiative balance globally. In other words, in theory we demand that 
\begin{equation}
\mathcal{F}_{-} = \mathcal{F}_{\rm conv, -} + \mathcal{F}_{\rm rad, -} = {\rm const}
\end{equation}
holds throughout the atmosphere, where $\mathcal{F}_{-}$ is the net flux, $\mathcal{F}_{\rm conv, -}$ is the net convective flux and $\mathcal{F}_{\rm rad, -}$ is the net radiative flux.  However, as the radiative heating and cooling only reacts to $\mathcal{F}_{\rm rad, -}$ and misses the contribution of $\mathcal{F}_{\rm conv, -}$, it tries to adjust $\mathcal{F}_{-}$ to a too low constant value. The result is that the model deceptively converges to a  radiative-convective solution, but wrongly exhibits a TOA net flux lower than the BOA net flux. This dilemma can be fixed by modifying the temperatures in the convective region as
\begin{equation}
T^\prime (P) = \Theta \left(\frac{P_0}{P}\right)^{-\kappa} f,
\end{equation}
which is eq. (\ref{eq:tprime}), rewritten with an additional factor $f$. This factor fine-tunes the temperatures up or down if $\mathcal{F}_{-}({\rm TOA})$ is too low or too high. For planets with a non-vanishing interior heat flux, $f$ reads
\begin{equation}
\label{eq:tweak1}
f = \left[\frac{X \cdot \mathcal{F}_{\rm rad, -}({\rm BOA})}{(X - 1) \cdot \mathcal{F}_{\rm rad, -}({\rm BOA}) + \mathcal{F}_{\rm rad, -}({\rm TOA})}\right]^\gamma ,
\end{equation}
where $\gamma$ and $X$ are dimensionless numbers. Once global equilibrium is reached, , i.e., $\mathcal{F}_{\rm rad, -}({\rm TOA}) = \mathcal{F}_{\rm rad, -}({\rm BOA})$, $f$ goes to unity and we recover the conventional eq. (\ref{eq:tprime}). The most straightforward version of eq. (\ref{eq:tweak1}) comes with $X = 1$, but if at some point during the iteration $\mathcal{F}_{\rm rad, -}({\rm TOA}) < 0$ occurs, the expression breaks down. Hence, choosing $X \gg 1$ provides numerical stability. In terms of $\gamma$ we have found through tweaking that a value $\sim 10^{-2}$ suffices to do the trick. When modeling planets with a vanishing interior heat flux, e.g., small rocky planets, eq. (\ref{eq:tweak1}) is not applicable as it returns zero. In this case we write $f$ as
\begin{equation}
f = \left[\frac{\mathcal{F}_{\rm rad, \downarrow}({\rm TOA})}{\mathcal{F}_{\rm rad, \uparrow}({\rm TOA})}\right]^\gamma ,
\end{equation}
where $\mathcal{F}_{\rm rad, \downarrow}({\rm TOA}))$ and $\mathcal{F}_{\rm rad, \uparrow}({\rm TOA})$ are the total downward and upward fluxes at TOA. 

Lastly, we assume global energy equilibrium if 
\begin{equation}
 \frac{\abs{\mathcal{F}_{\rm rad, -}({\rm BOA}) - \mathcal{F}_{\rm rad, -}({\rm TOA})}}{\mathcal{F}_{\rm rad, -}({\rm BOA})} < \epsilon
\end{equation}
in the case with $\mathcal{F}_{\rm rad, -}({\rm BOA}) \neq 0$ and if
\begin{equation}
 \frac{\abs{\mathcal{F}_{\rm rad, -}({\rm TOA})}}{\mathcal{F}_{\rm rad, \downarrow}({\rm TOA})} < \epsilon
\end{equation}
in the case with $\mathcal{F}_{\rm rad, -}({\rm BOA}) = 0$. We use $\epsilon = 10^{-4}$ in this work.

\section{Expressing the heat capacity in terms of the adiabatic coefficient}
\label{app:kappa}

Here we proceed to express the constant-pressure heat capacity $c_{\rm P}$ in a form similar to eq.~(2.11) of \citet{pierrehumbert10},
$c_{\rm P}=k_{\rm B}/(\mu m_{\textrm{u}}\kappa)$, but valid also at dissociation and ionization. Let us start with the usual expression
\begin{equation}
\label{eq:cP}
 c_{\rm P} = T \left(\frac{\partial S}{\partial T}\right)_P.
\end{equation}
By the triple-product rule,
\begin{equation}
\begin{split}
 \left(\frac{\partial S}{\partial T}\right)_P &= - \left(\frac{\partial P}{\partial T}\right)_S \left(\frac{\partial S}{\partial P}\right)_T \\
 &= - \left(\frac{\partial \ln P}{\partial \ln T}\right)_S \frac{P}{T} \left(\frac{\partial S}{\partial P}\right)_T. \label{eq:line two of dS/dT_P}
\end{split}
\end{equation}
The first factor in eq.~(\ref{eq:line two of dS/dT_P}) being nothing else than $1/\kappa$ by its definition (eq.~\ref{eq:kappadef}), inserting eq.~(\ref{eq:line two of dS/dT_P}) into eq.~(\ref{eq:cP}) leads to
\begin{equation}
\label{eq:cP App}
 c_{\rm P} = \frac{1}{\kappa} \left(-\frac{\partial S}{\partial \ln P}\right)_T 
= \frac{k_{\rm B}}{m_{\mathrm{u}}\kappa}\left(-\frac{\partial \tilde{S}}{\partial \ln P}\right)_T,
\end{equation}
where $\tilde{S}\equiv S/(k_{\rm B}/m_{\mathrm{u}})$ is the specific entropy in natural (dimensionless) units.
Equation~(\ref{eq:cP App}) is the same as eq.~(\ref{eq:cp}).

\clearpage

\FloatBarrier
\vspace{-0.0cm}
\bibliographystyle{apj}

\bibliography{mybib}

\end{document}